\documentclass[aps,prb,reprint,
amsmath,amssymb,eqsecnum]{revtex4-2}
\usepackage{amsmath,amssymb,color,braket}
\usepackage{bm}
\usepackage{bbm}
\usepackage{graphicx}
\usepackage{psfrag}
\usepackage{epsfig}
\usepackage{hyperref}
\newcommand{\bd}{\bm}

\begin{document}

\title{Functional renormalization group 
without functional integrals:
implementing Hilbert space projections
for strongly correlated electrons via 
Hubbard X-operators}

\author{Andreas R\"{u}ckriegel, Jonas Arnold, R\"{u}diger 
Kr\"{a}mer, and Peter Kopietz}
  
\affiliation{Institut f\"{u}r Theoretische Physik, Universit\"{a}t
  Frankfurt,  Max-von-Laue Stra{\ss}e 1, 60438 Frankfurt, Germany}

\date{May 11, 2023}

 \begin{abstract}

Exact functional renormalization group (FRG) 
flow equations for quantum systems 
can be  derived directly within 
an operator formalism 
without using functional integrals.    
This simple insight 
opens new possibilities 
for applying  FRG methods 
to models for strongly correlated electrons 
with projected Hilbert spaces, 
such as
quantum spin models, 
the $t$-$J$ model, 
or the Hubbard model 
at infinite on-site repulsion. 
By representing these models 
in terms of Hubbard X-operators,
we derive exact flow equations 
for the time-ordered correlation functions 
of the X-operators (X-FRG),
which allow us to calculate 
the electronic correlation functions 
in the projected Hilbert space of these models. 
The Hubbard-I approximation 
for the single-particle Green function 
of the Hubbard model
is recovered from a trivial truncation 
of the flow equations where
the two-point vertex
is approximated by its atomic limit.
We use our approach to calculate 
the quasi-particle residue and damping
in the ``hidden Fermi liquid'' state 
of the Hubbard model 
at infinite on-site repulsion 
where the Hamiltonian consists 
only of the projected kinetic energy.

\end{abstract}

\maketitle

\section{Introduction}

The functional renormalization group (FRG) has been used  
by a growing number of researchers to study 
a variety of problems in field theory and statistical physics, 
ranging from classical and quantum critical phenomena 
over correlated lattice models 
for interacting fermions or bosons, 
to quantum chromodynamics and quantum gravity, 
see Refs.~[\onlinecite{Berges02,Pawlowski07,
Kopietz10,Metzner12,Dupuis21}] for reviews.
Usually, 
derivations of  FRG flow equations  
use functional integral representations 
of suitably defined generating functionals
\cite{Wetterich93,Berges02,Pawlowski07,
Kopietz10,Metzner12,Dupuis21}.
However, functional integrals are really
not necessary for the derivation 
of FRG flow equations 
for quantum systems \cite{Pawlowski07,Krieg19}. 
It seems that this simple insight has not been 
sufficiently appreciated by researchers 
working in the field.
In this work 
we will show how to derive
FRG flow equations for quantum systems 
directly within a Hamiltonian formalism, 
avoiding the representation 
of correlation functions 
in terms of functional integrals. 
This should be helpful for teaching FRG methods 
to students who are not familiar 
with functional integrals. 
More importantly, 
this insight leads to new possibilities 
of applying FRG methods 
to quantum systems with projected Hilbert spaces 
or to systems whose Hamiltonian is defined 
in terms of a set of operators 
satisfying non-canonical commutation relations. 
An important example are quantum spin systems 
for which exact flow equations for the
spin correlation functions 
can be derived directly from the
representation of the 
corresponding generating functional 
as a time-ordered exponential~\cite{Krieg19}. 
Recent applications of our spin-FRG approach 
can be found in
Refs.~[\onlinecite{Krieg19,Tarasevych18,
Goll19,Goll20,Tarasevych21,Tarasevych22,
Rueckriegel22,Tarasevych22b}].

In this work 
we will show that this strategy can also be used 
to derive formally exact FRG flow equations 
for fermionic lattice models 
with projected Hilbert spaces 
such as  the $t$-$J$ model or the Hubbard model 
with infinitely strong on-site repulsion. 
Therefore, 
we express these models in terms of so-called 
Hubbard X-operators \cite{Ovchinnikov04,
Izyumov88,Izyumov05} 
which satisfy non-canonical commutation relations.  
For the $t$-$J$ model, 
this representation is convenient to
implement the exclusion of states 
with doubly occupied lattice sites 
from the relevant Hilbert space.
For the Hubbard model, 
this offers a non-perturbative route 
to study the physics 
at strong coupling \cite{Ovchinnikov04}.
Our X-FRG approach generalizes the 
spin-FRG~\cite{Krieg19,Tarasevych18,
Goll19,Goll20,Tarasevych21,Tarasevych22,
Rueckriegel22,Tarasevych22b} 
to fermionic lattice models 
and is based on the observation 
that within a Hamiltonian formalism 
we can derive exact flow equations 
for generating functionals 
even if the Hamiltonian is expressed 
in terms of operators satisfying 
non-canonical commutation relations.

The rest of this work is organized as follows: 
In Sec.~\ref{sec:without}, 
we show how the 
Wetterich equation \cite{Wetterich93} 
for canonical fermions and bosons can be derived 
without using functional integrals.
Our spin-FRG approach~\cite{Krieg19,
Tarasevych18,Goll19,Goll20,Tarasevych21,
Tarasevych22,Rueckriegel22,Tarasevych22b} 
and its extension to 
strongly correlated electronic systems 
using Hubbard X-operators 
presented in this work 
relies on a similar strategy. 
In Sec.~\ref{sec:models}, 
we define the 
Hubbard X-operators~\cite{Ovchinnikov04,
Izyumov88,Izyumov05} 
and express the Hubbard and $t$-$J$ models 
in terms of these operators. 
In Sec.~\ref{sec:XFRG} we use the method 
outlined in Sec.~\ref{sec:without}
to derive formally exact 
generalized Wetterich equations 
for the Hubbard and the $t$-$J$ models 
in X-operator representation.
In Sec.~\ref{sec:applications}, 
we present the first two applications of our method:  
First of all, we show how the so-called 
Hubbard-I approximation 
for the single-particle Green function 
of the Hubbard model emerges 
from the simplest possible truncation 
of our X-FRG flow equations. 
Moreover, 
from a level-1 truncation 
of the X-FRG flow equations, 
where the four-point vertex
is approximated by its initial value 
corresponding to vanishing hopping, 
we calculate the quasi-particle residue 
and the quasi-particle damping 
of the Hubbard model at 
infinite on-site repulsion.
Finally, in Sec.~\ref{sec:summary}, 
we summarize our results 
and point out possible future applications 
of our method.
In  three appendices, 
we present technical details 
of the calculation of 
time-ordered correlation functions 
of the X-operators in the atomic Hubbard model, 
where the hopping is switched off 
and the lattice sites are decoupled.

\section{FRG without functional integrals}
 \label{sec:without}

Before formulating the X-FRG approach, 
let us carefully explain 
why functional integrals are not necessary
for the derivation of exact FRG flow equations 
for canonical bosons or fermions.
In this section, 
we therefore 
consider a general second-quantized Hamiltonian
of a bosonic or fermionic many-body system,
\begin{equation}
{\cal{H}} =
\sum_{\bd{k}} \epsilon ( {\bd{k}} ) 
a^{\dagger}_{\bd{k}} a_{\bd{k}} 
+{\cal{H}}_{\rm int} ,
\end{equation}
where
$a_{\bd{k}}$ and $a^{\dagger}_{\bd{k}}$ 
are canonical bosonic or fermionic
annihilation and creation operators, 
and ${\cal{H}}_{\rm int}$ is 
an arbitrary many-body interaction.
The generating functional
${\cal{G}} [ \bar{j} , j ]$
of the connected imaginary-time 
ordered correlation functions 
is defined in terms of the following
trace of a time-ordered exponential 
over Fock space:
\begin{equation}
e^{ {\cal{G}} [ \bar{j} , j ] } = 
{\rm Tr}  \left\{  
e^{ - \beta {\tilde{\cal{H}}} }
{\cal{T}} 
e^{ 
\int_0^{\beta} d \tau  \sum_{\bd{k}} \left[ 
\bar{j}_{\bd{k}} ( \tau ) 
a^{\dagger}_{\bd{k}} (\tau ) + 
a^{\dagger}_{\bd{k}} ( \tau ) 
j_{\bd{k}} ( \tau ) \right]  
}
\right\} ,
\label{eq:Ggendef}
\end{equation}
where $\beta = 1/T$ is the inverse
temperature and ${\cal{T}}$ represents 
time-ordering in imaginary time.
The sources 
$j_{\bd{k}} ( \tau )$ and 
$\bar{j}_{\bd{k} } ( \tau )$ 
are complex numbers for bosons 
and Grassmann variables for fermions, 
and
\begin{equation}
{\tilde{\cal{H}}} = 
{\cal{H}} - \mu \sum_{\bd{k}} 
a^{\dagger}_{\bd{k}} a_{\bd{k}}
=  
\sum_{\bd{k}} \left[ 
\epsilon ( {\bd{k}} ) - \mu 
\right] 
a^{\dagger}_{\bd{k}} a_{\bd{k}} 
+ {\cal{H}}_{\rm int}
\end{equation}
is the grand canonical Hamiltonian, 
where $\mu$ is the chemical potential.
The time evolution of the operators 
$a_{\bd{k}} ( \tau )$ and
$ a^{\dagger}_{\bd{k}} ( \tau )$ 
is in the imaginary time Heisenberg picture,
i.e.,
$a_{\bd{k}} ( \tau ) = 
e^{ \tilde{\cal{H}} \tau } a_{\bd{k}}  
e^{ - \tilde{\cal{H}} \tau}$ 
and
$a^{\dagger}_{\bd{k}} ( \tau ) = 
e^{ \tilde{\cal{H}} \tau } a^{\dagger}_{\bd{k}} 
e^{ - \tilde{\cal{H}} \tau}$. 

For our purpose, it is more convenient 
to rewrite the 
generating functional~\eqref{eq:Ggendef} 
in the interaction picture \cite{Shankar94}.  
Therefore,
we express our grand canonical 
Hamiltonian $\tilde{\cal{H}}$ 
as a sum of two terms:
\begin{equation}
\tilde{\cal{H}} = 
{\cal{H}}_1 + {\cal{H}}_2 . 
\label{eq:H01}
\end{equation}
This decomposition is not unique 
and below we will use this freedom  
to derive FRG flow equations. 
In the interaction picture 
with respect to ${\cal{H}}_1$
the generating functional~\eqref{eq:Ggendef} 
can then be written as
\begin{align}
e^{{ \cal{G}} [ \bar{j} , j ] }
& = 
{\rm Tr} \biggl\{  
e^{ - \beta {\cal{H}}_1  } {\cal{T}} 
\Bigl[ 
e^{ - \int_0^{\beta} d \tau  
{\cal{H}}_2 ( \tau ) }
\nonumber\\
& \phantom{aaaaa}
\times e^{
\int_0^{\beta} d \tau \sum_{\bd{k}} 
\left[ 
\bar{j}_{\bd{k}} ( \tau )  
a_{\bd{k}} ( \tau ) + 
a^{\dagger}_{\bd{k}} ( \tau ) 
j_{\bd{k}} ( \tau ) 
\right] 
} 
\Bigr] \biggr\},
\label{eq:Ggeninter}
\end{align}
where now the time-dependence 
of all operators on the right-hand side 
is generated by ${\cal{H}}_1$, 
i.e.,
\begin{subequations}
\begin{align}
{\cal{H}}_2 ( \tau ) & =  
e^{{\cal{H}}_1 \tau } {\cal{H}}_2  
e^{ - {\cal{H}}_1 \tau } ,
\\
a_{\bd{k}} ( \tau ) & =  
e^{{\cal{H}}_1 \tau } a_{\bd{k}}  
e^{ - {\cal{H}}_1 \tau} ,
\\
a^{\dagger}_{\bd{k}} ( \tau ) & =  
e^{ {\cal{H}}_1 \tau } a^{\dagger}_{\bd{k}}  
e^{ - {\cal{H}}_1 \tau} .
\end{align}
\end{subequations}
By taking derivatives 
of the right-hand side 
of Eq.~\eqref{eq:Ggeninter} 
with respect to the sources
$j_{\bd{k}} ( \tau )$ and 
$\bar{j}_{\bd{k}} ( \tau )$ 
and subsequently setting 
the sources equal to zero, 
we generate the usual expressions 
for the imaginary-time ordered 
connected correlation functions 
in the interaction picture. 

To derive formally exact FRG flow equations, 
we add a scale-dependent regulator 
to our Hamiltonian,
\begin{align}
\tilde{{\cal{H}}}_{\Lambda} 
& = 
\tilde{{\cal{H}}} + 
\sum_{\bd{k}} R_{\Lambda} ( \bd{k} )  
a^{\dagger}_{\bd{k}} a_{\bd{k}}
\nonumber\\
& = 
\sum_{\bd{k}} \left[ 
\epsilon ( \bd{k} ) - \mu  
+ R_{\Lambda} ( \bd{k} ) 
\right]
a^{\dagger}_{\bd{k}} a_{\bd{k}} 
+ {\cal{H}}_{\rm int} .
\label{eq:Hregulator}
\end{align}
The regulator  $R_{\Lambda} ( \bd{k} )$ depends 
on a continuous scale parameter $\Lambda$. 
It is chosen such that at
the initial scale $ \Lambda = \Lambda_0 $
the model can be solved in a controlled way,
while at the final scale $ \Lambda = \Lambda_1 $
the regulator vanishes,
$R_{ \Lambda = \Lambda_1 } ( \bd{k} ) =0$,
so that the original system is recovered. 
Using the interaction representation \eqref{eq:Ggeninter} with
$ {\cal{H}}_1 = {\cal{H}}_{\rm int}$ and
${\cal{H}}_2 = \sum_{\bd{k}} [ \epsilon ( \bd{k} ) - \mu  +   R_{\Lambda} ( \bd{k} ) ]
 a^{\dagger}_{\bd{k}} a_{\bd{k}}$ 
we obtain the following representation of the scale-dependent generating functional 
${\cal{G}}_{\Lambda} [ \bar{j} , j ]$
of the connected correlation functions of our deformed theory,
\begin{align}
e^{ {\cal{G}}_{\Lambda} [ \bar{j} , j ] }
& = 
{\rm Tr} \biggl\{  
e^{ - \beta {\cal{H}}_{\rm int}  } 
{\cal{T}} 
\Bigl[ 
e^{ - \int_0^{\beta} d \tau \sum_{\bd{k}} 
\left[ 
\epsilon ( \bd{k} ) - \mu + 
R_{\Lambda} ( \bd{k} ) 
\right] 
a^{\dagger}_{\bd{k}} ( \tau ) 
a_{\bd{k}} ( \tau )  }
\nonumber\\
& \phantom{aaaaa}
\times 
e^{  
\int_0^{\beta} d \tau \sum_{\bd{k}} 
\left[ 
\bar{j}_{\bd{k}} ( \tau )  
a_{\bd{k}} ( \tau ) + 
a^{\dagger}_{\bd{k}} ( \tau ) 
j_{\bd{k}} ( \tau ) 
\right] 
} 
\Bigr] 
\biggr\} .
\label{eq:GgenA}
\end{align}
Alternatively, 
we may choose 
${\cal{H}}_1 = \tilde{\cal{H}}$ and
$ {\cal{H}}_2 = \sum_{\bd{k}} 
R_{\Lambda} ( \bd{k} ) 
a^{\dagger}_{\bd{k}} a_{\bd{k}}$, 
which yields
\begin{align}
e^{ {\cal{G}}_{\Lambda} [ \bar{j} , j ] }
& = {\rm Tr} 
\biggl\{  
e^{ - \beta {\tilde{\cal{H}}} }  
{\cal{T}} 
\Bigl[ 
e^{
- \int_0^{\beta} d \tau   
\sum_{\bd{k}} R_{\Lambda} ( \bd{k} ) 
a^{\dagger}_{\bd{k}} ( \tau ) 
a_{\bd{k}} ( \tau ) 
}
\nonumber\\
& \phantom{aaaaa} 
\times e^{ 
\int_0^{\beta} d \tau \sum_{\bd{k}} 
\left[ 
\bar{j}_{\bd{k}} ( \tau )  
a_{\bd{k}} ( \tau ) + 
a^{\dagger}_{\bd{k}} ( \tau ) 
j_{\bd{k}} ( \tau ) 
\right] 
} 
\Bigr] 
\biggr\} ,
\label{eq:GgenB}
\end{align}
where the time-dependence is 
in the Heisenberg picture with respect to
the regulator-independent
Hamiltonian ${\tilde{\cal{H}}}$.
Using either Eq.~\eqref{eq:GgenA} 
or \eqref{eq:GgenB}, 
we may take the derivative with 
respect to $\Lambda$ of both sides 
and represent the operators 
$a_{\bd{k}} ( \tau )$ and 
$a^{\dagger}_{\bd{k}} ( \tau )$ 
in the time-ordered expectation value 
as derivatives with respect to the sources, 
which can then be pulled out 
of the time-ordering bracket. 
Explicitly, 
for the representation \eqref{eq:GgenB}
the chain of identities is
\begin{widetext}
\begin{subequations}
\label{eq:sourcetrick}
\begin{align}
e^{ {\cal{G}}_{\Lambda} [ \bar{j} , j ] }  \partial_{\Lambda} 
{\cal{G}}_{\Lambda} [ \bar{j} , j ]
& = 
- \int_0^{\beta} d \tau \sum_{\bd{k}} 
\left[ 
\partial_{\Lambda} R_{\Lambda} ( \bd{k} ) 
\right] 
{\rm Tr} 
\biggl\{  
e^{ - \beta {\tilde{\cal{H}}}  }  
{\cal{T}} 
\Bigl[ 
e^{
- \int_0^{\beta} d \tau   
\sum_{\bd{k}} R_{\Lambda} ( \bd{k} ) 
a^{\dagger}_{\bd{k}} ( \tau ) 
a_{\bd{k}} ( \tau )   
}  
\nonumber\\
& \phantom{aaaaaaaaaaaaaaaaaaaaaaaa}
\times
a^{\dagger}_{\bd{k}} ( \tau ) 
a_{\bd{k}} ( \tau )    
e^{ 
\int_0^{\beta} d \tau \sum_{\bd{k}} 
\left[ 
\bar{j}_{\bd{k}} ( \tau )  
a_{\bd{k}} ( \tau ) + 
a^{\dagger}_{\bd{k}} ( \tau ) 
j_{\bd{k}} ( \tau ) 
\right] 
} 
\Bigr] 
\biggr\}
\\
& =  
- \int_0^{\beta} d \tau \sum_{\bd{k}} 
\left[ 
\partial_{\Lambda} R_{\Lambda} ( \bd{k} ) 
\right] 
{\rm Tr} 
\biggl\{  
e^{ - \beta {\tilde{\cal{H}}} }  
{\cal{T}} 
\Bigl[ 
e^{- \int_0^{\beta} d \tau   
\sum_{\bd{k}} R_{\Lambda} ( \bd{k} ) 
a^{\dagger}_{\bd{k}} ( \tau ) 
a_{\bd{k}} ( \tau ) }  
\nonumber\\
& \phantom{aaaaaaaaaaaaaaaaaaaaaaaa} 
\times
\zeta 
\frac{ \delta }{ 
\delta j_{\bd{k}} ( \tau ) } \frac{ \delta }{ 
\delta \bar{j}_{\bd{k}} ( \tau ) }    
e^{ 
\int_0^{\beta} d \tau \sum_{\bd{k}} 
\left[ 
\bar{j}_{\bd{k}} ( \tau )  
a_{\bd{k}} ( \tau ) + 
a^{\dagger}_{\bd{k}} ( \tau ) 
j_{\bd{k}} ( \tau ) 
\right] 
} 
\Bigr] 
\biggr\} \\
& = 
- \zeta 
\int_0^{\beta} d \tau \sum_{\bd{k}} 
\left[ 
\partial_{\Lambda} R_{\Lambda} ( \bd{k} ) 
\right]
\frac{ \delta }{ 
\delta j_{\bd{k}} ( \tau ) } 
\frac{ \delta }{
\delta \bar{j}_{\bd{k}} ( \tau ) }
{\rm Tr} 
\biggl\{  
e^{ - \beta {\tilde{\cal{H}}} }  
{\cal{T}} 
\Bigl[ 
e^{
- \int_0^{\beta} d \tau   
\sum_{\bd{k}} R_{\Lambda} ( \bd{k} ) 
a^{\dagger}_{\bd{k}} ( \tau ) 
a_{\bd{k}} ( \tau ) }  
\nonumber\\
& 
\phantom{aaaaaaaaaaaaaaaaaaaaaaaaaaaaaaaaaaaa\,} 
\times
e^{ 
\int_0^{\beta} d \tau \sum_{\bd{k}} 
\left[ 
\bar{j}_{\bd{k}} ( \tau )  
a_{\bd{k}} ( \tau ) + 
a^{\dagger}_{\bd{k}} ( \tau ) 
j_{\bd{k}} ( \tau ) 
\right] 
} 
\Bigr] 
\biggr\}
\\
& = 
- \zeta 
\int_0^{\beta} d \tau \sum_{\bd{k}} 
\left[ 
\partial_{\Lambda} R_{\Lambda} ( \bd{k} ) 
\right] 
 \frac{ \delta }{
 \delta j_{\bd{k}} ( \tau )} 
 \frac{ \delta }{
 \delta \bar{j}_{\bd{k}} ( \tau ) } 
 e^{ {\cal{G}}_{\Lambda} [ \bar{j} , j ] } ,
\end{align}
\end{subequations}
where $\zeta =1$ for bosons 
and $\zeta =-1$ for fermions.
Carrying out the functional derivatives, 
we obtain the following exact flow equation 
for the generating functional 
of connected correlation functions:
\begin{equation}
\partial_{\Lambda} 
{\cal{G}}_{\Lambda} [ \bar{j} , j ] = 
- \zeta \int_0^{\beta} d \tau \sum_{\bd{k}} 
\left[ 
\partial_{\Lambda} R_{\Lambda} ( \bd{k} ) 
\right] 
\left\{ 
\frac{ 
\delta^2 {\cal{G}}_{\Lambda} [ \bar{j} , j ] 
}{
\delta j_{\bd{k}} ( \tau ) 
\delta \bar{j}_{\bd{k}} ( \tau ) 
}
+     
\frac{ 
\delta {\cal{G}}_{\Lambda} [ \bar{j} , j ] 
}{
\delta j_{\bd{k}} ( \tau ) 
}
\frac{ 
\delta {\cal{G}}_{\Lambda} [ \bar{j} , j ] 
}{
\delta \bar{j}_{\bd{k}} ( \tau ) 
} 
\right\} .
\label{eq:Gflowfinal}
\end{equation}
The fermionic version ($\zeta =-1$) 
of this flow equation has already been derived 
in Ref.~[\onlinecite{Tarasevych18}]. 
To derive the corresponding 
Wetterich equation \cite{Wetterich93,Kopietz10,
Tarasevych18}, 
we introduce the generating functional 
of the cutoff-dependent irreducible vertices 
via the usual subtracted Legendre transformation,
\begin{equation}
\Gamma_{\Lambda} [ \bar{\psi} , \psi ] =
\int_0^{\beta} d \tau \sum_{\bd{k}}
\left[ 
\bar{j}_{\bd{k}} ( \tau ) 
\psi_{\bd{k}} ( \tau ) + 
\bar{\psi}_{\bd{k}} ( \tau ) 
j_{\bd{k}} ( \tau ) 
\right]
- {\cal{G}}_{\Lambda} [ \bar{j} , j ] 
- \int_0^{\beta} d \tau \sum_{\bd{k}} 
R_{\Lambda} ( \bd{k} )
\bar{\psi}_{\bd{k}} ( \tau ) 
\psi_{\bd{k}} ( \tau ) ,
\label{eq:Legendresubtract}
\end{equation}
where on the right-hand side 
the sources $\bar{j}$ and $j$ 
should be expressed in terms 
of the operator expectation values 
$\psi$ and $\bar{\psi}$ 
by inverting the relations
\begin{equation}
\psi_{\bd{k} } ( \tau ) =  
\braket{ {\cal{T}} a_{\bd{k}} ( \tau ) } =
\frac{ 
\delta {\cal{G}}_{\Lambda} [ \bar{j} , j ] 
}{
\delta \bar{j}_{\bd{k}} ( \tau ) 
} ,
\; \; \;  \; \; \; 
\bar{\psi}_{\bd{k} } ( \tau ) =   
\braket{ 
{\cal{T}} a^{\dagger}_{\bd{k}} ( \tau ) 
} =
\zeta 
\frac{ 
\delta {\cal{G}}_{\Lambda} [ \bar{j} , j ] 
}{
\delta {j}_{\bd{k}} ( \tau ) 
} .
\end{equation}
Here, 
the symbol $ \braket{ {\cal{T}} A ( \tau ) } $ 
denotes the time-ordered expectation value 
in the presence of sources, 
i.e.,
\begin{equation}
\braket{ {\cal{T}} A ( \tau ) } = 
\frac{ 
{\rm Tr} \biggl\{  
e^{ - \beta {\tilde{\cal{H}}} }  
{\cal{T}} 
\Bigl[ 
A ( \tau )
e^{
- \int_0^{\beta} d \tau \sum_{\bd{k}} 
R_{\Lambda} ( \bd{k} ) 
a^{\dagger}_{\bd{k}} ( \tau ) 
a_{\bd{k}} ( \tau ) }
e^{ 
\int_0^{\beta} d \tau \sum_{\bd{k}} 
\left[ 
\bar{j}_{\bd{k}} ( \tau )  
a_{\bd{k}} ( \tau ) + 
a^{\dagger}_{\bd{k}} ( \tau ) 
j_{\bd{k}} ( \tau ) 
\right] 
} 
\Bigr] 
\biggr\}
}{
{\rm Tr} 
\biggl\{  
e^{ - \beta {\tilde{\cal{H}}} }  
{\cal{T}} 
\Bigl[ 
e^{
- \int_0^{\beta} d \tau \sum_{\bd{k}} 
R_{\Lambda} ( \bd{k} ) 
a^{\dagger}_{\bd{k}} ( \tau ) 
a_{\bd{k}} ( \tau ) }
e^{ 
\int_0^{\beta} d \tau \sum_{\bd{k}} 
\left[ 
\bar{j}_{\bd{k}} ( \tau )  
a_{\bd{k}} ( \tau ) + 
a^{\dagger}_{\bd{k}} ( \tau ) 
j_{\bd{k}} ( \tau ) 
\right] 
} 
\Bigr] 
\biggr\} 
} .
\end{equation}
\end{widetext}
Taking the derivative 
$\partial_{\Lambda} 
\Gamma_{\Lambda} [ \bar{\psi } , \psi ]$ 
and substituting for 
$\partial_{\Lambda} 
{\cal{G}}_{\Lambda} [ \bar{j} , j ]$   
the flow equation \eqref{eq:Gflowfinal}, 
we find
\begin{equation}
\partial_{\Lambda} 
\Gamma_{\Lambda} [ \bar{\psi} , \psi ]
= 
\zeta \int_0^{\beta} d \tau \sum_{\bd{k}} 
\left[ 
\partial_{\Lambda} R_{\Lambda} ( \bd{k} ) 
\right]
\frac{ 
\delta^2 {\cal{G}}_{\Lambda} [ \bar{j} , j ] 
}{
\delta j_{\bd{k}} ( \tau ) 
\delta \bar{j}_{\bd{k}} ( \tau ) } .
\label{eq:GammaG}
\end{equation}
To derive the usual form 
of the Wetterich equation \cite{Wetterich93},
we adopt the super-field notation 
developed in Refs.~[\onlinecite{Schuetz05,
Kopietz10}]:
for our model describing 
spinless fermions or bosons 
this amounts to introducing 
a two-component field
\begin{equation}
( \Phi_{\alpha} ) 
= 
\begin{pmatrix}
\Phi_{ (\psi  \bd{k} \tau ) } \\
\Phi_{( \bar{\psi} \bd{k} \tau ) } 
\end{pmatrix} 
=
\begin{pmatrix} 
\psi_{\bd{k}} ( \tau ) \\ 
\bar{\psi}_{\bd{k}} ( \tau )
\end{pmatrix} ,
\end{equation}
where the super-label 
$\alpha = ( p \bd{k} \tau )$ 
denotes all parameters which are necessary
to completely specify the field combination. 
Here, 
$p = \psi , \bar{\psi}$ 
keeps track of the field type.
As shown in Ref.~[\onlinecite{Tarasevych18}], 
the second functional derivative
in Eq.~\eqref{eq:GammaG} can then 
be expressed in terms of 
the second derivative matrix
of $\Gamma_\Lambda [ \bar{\psi} , \psi ]$ 
as follows:
\begin{align}
& 
\frac{ 
\delta^2 {\cal{G}}_{\Lambda} [ \bar{j} , j ] 
}{
\delta j_{\bd{k}} ( \tau ) 
\delta \bar{j}_{\bd{k}} ( \tau ) 
}
\nonumber\\ 
= {} & 
\left[  
\left( 
\frac{\delta}{\delta \Phi} \otimes 
\frac{\delta}{\delta \Phi} \right)^T 
\Gamma_{\Lambda}  [ \bar{\psi} , \psi ] + \mathbf{R}_\Lambda 
\right]^{-1}_{ 
\alpha = (\psi \bd{k} \tau ), 
\alpha^{\prime} =( \bar{\psi} \bd{k} \tau ) 
} ,
\label{eq:GGammaident}
\end{align}
where the matrix elements 
of the second derivative operators
$\frac{\delta}{\delta \Phi} \otimes 
\frac{\delta}{\delta \Phi} $ 
in super-field space is defined by 
\begin{subequations}
\begin{align}
\left[ 
\frac{\delta}{\delta \Phi} \otimes 
\frac{\delta}{\delta \Phi} 
\right]_{ \alpha \alpha' }
& = 
\frac{ \delta }{ \delta \Phi_{\alpha} } 
\frac{\delta}{\delta \Phi_{\alpha'} } ,
\\
\left[ 
\frac{\delta}{\delta \Phi} \otimes 
\frac{\delta}{\delta \Phi} 
\right]^T_{ \alpha \alpha'}
& = 
\frac{\delta}{\delta \Phi_{\alpha'}} 
\frac{\delta}{\delta \Phi_{\alpha}} 
= \zeta  
\frac{\delta}{\delta \Phi_{\alpha}} 
\frac{\delta}{\delta \Phi_{\alpha'}} .
\label{eq:derivtrans}
\end{align}
\end{subequations}
Note that for fermions 
where the $ \Phi_{\alpha} $ anticommute 
the transposition 
in Eq.~\eqref{eq:derivtrans} generates 
an extra factor of $\zeta = -1$.
The regulator matrix $\mathbf{R}_{\Lambda}$ 
in superfield space in Eq.~\eqref{eq:GGammaident}
is defined by writing the regulator term
in Eq.~\eqref{eq:Legendresubtract} 
in (anti)-symmetrized 
superfield notation \cite{Tarasevych18},
\begin{equation}
\int_0^{\beta} d \tau \sum_{\bd{k}} 
R_{\Lambda} ( \bd{k} )
\bar{\psi}_{\bd{k}} ( \tau ) 
\psi_{\bd{k}} ( \tau ) 
=
\frac{1}{2} 
\int_\alpha \int_{\beta} \Phi_{\alpha}
[ \mathbf{R}_{\Lambda} ]_{\alpha \alpha' }
\Phi_{\alpha'} ,
\label{eq:regulatorsym}
\end{equation}
where 
$\int_{\alpha} = 
\sum_{ p = \psi , \bar{\psi} } 
\sum_{\bd{k}}   
\int_0^{\beta} d \tau$ 
denotes summation or integration 
over all components of the superfield label.
Substituting Eq.~\eqref{eq:GGammaident} 
into the flow equation \eqref{eq:GammaG},
we finally obtain the Wetterich equation 
for the average effective action,
$\Gamma_{\Lambda} [ \Phi ] = 
\Gamma_{\Lambda} [ \bar{\psi} , \psi ]$ 
in the form
\begin{align}
& 
\partial_{\Lambda} 
\Gamma_{\Lambda} [ \Phi] = 
\nonumber\\
&
\frac{\zeta}{2}
{\rm Tr} 
\left\{ 
( \partial_{\Lambda} \mathbf{R}_{\Lambda} ) \left[
\left( 
\frac{\delta}{\delta \Phi} \otimes 
\frac{\delta}{\delta \Phi} \right)^T 
\Gamma_{\Lambda} [ \Phi ] + 
\mathbf{R}_{\Lambda} 
\right]^{-1} 
\right\} ,
\label{eq:Wetterich}
\end{align}
where the trace is over all labels 
contained in the superfield label $\alpha$.
Note that nowhere in our derivation of
the Wetterich equation~\eqref{eq:Wetterich} 
we have used functional integrals.
Our derivation of the Wetterich equation 
for quantum spin systems presented 
in Ref.~[\onlinecite{Krieg19}]
relies on the same sequence  of manipulations. 
Evidently, 
this strategy can also be used 
to derive (generalized) Wetterich equations 
for other quantum systems 
whose Hamiltonian is defined 
in terms of operators 
satisfying non-canonical commutation relations.

Since in Eq.~\eqref{eq:Hregulator} 
we have introduced the regulator
$R_{\Lambda} ( \bd{k} )$ 
directly in the Hamiltonian, 
it seems at first sight
that the flexibility 
of introducing a frequency-dependent 
regulator in a functional integral approach 
is lost within our operator formalism. 
This is not true, 
however,
because we can alternatively 
introduce the regulator directly 
in the generating functional
${\cal{G}}_{\Lambda} [ \bar{j} , j ]$ 
of connected correlation functions, 
so that the original generating functional in Eq.~\eqref{eq:GgenB} is replaced 
by the deformed functional
\begin{align}
e^{ {\cal{G}}_{\Lambda} [ \bar{j} , j ] }
& = 
{\rm Tr} 
\biggl\{  
e^{ - \beta {\tilde{\cal{H}}} }  
{\cal{T}} 
\Bigl[ 
e^{
- \int_0^{\beta} d \tau d \tau^{\prime} 
\sum_{\bd{k}} 
R_{\Lambda} ( \bd{k} , \tau - \tau^{\prime} ) 
a^{\dagger}_{\bd{k}} ( \tau ) 
a_{\bd{k}} ( \tau^{\prime} ) 
}
\nonumber\\
& \phantom{aaaaa}  
\times 
e^{ 
\int_0^{\beta} d \tau \sum_{\bd{k}} 
\left[ 
\bar{j}_{\bd{k}} ( \tau )  
a_{\bd{k}} ( \tau ) + 
a^{\dagger}_{\bd{k}} ( \tau ) 
j_{\bd{k}} ( \tau ) 
\right] 
} 
\Bigr] 
\biggr\} ,
\label{eq:GgenC}
\end{align}
where  
$R_{\Lambda} ( \bd{k} , \tau - \tau^{\prime} )$ 
is the imaginary-time Fourier transform
of  a general frequency-dependent regulator  
$ R_{\Lambda} ( \bd{k} , i \omega )$; 
i.e., 
\begin{equation}
R_{\Lambda} ( \bd{k} , \tau - \tau^{\prime} )  
= 
\frac{1}{\beta} \sum_{\omega}
e^{ - i \omega ( \tau - \tau^{\prime} ) } R_{\Lambda} ( \bd{k} , i \omega ) .
\end{equation}

\section{Hubbard X-operators}
\label{sec:models}

\subsection{Hubbard model}
\label{subsec:hubmodel}
The Hubbard model is defined 
by the following second quantized 
lattice Hamiltonian~\cite{Fulde95,Fazekas99}:
\begin{equation}
{\cal{H}} 
= 
\sum_{ij \sigma} t_{ij} 
c^{\dagger}_{i \sigma} 
c_{j \sigma} + 
U \sum_i 
n_{i \uparrow} n_{i \downarrow} ,
\label{eq:Hubbard}
\end{equation}
where $c_{i \sigma}$ is 
the canonical annihilation operator 
of an electron with spin projection 
$\sigma = \uparrow, \downarrow$ 
at lattice site $\bd{r}_i$. 
The corresponding creation operator 
is denoted by $c^{\dagger}_{i \sigma}$ and 
$n_{i \sigma} = 
c^{\dagger}_{i \sigma} c_{i \sigma}$ 
is the occupation number operator. 
For each lattice site,
the Hilbert space is spanned by four states 
$ \ket{ i, 0 } $, 
$ \ket{ i, \uparrow } = 
c^{\dagger}_{i \uparrow} \ket{ i,  0 } $, 
$ \ket{ i, \downarrow } = 
c^{\dagger}_{i\downarrow} \ket{ i, 0 } $, and 
$ \ket{ i,2 } = 
c^{\dagger}_{i\downarrow} c^{\dagger}_{i\uparrow} 
\ket{ i, 0 }  =  
- c^{\dagger}_{i\uparrow} 
c^{\dagger}_{i\downarrow} 
\ket{ i, 0 } $ 
that describe lattice sites 
which are either empty, 
occupied by a single electron 
with spin $\sigma = \uparrow, \downarrow$, 
or occupied by two electrons with opposite spin. 
We denote the states by $ \ket{ i, a } $ 
with  $a \in \{ 0,\uparrow, \downarrow, 2 \}$. 
The X-operators $X_i^{ab}$ act on 
the single-site Fock space
and can be defined 
as follows \cite{Ovchinnikov04,
Izyumov88,Izyumov05}:
\begin{equation}
X_i^{ab} = \ket{ i, a } \bra{ i, b } .
\label{eq:Xdef}
\end{equation}
This set of $16$ operators describes
all possible transitions 
between the four states 
of the single-site Fock space.
The X-operators can be further 
subdivided into a subset of
eight Fermi type operators and 
eight Bose type operators.
The Fermi type X-operators can be expressed 
in terms of an odd number 
of fermionic annihilation or creation operators,
\begin{subequations}
\label{eq:Fermi}
\begin{align}
X_i^{\sigma 0} & = 
c^{\dagger}_{ i \sigma} 
( 1 - n_{ i \bar{\sigma}} ) , 
\\
X_i^{0 \sigma} & = 
(X_i^{\sigma 0 } )^{\dagger} = 
( 1- n_{ i \bar{\sigma} }) c_{ i \sigma} ,
\\
X_i^{2 \sigma} & = 
\sigma c^{\dagger}_{ i \bar{\sigma}} 
n_{ i \sigma} ,
\\
X_i^{\sigma 2 } & = 
(X_i^{ 2 \sigma } )^{\dagger} = 
\sigma n_{ i \sigma} c_{ i \bar{\sigma} } ,
\end{align}
\end{subequations}
where  $\sigma = \uparrow, \downarrow = +,-$
labels the two spin projections and
$\bar{\sigma} = - \sigma$ 
represents the spin projection 
opposite to $\sigma$.
Note that for Fermi type X-operators
$X_{i}^{ ab }$ the number of electrons
in the two states $ \ket{ i,a }$ and 
$\ket{ i , b }$ differs by an odd number.
Equation~\eqref{eq:Fermi} implies 
that the canonical Fermi operators
$c_{ i \sigma}$ and $c^{\dagger}_{i \sigma}$ 
can be expressed in terms of 
the Fermi type X-operators as follows:
\begin{equation}
c_{i \sigma} = 
X_i^{0 \sigma} - 
\sigma X_i^{\bar{\sigma} 2} , 
\; \; \;  \; \; \;
c^{\dagger}_{ i \sigma} =  
X_i^{ \sigma 0} - 
\sigma X_i^{2 \bar{\sigma}} .
\label{eq:cX}
\end{equation}
The remaining eight X-operators $X_i^{ab}$ 
are of the Bose type. 
This means that the states 
$ \ket{ i , a } $ and $ \ket{ i , b } $ 
differ by an even number of electrons;  
these operators can consequently 
be expressed in terms of 
an even number of 
annihilation or creation operators,
\begin{subequations}
\label{eq:Bose}
\begin{align}
X_i^{\sigma \bar{\sigma}} & = 
c^{\dagger}_{ i \sigma} c_{i \bar{\sigma}} , 
\; \; \; \sigma = \uparrow, \downarrow ,
\\
X_i^{ \sigma \sigma} & =   
n_{ i {\sigma} } ( 1 - n_{i \bar{\sigma}} ) ,
\; \; \; \sigma = \uparrow, \downarrow ,
\\
X_i^{00} & = 
( 1 -  n_{ i \uparrow} ) 
( 1 - n_{ i \downarrow}) ,
\\
X_i^{22} & =  
n_{ i \uparrow} 
n_{ i \downarrow} ,
\label{eq:X22}
\\
X_i^{20} & = 
c^{\dagger}_{ i \downarrow } 
c^{\dagger}_{ i \uparrow} ,
\\
X_i^{02} & =   
( X_i^{20} )^{\dagger} =    
c_{ i \uparrow } c_{ i \downarrow} .
\end{align}
\end{subequations}
From the definition \eqref{eq:Xdef} we see 
that the product of two $X$-operators 
associated with the same lattice site
can be expressed in terms of another X-operator,
\begin{equation}
X_i^{ab} X_i^{cd} = \delta_{bc} X_i^{ad} .
\label{eq:product}
\end{equation}
This product rule together with the
anticommutation relations of the 
fermionic creation and annihilation operators
at different sites defines the algebra 
of the X-operators.
Any pair of Fermi type X-operators satisfies 
the anti-commutation relations
\begin{equation}
[ X_i^{ab} , X_j^{cd} ]_{+} = 
\delta_{ij} \left( 
\delta_{bc} X_i^{ad} + 
\delta_{da}  X_i^{cb} \right) .
\label{eq:anticom}
\end{equation}
On the other hand, 
a pair of Bose type X-operators or 
a pair consisting of one Bose type X-operator 
and one Fermi type X-operator
satisfies the commutation relations
\begin{equation}
[ X_i^{ab} , X_j^{cd} ] = 
\delta_{ij} \left( 
\delta_{bc} X_i^{ad}  - 
\delta_{da} X_i^{cb} \right) .
\label{eq:comcom}
\end{equation}
Finally, 
the completeness of the 
four-dimensional single-site Fock space
implies the sum rule
\begin{equation}
\sum_{a} X^{aa}_i = 
\sum_{a} \ket{ i, a } \bra{ i, a } = 1 .
\end{equation}
In principle one could 
require that for different sites 
all X-operators commute \cite{Foglio97}, 
but then the representation \eqref{eq:Fermi}
of the Fermi type X-operators 
in terms of canonical fermions 
would not be valid.
The underlying reason for choosing the 
anticommutation relations \eqref{eq:anticom}
for the Fermi type X-operators
is the physical requirement
that the many-body electronic wave function
has to be totally antisymmetric
with respect to exchange of any two electrons.
The anticommutation relations \eqref{eq:anticom}
together with the 
commutation relations \eqref{eq:comcom} 
define the semi-simple doubly graded 
(supersymmetrical) Lie algebra 
Spl$(1,2)$ \cite{Wiegmann88}.
The representations \eqref{eq:Fermi} 
and \eqref{eq:Bose} associate the 
type assignments of the X-operators 
with the number of canonical fermions: 
Fermi type X-operators 
can be expressed in terms of an odd number 
of canonical Fermi operators, 
while Bose type X-operators can be expressed 
in terms of an even number of 
canonical Fermi operators. 
Note that this assignment does not agree 
with the classification scheme 
given by Tsvelik \cite{Tsvelik95}.

Using Eqs.~\eqref{eq:X22} and \eqref{eq:cX}, 
we can express the
Hubbard Hamiltonian \eqref{eq:Hubbard}
in terms of the X-operators as follows:
\begin{equation}
{\cal{H}} =   
U \sum_i X_i^{22} + 
{\cal{H}}_2 ,
\end{equation}
where the first part 
represents the on-site interaction
and the kinetic energy represented by 
the second part ${\cal{H}}_2$ 
can be written as
\begin{align}
& 
{\cal{H}}_2 =    
\sum_{ij \sigma} t_{ij} 
c^{\dagger}_{i \sigma} c_{j \sigma}
\nonumber\\
& =    
\sum_{ij \sigma} t_{ij}
\left(
X_i^{ \sigma 0} - \sigma X_i^{2 \bar{\sigma}}
\right)
\left( 
X_j^{0 \sigma} - \sigma X_j^{\bar{\sigma} 2}
\right)
\nonumber
\\
& = 
\sum_{ij \sigma} t_{ij} 
\left( 
X_i^{ \sigma 0} X_j^{0 \sigma} +   
X_i^{2 \bar{\sigma}} X_j^{\bar{\sigma} 2} + 
\bar{\sigma} 
X_i^{ \sigma 0} X_j^{\bar{\sigma} 2} +
\bar{\sigma} 
X_i^{2 \bar{\sigma}} X_j^{0 \sigma} 
\right)
\nonumber\\
& = 
\sum_{ij \sigma} t_{ij} 
\bd{\psi}^{\dagger}_{ i \sigma} 
\begin{pmatrix} 
1 & 1 \\ 
1 & 1 
\end{pmatrix} 
\bd{\psi}_{ j  \sigma} 
= 
\sum_{ij \sigma}  
\bd{\psi}^{\dagger}_{ i \sigma} \hat{t}_{ij} 
\bd{\psi}_{ j  \sigma} .
\label{eq:Vdef}
\end{align} 
Here, 
we have introduced the two-flavor 
Fermi-like operators
\begin{equation}
\bd{\psi}_{ i \sigma} =
\begin{pmatrix} 
X_i^{0 \sigma} \\
\bar{\sigma} X_i^{ \bar{\sigma} 2} 
\end{pmatrix} , 
\; \; \;  \; \; \;
\bd{\psi}^{\dagger}_{ i \sigma} = 
\left(
X_i^{\sigma 0 } ,  
\bar{\sigma} X_i^{2 \bar{\sigma}} 
\right) ,
\label{eq:psi2flavor}
\end{equation}
and the $2 \times 2$ hopping matrix 
in flavor space,
\begin{equation}
\hat{t}_{ij} = t_{ij}  
\begin{pmatrix} 
1 & 1 \\ 
1 & 1 
\end{pmatrix} .
\end{equation}
Alternatively, 
introducing the four-component operators
\begin{equation}
\bd{\phi}_{ i \sigma} =
\begin{pmatrix} 
X_i^{ 0 \sigma} \\ 
\bar{\sigma} X_i^{ \bar{\sigma}  2} \\
X_i^{  \sigma 0} \\ 
\bar{\sigma} X_i^{ 2 \bar{\sigma}} 
\end{pmatrix} ,
\end{equation}
the kinetic energy operator can be written in the manifestly antisymmetric form
\begin{equation}
{\cal{H}}_2 
= \frac{1}{2} \sum_{i j \sigma}
\bd{\phi}_{ i  \sigma}^T 
\left( \begin{array}{ cc | cc} 
0 & 0 & -1 & -1 \\
0 & 0 & -1 & -1 \\
\hline
1 & 1 & 0 & 0 \\
1 & 1 & 0 & 0 
\end{array} \right) 
\bd{\phi}_{ j \sigma} .
\end{equation}
Finally, 
the particle number operator 
can be written as 
\begin{equation}
{\cal{N}} = 
\sum_{ i \sigma} n_{ i \sigma} = 
\sum_i \left( 
X_i^{\uparrow \uparrow} + 
X_i^{ \downarrow \downarrow} + 
2 X_i^{22} \right) ,
\end{equation}
so that the grand canonical Hamiltonian 
of the Hubbard model is of the form
$\tilde{\cal{H}} = 
{\cal{H}}_1 + {\cal{H}}_2$ 
as given in Eq.~\eqref{eq:H01}, 
with ${\cal{H}}_2$ given in Eq.~\eqref{eq:Vdef}
and
\begin{equation}
{\cal{H}}_1 = \sum_{i a} \epsilon_a X_i^{aa} ,
\label{eq:H1def}
\end{equation}
with 
$ \epsilon_0 =0$, 
$\epsilon_{\uparrow} = 
\epsilon_{\downarrow} = - \mu$, and
$\epsilon_2 = U - 2 \mu$. 
In the presence of 
an external magnetic field 
the Zeemann energy $h$ 
lifts the degeneracy between 
$\epsilon_{\uparrow} $ and 
$\epsilon_{\downarrow}$ so that
$\epsilon_{\sigma} = - \mu - \sigma h /2$.

\subsection{$t$-$J$ model}
\label{subsec: tjmodel}

The $t$-$J$ model is an effective model 
for electrons hopping on a lattice 
with strong on-site interactions 
such that states with 
doubly occupied sites 
are not accessible.
The Hamiltonian of the $t$-$J$ model acts 
on a projected Hilbert space 
defined by the Fock space spanned by 
the three states 
$ \ket{ i,0 } $, 
$ \ket{ i , \uparrow } = 
c^{\dagger}_{ i \uparrow} \ket{ i , 0 } $, and  
$ \ket{ i , \downarrow } = 
c^{\dagger}_{ i \downarrow} \ket{ i , 0 }$.
The Hamiltonian of the
$t$-$J$ model is \cite{Fulde95,Fazekas99}
\begin{equation}
{\cal{H}}_{tJ} =   
\sum_{ ij \sigma } t_{ij} 
\tilde{c}^{\dagger}_{i \sigma}
\tilde{c}_{j \sigma}
+ \frac{1}{2} \sum_{ ij } J_{ij} 
\left( 
\bd{S}_i \cdot \bd{S}_j - 
\frac{\tilde{n}_i \tilde{n}_j}{4} 
\right) ,
\end{equation}
where
\begin{subequations}
\begin{align}
\tilde{c}^{\dagger}_{ i \sigma} & =    
c^{\dagger}_{i \sigma} 
( 1 - n_{ i  \bar{\sigma} } ) = 
X_i^{\sigma 0} , 
\\
\tilde{c}_{ i \sigma} & = 
( 1 - n_{ i  \bar{\sigma} } )  
c_{i \sigma} = 
X_i^{0 \sigma} 
\label{eq:cprojected}
\end{align}
\end{subequations}
are projected Fermi operators 
acting on the restricted Hilbert space 
of states without doubly occupied lattice sites,
and the spin operators 
and the projected particle number operators 
are defined by
\begin{align}
\bd{S}_i & =  
\frac{1}{2} \left( 
\tilde{c}_{i \uparrow}^{\dagger} , 
\tilde{c}_{ i \downarrow}^{\dagger} 
\right)
\bd{\sigma} 
\begin{pmatrix} 
\tilde{c}_{ i \uparrow} \\ 
\tilde{c}_{i \downarrow} 
\end{pmatrix} ,
\\
\tilde{n}_i & = 
\sum_{\sigma} 
\tilde{c}^{\dagger}_{ i \sigma} 
\tilde{c}_{i \sigma} 
= 
\sum_{\sigma} 
n_{ i \sigma} ( 1 - n_{i \bar{\sigma}} ) 
= 
\sum_{\sigma} X_i^{\sigma \sigma} .
\end{align}
Here, 
the components of the vector $\bd{\sigma}$ 
are the usual Pauli matrices.
Using the identities
\begin{subequations}
\begin{align}
S_i^+ & = 
S^x_i + i S^y_i = 
\tilde{c}^{\dagger}_{ i \uparrow} 
\tilde{c}_{i \downarrow} = 
c^{\dagger}_{ i \uparrow}
c_{i \downarrow} = 
X_i^{+-} ,
\\
S_i^- & = 
S^x_i - i S^y_i = 
\tilde{c}^{\dagger}_{ i \downarrow} 
\tilde{c}_{i \uparrow} = 
c^{\dagger}_{ i \downarrow} c_{i \uparrow} =
X_i^{-+} ,
\\
S_i^z & = \frac{1}{2} 
\left[ 
\tilde{n}_{ i \uparrow} - 
\tilde{n}_{ i \downarrow}  
\right]
= \frac{1}{2} 
\left[ 
n_{i \uparrow} ( 1 - n_{ i \downarrow} ) - 
n_{ i \downarrow} ( 1- n_{i \uparrow} ) 
\right]
\nonumber\\
& =  
\frac{1}{2} \left[ 
n_{i \uparrow} - 
n_{ i \downarrow} 
\right] 
= 
\frac{1}{2} \sum_{\sigma} \sigma 
X_i^{\sigma \sigma} ,
\end{align}
\end{subequations}
and
\begin{align}
S^z_i S^z_j - 
\frac{ \tilde{n}_i \tilde{n}_j }{4} 
& =
- \frac{1}{2} \Bigl[ 
n_{ i \uparrow} ( 1 - n_{ i \downarrow} ) 
n_{ j \downarrow} ( 1 - n_{i \uparrow} )
\nonumber\\
& \phantom{aaaaa}
 + 
n_{ i \downarrow} ( 1 - n_{ i \uparrow} ) 
n_{ j \uparrow} ( 1 - n_{i \downarrow} ) 
\Bigr]
\nonumber\\
& = 
- \frac{1}{2} \sum_{\sigma}  
X_i^{\sigma \sigma} 
X_{j}^{ \bar{\sigma} \bar{\sigma}} ,
\end{align}
we see that
in terms of X-operators 
the $t$-$J$ Hamiltonian 
assumes the form
\begin{align}
{\cal{H}}_{tJ} 
= {} &  
\sum_{ ij \sigma } t_{ij} 
X_i^{\sigma 0} X_j^{0 \sigma} 
\nonumber\\
& 
+ \frac{1}{4} \sum_{ ij \sigma } J_{ij} 
\left( 
X_i^{\sigma \bar{\sigma}} 
X_j^{\bar{\sigma} \sigma} - 
X_i^{\sigma \sigma} 
X_j^{ \bar{\sigma} \bar{\sigma}} 
\right) .
\label{eq:HtJbest}
\end{align}
Using the completeness relation 
in the projected Hilbert space,
\begin{equation}
X_i^{00} + 
X_i^{\uparrow \uparrow} + 
X_i^{\downarrow \downarrow} = 1 ,
\label{eq:opid}
\end{equation}
we may write
\begin{align}
X_i^{00} X_j^{00} 
= {} &  
1 - \sum_{\sigma} 
\left( 
X_i^{\sigma \sigma} + 
X_j^{\sigma \sigma} 
\right) 
\nonumber\\
& + \sum_{\sigma} 
\left( 
X_i^{\sigma \sigma} 
X_j^{\sigma \sigma} + 
X_i^{\sigma \sigma}
X_j^{\bar{\sigma} \bar{\sigma}} 
\right) ,
\end{align}
and hence
\begin{align}
\sum_{\sigma}  
X_i^{\sigma \sigma} 
X_j^{\bar{\sigma} \bar{\sigma}}  
= {} &  
X_i^{00} X_j^{00} - \sum_{\sigma}  
X_i^{\sigma \sigma} 
X_j^{{\sigma} {\sigma}}
\nonumber\\
& 
+ \sum_{\sigma} 
\left( 
X_i^{\sigma \sigma} + 
X_j^{\sigma \sigma} 
\right) 
- 1 .
\end{align}
Using the anticommutation relation
$ X_j^{\sigma 0} X_i^{0 \sigma} = 
- X_i^{0 \sigma} X_j^{\sigma 0 } $
for $ i \neq j $
we can alternatively write 
the $t$-$J$ Hamiltonian 
in terms of X-operators as follows:
\begin{align}
{\cal{H}}_{tJ} 
= {} &   
\frac{1}{2} \sum_{ ij \sigma } t_{ij} 
\left(  
X_i^{0 \sigma}   
X_j^{\sigma 0} -    
X_i^{\sigma 0} 
X_j^{0 \sigma} 
\right)
\nonumber\\
& 
+ \frac{1}{4} \sum_{ ij } J_{ij} 
\Bigl[  
- X_i^{00} X_j^{00} +   
\sum_{\sigma \sigma^{\prime}} 
X_i^{\sigma \sigma^{\prime}} 
X_j^{\sigma^{\prime} \sigma}  
\Bigr]
\nonumber\\
& 
+ \frac{1}{4} \sum_{ ij } J_{ij} 
\left( 
X_i^{00} - \sum_\sigma 
X^{\sigma \sigma}_i 
\right) ,
\label{eq:tJ2}
\end{align}
where in the last term we have used 
the operator identity \eqref{eq:opid}
to write
\begin{align}
& \sum_{ ij } J_{ij} \Bigl[ 
1 - \sum_{\sigma} 
\left( 
X_i^{\sigma \sigma} +
X_j^{\sigma \sigma} 
\right) 
\Bigr] 
\nonumber\\
= {} & 
\sum_{ ij } J_{ij} 
\left( 
X_i^{00} - \sum_\sigma 
X^{\sigma \sigma}_j 
\right)
\nonumber\\
= {} &   
\sum_{ ij } J_{ij} 
\left( 
X_i^{00} - \sum_\sigma 
X^{\sigma \sigma}_i 
\right) .
\end{align}
For completeness let us mention that   
for nearest neighbor hopping and 
exchange where
$t_{ij} = - t$ and $J_{ij} = J$ 
for all pairs of nearest neighbors
on a square lattice,
there is a special supersymmetric point 
$t = J/2$ where
the $t$-$J$ Hamiltonian \eqref{eq:tJ2} 
exhibits an enhanced symmetry and
can be written as \cite{Wiegmann88,
Foerster89,footnotenotation}   
\begin{equation}
{\cal{H}}^{\rm SUSY}_{tJ} =      
\frac{J}{2} \sum_{ \braket{ ij }  } 
\left[ 
\sum_{ ab } \zeta_b X_i^{ab} X_j^{ba} - 
\sum_a \zeta_a X_i^{aa} 
\right] ,
\label{eq:tJSUZY}
\end{equation}
where $a,b \in \{ 0, + , - \}$ and 
we have introduced the 
statistics factors $\zeta_0 = -1$ and
$\zeta_+ = \zeta_- = 1$.

\subsection{$t$ model}
\label{sec:tdef}

In the extreme strong-coupling limit 
$J \rightarrow 0$, 
the $t$-$J$ Hamiltonian reduces to 
the $t$ model consisting only of 
the projected kinetic energy,
\begin{equation}
{\cal{H}}_t =  
\sum_{ ij \sigma } t_{ij}  
\tilde{c}^{\dagger}_{i \sigma}
\tilde{c}_{j \sigma} 
=  
\sum_{ ij \sigma} t_{ij}  
X_i^{\sigma 0} X_j^{0 \sigma} ,
\label{eq:Htdef}
\end{equation}
which can be obtained  
from the Hubbard model in the limit 
of infinite on-site repulsion $U$.
Due to the projected Hilbert space 
the correlation functions of 
the $t$ model are non-trivial; 
see  Ref.~[\onlinecite{Gehlhoff95}]
for a calculation of the
density-density and spin-spin 
correlation functions of the $t$ model
within a $1/N$ expansion. 
Moreover, 
the single-particle Green function 
of the projected fermions 
is also expected to be dominated by 
the strong kinematic constraints due to 
the Hilbert space projection. 
In fact, 
according to the late 
P.~W.~Anderson \cite{Anderson08,
Anderson09,Casey11}, 
in two dimensions 
the projected kinetic energy 
represented by the $t$ model is 
an effective Hamiltonian describing 
the normal metallic state of the
cuprate superconductors, 
which he called a ``hidden Fermi liquid''.
In Sec.~\ref{sec:hidden} we will use our X-FRG 
approach to calculate the quasi-particle damping
in the hidden Fermi liquid.

\section{Exact X-FRG flow equations}
\label{sec:XFRG}

The grand canonical Hamiltonian 
of each of the models
introduced in Sec.~\ref{sec:models} 
can be written in the form
$\tilde{\cal{H}} = {\cal{H}}_1 + {\cal{H}}_2$, 
where the single-site term ${\cal{H}}_1$ 
is linear in the $X$-operators and 
the hopping term ${\cal{H}}_2$ is quadratic and
involves pairs of $X$-operators 
at different lattice sites.
Following the same strategy 
as in the derivation of FRG flow equations for 
quantum spin systems \cite{Krieg19}, 
we now replace the inter-site hopping $t_{ij}$ 
in ${\cal{H}}_2$ by a deformed hopping 
$t_{ij,\Lambda}$ depending on 
a continuous parameter $\Lambda$, 
which we parametrize as
\begin{equation}
t_{ij , \Lambda} = t_{ij} + R_{ij, \Lambda}^t .
\end{equation}
The regulator function $R_{ij, \Lambda}^t$ 
should be chosen such that
for some initial value $\Lambda = 0$  
the deformed model simplifies so 
that its correlation functions 
can be calculated in a controlled way, 
while $t_{ij,\Lambda} \rightarrow t_{ij}$ 
for $\Lambda \rightarrow 1$, 
so that in this limit 
we recover our original model.
For the $t$-$J$ model, 
we similarly replace 
the exchange coupling by a deformed coupling
\begin{equation}
J_{ij , \Lambda} = J_{ij} + R_{ij, \Lambda}^J .
\end{equation}
The deformed grand canonical Hamiltonian 
is then 
\begin{equation}
\tilde{\cal{H}}_{\Lambda} = 
{\cal{H}}_1 + {\cal{H}}_{ 2 , \Lambda} ,
\end{equation}
where ${\cal{H}}_1$ is 
of the form \eqref{eq:H1def}. 
For the Hubbard model, 
the deformed hopping term is given by 
the deformed kinetic energy,
\begin{align}
{\cal{H}}_{ 2 , \Lambda}
& = 
\sum_{ij \sigma} t_{ij, \Lambda} 
\left( 
X_i^{ \sigma 0}   
X_j^{0 \sigma} +   
X_i^{2 \bar{\sigma}}  
X_j^{\bar{\sigma} 2}
\right.
\nonumber\\
& \phantom{aaaaaaaaaa} 
\left. 
+ \bar{\sigma}   
X_i^{ \sigma 0} 
X_j^{\bar{\sigma} 2} 
+ \bar{\sigma} 
X_i^{2 \bar{\sigma}} 
X_j^{0 \sigma} 
\right).
\end{align}
In the case of the $t$-$J$ model, 
we should omit all X-operators
involving doubly occupied sites and 
add the exchange interaction involving 
Bose type X-operators; 
using the representation \eqref{eq:HtJbest}
we find for the hopping contribution
to the deformed $t$-$J$ Hamiltonian
\begin{align}
{\cal{H}}_{ 2 , \Lambda}
= {} & 
\sum_{ij \sigma} t_{ij, \Lambda}  
X_i^{ \sigma 0}   X_j^{0 \sigma}
\nonumber\\
& 
+ \frac{1}{4} \sum_{ ij \sigma } 
J_{ ij , \Lambda}
\left( 
X_i^{\sigma \bar{\sigma}} 
X_j^{\bar{\sigma} \sigma} - 
X_i^{\sigma \sigma} 
X_j^{ \bar{\sigma} \bar{\sigma}} 
\right) .
\end{align}
For both models, 
the deformed generating functional 
of connected imaginary-time ordered 
X-operator correlation functions 
can be written as a trace of a 
time-ordered exponential,
\begin{equation}
e^{{\cal{G}}_{\Lambda} [ J ] }
=  
{\rm Tr} \left[ 
e^{ - \beta {\cal{H}}_1 } {\cal{T}} 
e^{ 
- \int_0^{\beta} d \tau
{\cal{H}}_{2,\Lambda} ( \tau ) 
}
e^{ 
\int_0^{\beta} d \tau \sum_{i, p} 
J^p_i ( \tau ) X^p_i ( \tau ) 
}  
\right] ,
\label{eq:GXdef}
\end{equation}
where ${\cal{H}}_{ 2 , \Lambda}$ 
is obtained from the inter-site part
${\cal{H}}_2$ of the Hamiltonian 
by replacing the hopping and
the exchange coupling by 
the deformed quantities, 
and the time-dependence of all operators 
is in the interaction picture 
with respect to ${\cal{H}}_1$.
The label $p$ enumerates all types 
of X-operators; 
i.e., 
$p = 1, \ldots , 16$
for the Hubbard model and 
$p = 1, \ldots , 9$ for the $t$-$J$ model.
Because Bose (Fermi) type X-Operators
(anti-)commute at different lattice sites,
the sources $J^p_i ( \tau )$ 
are Grassmann variables 
for Fermi type X-operators and 
complex variables 
for Bose type $X$-operators.
For the same reason,
the time-ordering of 
Bose (Fermi) type X-operators 
is defined as for canonical bosons (fermions).

\subsection{Flow equation for the Hubbard model}
\label{subsec:HubbardFRG}

To derive an exact flow equation for 
${\cal{G}}_{\Lambda} [ J ]$
we proceed as in Eq.~\eqref{eq:sourcetrick} 
and take a derivative of
both sides of Eq.~\eqref{eq:GXdef} 
with respect to the 
deformation parameter $\Lambda$.
For the Hubbard model, 
we obtain
\begin{widetext}
\begin{subequations}
\begin{align}
e^{{\cal{G}}_{\Lambda} [ J ] } 
\partial_{\Lambda} {\cal{G}}_{\Lambda} [ J ]
& =   
- \int_0^{\beta} d \tau  \sum_{ij \sigma} 
\left( 
\partial_{\Lambda} t_{ ij, \Lambda} 
\right)
{\rm Tr} \Biggl\{
e^{ - \beta {\cal{H}}_1 } {\cal{T}} 
\biggl[ 
e^{ 
- \int_0^{\beta} d \tau
{\cal{H}}_{2,\Lambda} ( \tau ) 
} 
\Bigl(  
X_i^{ \sigma 0} ( \tau )  
X_j^{0 \sigma} ( \tau ) +  
X_i^{2 \bar{\sigma}} ( \tau)  
X_j^{\bar{\sigma} 2} (\tau )
\nonumber\\
&  \hspace{58mm}
+ \bar{\sigma}  
X_i^{{\sigma}0} ( \tau) 
X_j^{\bar{\sigma} 2} (\tau ) 
+ \bar{\sigma} 
X_{ i }^{ 2 \bar{\sigma}} ( \tau )
X_j^{0 \sigma} ( \tau ) 
\Bigr) 
e^{ 
\int_0^{\beta} d \tau \sum_{i, p} 
J^p_i ( \tau ) X^p_i ( \tau ) 
}   
\biggr] 
\Biggr\}
\\
& =  
- \int_0^{\beta} d \tau \sum_{ij \sigma} 
\left( 
\partial_{\Lambda} t_{ ij, \Lambda} 
\right)
{\rm Tr} \Biggl\{
e^{ - \beta {\cal{H}}_1 } {\cal{T}} 
\biggl[ 
e^{ 
- \int_0^{\beta} d \tau
{\cal{H}}_{2,\Lambda} ( \tau ) 
}  
\biggl(    
\frac{ \delta}{\delta J_i^{ \sigma 0} ( \tau )}
\frac{\delta}{\delta  J_j^{0 \sigma} ( \tau )}
+  
\frac{ \delta}{
\delta J_i^{2 \bar{\sigma}} ( \tau) } 
\frac{\delta}{ 
\delta J_j^{\bar{\sigma} 2} (\tau )}
\nonumber\\
& \hspace{50.6mm}
+ \bar{\sigma}  
\frac{ \delta }{ 
\delta J_i^{{\sigma}0} ( \tau)} 
\frac{ \delta}{
\delta J_j^{\bar{\sigma} 2} (\tau )} 
+ \bar{\sigma} 
\frac{ \delta}{ 
\delta J_{ i }^{ 2 \bar{\sigma}} ( \tau )}
\frac{ \delta}{
\delta J_j^{0 \sigma} ( \tau )} \biggr) 
e^{ 
\int_0^{\beta} d \tau \sum_{i, p} 
J^p_i ( \tau ) X^p_i ( \tau ) 
}   
\biggr] 
\Biggr\}
\\
& \hspace{-22mm}  =  
- \int_0^{\beta} d \tau \sum_{ij \sigma} 
\left( 
\partial_{\Lambda} t_{ ij, \Lambda} 
\right)
\biggl(    
\frac{ \delta}{\delta J_i^{ \sigma 0} ( \tau )}
\frac{\delta}{\delta  J_j^{0 \sigma} ( \tau )}
+  
\frac{ \delta}{
\delta J_i^{2 \bar{\sigma}} ( \tau) } 
\frac{\delta}{ 
\delta J_j^{\bar{\sigma} 2} (\tau )}
+ \bar{\sigma}  
\frac{ \delta }{
\delta J_i^{{\sigma}0} ( \tau)} 
\frac{ \delta}{
\delta J_j^{\bar{\sigma} 2} (\tau )} 
+ \bar{\sigma} 
\frac{ \delta}{ 
\delta J_{ i }^{ 2 \bar{\sigma}} ( \tau )}
\frac{ \delta}{
\delta J_j^{0 \sigma} ( \tau )} 
\biggr) 
e^{{\cal{G}}_{\Lambda} [ J ] } .
\end{align}
\end{subequations}
We conclude that 
the generating functional of 
the connected imaginary-time ordered
X-operator correlation functions 
of the Hubbard model satisfies 
the exact flow equation
\begin{align}
\partial_{\Lambda} {\cal{G}}_{\Lambda} [ J ]
& =
- \int_0^{\beta} d \tau \sum_{ij \sigma} 
\left( 
\partial_{\Lambda} t_{ ij, \Lambda} 
\right)
\biggl\{
\frac{ \delta^2  {\cal{G}}_{\Lambda} [ J ]
}{
\delta J_i^{ \sigma 0} ( \tau ) 
\delta  J_j^{0 \sigma} ( \tau )
}
+ 
\frac{ \delta  {\cal{G}}_{\Lambda} [ J ] }{
\delta J_i^{ \sigma 0} ( \tau )} 
\frac{ \delta  {\cal{G}}_{\Lambda} [ J ] }{ 
\delta  J_j^{0 \sigma} ( \tau )} 
+
\frac{ \delta^2  {\cal{G}}_{\Lambda} [ J ]
}{
\delta J_i^{ 2 \bar{\sigma} } ( \tau ) 
\delta  J_j^{ \bar{\sigma} 2} ( \tau )
}
+   
\frac{ \delta  {\cal{G}}_{\Lambda} [ J ] }{
\delta J_i^{ 2 \bar{\sigma} } ( \tau )} 
\frac{ \delta  {\cal{G}}_{\Lambda} [ J ] }{ 
\delta  J_j^{  \bar{\sigma} 2 } ( \tau )} 
\nonumber\\
& \phantom{aaaaaaaaaaaaa}
+ \bar{\sigma} 
\frac{ \delta^2  {\cal{G}}_{\Lambda} [ J ]
}{
\delta J_i^{  {\sigma} 0  } ( \tau ) 
\delta  J_j^{ \bar{\sigma} 2} ( \tau )
}
+ \bar{\sigma} 
\frac{ \delta  {\cal{G}}_{\Lambda} [ J ]}{
\delta J_i^{ {\sigma} 0 } ( \tau )} 
\frac{ \delta  {\cal{G}}_{\Lambda} [ J ] }{ 
\delta  J_j^{  \bar{\sigma} 2 } ( \tau )} 
+ \bar{\sigma} 
\frac{ \delta^2  {\cal{G}}_{\Lambda} [ J ]
}{
\delta J_i^{  2 \bar{\sigma} } ( \tau ) 
\delta  J_j^{ 0 {\sigma} } ( \tau )}
+ \bar{\sigma} 
\frac{ \delta  {\cal{G}}_{\Lambda} [ J ]}{
\delta J_i^{ 2 \bar{\sigma}  } ( \tau )} 
\frac{ \delta  {\cal{G}}_{\Lambda} [ J ] }{ 
\delta  J_j^{  0 {\sigma}  } ( \tau )} 
\biggr\} .
\label{eq:GXhubflow}
\end{align}
\end{widetext}
Next, 
we introduce the 
subtracted Legendre transform of 
${\cal{G}}_{\Lambda} [ J ]$ via
\begin{align}
& \Gamma_{\Lambda} [  \bar{X}  ]  = 
\int_0^{\beta} d \tau \sum_{ i, p} 
J_i^p (  \tau ) \bar{X}_i^p ( \tau ) 
- {\cal{G}}_{\Lambda} [ J ]
\nonumber\\
& 
- \int_0^{\beta} d \tau \sum_{ ij \sigma} 
R^{t}_{ ij, \Lambda}
\Bigl[  
\bar{X}_i^{ \sigma 0} ( \tau )  
\bar{X}_j^{0 \sigma} ( \tau ) +  
\bar{X}_i^{2 \bar{\sigma}} ( \tau)  
\bar{X}_j^{\bar{\sigma} 2} (\tau )
\nonumber\\
&  \phantom{aaaaaaa}
+ \bar{\sigma}  
\bar{X}_i^{{\sigma}0} ( \tau) 
\bar{X}_j^{\bar{\sigma} 2} (\tau ) 
+ \bar{\sigma} 
\bar{X}_{ i }^{ 2 \bar{\sigma}} ( \tau )
\bar{X}_j^{0 \sigma} ( \tau ) 
\Bigr] ,
\label{eq:GammaXdef}
\end{align}
where on the right-hand side 
the sources $J_i^p ( \tau )$ 
should be expressed in terms of the
expectation values $\bar{X}_i^p ( \tau )$
by inverting the relation
\begin{equation}
\bar{X}_i^p ( \tau ) = 
\braket{ {\cal{T}} X_i^p ( \tau ) } = 
\frac{ \delta {\cal{G}}_{\Lambda} [ J ] 
}{ \delta J_i^p ( \tau ) } .
\label{eq:XJrelation}
\end{equation}
Using the flow equation \eqref{eq:GXhubflow} 
we find that the functional
$ \Gamma_{\Lambda} [  \bar{X}  ]$ 
satisfies the exact flow equation
\begin{align}
& \partial_{\Lambda} 
\Gamma_{\Lambda} [  \bar{X}  ]  =  
\int_0^{\beta} d \tau  \sum_{ij \sigma} 
\left( 
\partial_{\Lambda} R^t_{ ij, \Lambda} 
\right)
\nonumber\\
& 
\times 
\biggl[
\frac{ \delta^2  {\cal{G}}_{\Lambda} [ J ]
}{
\delta J_i^{ \sigma 0} ( \tau ) 
\delta  J_j^{0 \sigma} ( \tau )
}
+ \frac{ \delta^2  {\cal{G}}_{\Lambda} [ J ]}
{
\delta J_i^{ 2 \bar{\sigma} } ( \tau ) 
\delta  J_j^{ \bar{\sigma} 2} ( \tau )
}
\nonumber\\
& \phantom{aaa}
+ \bar{\sigma} 
\frac{ \delta^2  {\cal{G}}_{\Lambda} [ J ]
}{
\delta J_i^{  {\sigma} 0  } ( \tau ) 
\delta  J_j^{ \bar{\sigma} 2} ( \tau )
}
+ \bar{\sigma} 
\frac{ \delta^2  {\cal{G} }_{\Lambda} [ J ]
}{
\delta J_i^{  2 \bar{\sigma}   } ( \tau ) 
\delta  J_j^{ 0 {\sigma} } ( \tau )
}
\biggr] .
\end{align}
To compactify the notation, 
we now collect all
X-operators into a 16-component vector
\begin{equation}
( \Phi_{\alpha} ) = 
\begin{pmatrix}
{\Phi}_{\uparrow}  \\  
{\Phi}_{\downarrow} \\
{\Phi}_0 \\ 
{\Phi}_2  
\end{pmatrix} ,
\label{eq:superfield}
\end{equation}
where the four-component vectors 
$\Phi_{\uparrow}$ and $\Phi_{\downarrow}$ 
contain the expectation values 
of the Fermi type X-operators,
\begin{equation}
{\Phi}_{\sigma} = 
\begin{pmatrix} 
\bar{X}^{0 \sigma} \\ 
\bar{\sigma} \bar{X}^{\bar{\sigma} 2} \\
\bar{X}^{\sigma 0} \\ 
\bar{\sigma} \bar{X}^{ 2 \bar{\sigma}} \end{pmatrix} , 
\; \; \;  \; \; \; 
\sigma = \uparrow, \downarrow,
\end{equation}
while the four-component vectors 
$\Phi_0$ and $\Phi_2$ contain 
the expectation values of the 
Bose type X-operators as follows:
\begin{equation}
{\Phi}_0 = 
\begin{pmatrix} 
\bar{X}^{ \uparrow \uparrow} \\ 
\bar{X}^{\downarrow \downarrow} \\
\bar{X}^{\uparrow \downarrow} \\ 
\bar{X}^{ \downarrow \uparrow} 
\end{pmatrix} , 
\; \; \;  \; \; \;
{\Phi}_2 = 
\begin{pmatrix} 
\bar{X}^{ 00 } \\ 
\bar{X}^{02} \\
\bar{X}^{20} \\ 
\bar{X}^{ 22} \end{pmatrix} .
\end{equation}
For simplicity we have omitted the 
lattice site and imaginary time labels.
The derivation of the Wetterich equation 
follows precisely the same steps as the
derivation for canonical bosons or fermions
outlined in Sec.~\ref{sec:without}.
The only difference is that now 
the superfield $\Phi_\alpha$ has 
fermionic and bosonic components, 
so that Eq.~\eqref{eq:Wetterich} 
should be replaced by
\begin{align}
& 
\partial_{\Lambda} 
\Gamma_{\Lambda} [ \Phi] = 
\nonumber\\
&
\frac{1}{2}
{\rm Tr} \left\{  
\mathbf{Z} 
\left( 
\partial_{\Lambda} \mathbf{R}_{\Lambda} 
\right) 
\left[
\left( 
\frac{\delta}{\delta \Phi} \otimes 
\frac{\delta}{\delta \Phi} 
\right)^T 
\Gamma_{\Lambda} [ \Phi ]
+ \mathbf{R}_{\Lambda} 
\right]^{-1} 
\right\} ,
\label{eq:Wetterich2}
\end{align}
where the matrix elements of the 
statistics matrix \cite{Kopietz10} are
$[ \mathbf{Z} ]_{\alpha \beta } = 
\delta_{\alpha \beta } \zeta_{\alpha}$ with
$\zeta_{\alpha} = 1$ if the 
label $\alpha$ corresponds to  
a Bose type X-operator, 
and $\zeta_\alpha = -1$ if $\alpha$ 
corresponds to a Fermi type X-operator.
In analogy with Eq.~\eqref{eq:regulatorsym},
the regulator matrix $ \mathbf{R}_{\Lambda}$ 
is defined by writing the regulator term 
in Eq.~\eqref{eq:GammaXdef} in 
(anti)-symmetrized superfield notation,
\begin{align}
& 
\int_0^{\beta} d \tau \sum_{ ij \sigma} 
R^{t}_{ ij, \Lambda}
\Bigl[  
\bar{X}_i^{ \sigma 0} ( \tau )  
\bar{X}_j^{0 \sigma} ( \tau ) +  
\bar{X}_i^{2 \bar{\sigma}} ( \tau )  
\bar{X}_j^{\bar{\sigma} 2} (\tau )
\nonumber
\\
&  \hspace{10mm}
+ \bar{\sigma}  
\bar{X}_i^{{\sigma}0} ( \tau ) 
\bar{X}_j^{\bar{\sigma} 2} ( \tau ) 
+ \bar{\sigma} 
\bar{X}_{ i }^{ 2 \bar{\sigma}} ( \tau )
\bar{X}_j^{0 \sigma} ( \tau ) \Bigr] 
\nonumber\\
& =
\frac{1}{2} \int_\alpha \int_{\alpha'} 
\Phi_{\alpha}
[ \mathbf{R}_{\Lambda} ]_{\alpha \alpha' } 
\Phi_{\alpha'} .
\label{eq:regulatorsym2}
\end{align}

Although the X-FRG 
Wetterich equation \eqref{eq:Wetterich2} 
is formally exact, 
in this form
it is only of academic interest 
because in practice we have to impose 
the initial condition of decoupled sites 
where the deformed hopping $t_{ij, \Lambda =0}$ 
vanishes at the initial scale $\Lambda =0$.
Unfortunately, 
in this limit the
Legendre transform 
$\Gamma_{\Lambda =0} [ \bar{X} ]$ 
does not exist because for 
decoupled lattice sites the 
diagonal X-operators $X_{i}^{ a a } ( \tau )$ 
are constants of motion. 
This follows from the fact that 
in the interaction picture 
with respect to the single-site part
${\cal{H}}_1 = \sum_{ i a} \epsilon_a X^{aa}_i$ 
of the Hamiltonian [see Eq.~\eqref{eq:H1def}] 
the X-operators satisfy the following 
equations of motion
in imaginary time:
\begin{equation}
\frac{ \partial X_i^{ab} ( \tau ) }{
\partial \tau } = 
[ {\cal{H}}_1 , X_{i}^{ ab} ( \tau ) ]
= 
( \epsilon_a - \epsilon_b ) X_i^{ab} ( \tau ) ,
\end{equation}
with the solution
\begin{equation}
X_i^{ab} ( \tau ) =  
e^{ ( \epsilon_a - \epsilon_b ) \tau }  
X_i^{ab} ( 0 ) .
\label{eq:Xabdyn}
\end{equation}
Obviously, 
the diagonal operators $X_i^{aa} ( \tau )$ 
are independent of $\tau$; 
moreover, 
in  the absence of a magnetic field 
also the spin-flip operators 
$X_i^{ \sigma \bar{\sigma} } ( \tau )$ 
are conserved. 
The lack of intrinsic dynamics of 
some of the X-operators implies 
that the relations \eqref{eq:XJrelation} 
cannot be inverted because 
the matrix of second derivatives of 
${\cal{G}}_{\Lambda} [ J ]$  
is not invertible at the
initial scale $\Lambda =0$ 
where the hopping is switched off.
A similar problem is encountered in 
the spin-FRG approach \cite{Krieg19}, 
where for vanishing exchange interactions 
the components of the individual spins 
parallel to the magnetic field 
do not have any dynamics.
We can avoid this problem by performing  
only a partial Legendre transform
in the sector where the 
above second derivative matrix  
can be inverted. 
This can be implemented in two ways:

\begin{enumerate}

\item 
The simplest strategy is 
to include in the source term
$\int_0^{\beta} d \tau \sum_{ i , p}  
J^p_{i} ( \tau ) X_i^p ( \tau )$
in Eq.~\eqref{eq:GXdef}
only those X-operators 
which have an intrinsic dynamics 
without hopping.
The disadvantage of this approach is 
that correlation functions of the
other X-operators are not easily accessible. 
If these fluctuations are singular, 
the resulting vertices of the
X-operators which are retained are 
non-local so that simple approximations fail.
Fortunately, 
for the Hubbard model all 
Fermi type X-operators do have  
intrinsic dynamics even for vanishing hopping, 
so that this strategy can be used to calculate
the electronic single-particle spectral function 
of the Hubbard model. 
In fact, 
we will show in Sec.~\ref{sec:Hubbard1} 
that this strategy reproduces 
the Hubbard-I approximation for 
the single-particle spectral function 
of the Hubbard model in a very simple way.
Note also that there is a large degree of 
redundancy contained in the  X-operators: 
for example,
the Bose type X-operator 
$X^{\sigma \sigma^\prime}_i$ can be written 
as a composite operator involving 
a product of two Fermi type X-operators,
$X^{\sigma \sigma^{\prime}}_i = 
X^{\sigma 0}_i  X^{ 0 \sigma^{\prime}}_i$.
Hence, 
even if we include only Fermi type X-operators 
in our  Legendre transform,
correlation functions of the 
Bose type X-operators can still be obtained 
as higher-order correlation functions 
of Fermi type X-operators.

\item 
To explicitly take into account also the 
fluctuations associated with the
X-operators which do not have a dynamics 
for vanishing hopping,
we can try to construct a suitable 
hybrid functional 
$\Phi_{\Lambda} [ \bar{X}^{\prime} ; \bd{\eta} ]$ 
which is defined via the 
usual Legendre transform with respect to 
all X-operators 
$X^{\prime}$ which do have 
intrinsic single-site dynamics, 
and includes the effect of the 
remaining set of X-operators
via certain auxiliary fields $\bd{\eta}$. 
For quantum spin systems we have 
explicitly constructed suitable 
hybrid functionals in our 
previous works \cite{Goll19,Goll20,
Tarasevych21,Rueckriegel22,Tarasevych22}.
Although it is not clear how to generalize 
this strategy for the Hubbard model, 
for the $t$-$J$ model this can be done
in the same way as for quantum spin systems \cite{Goll19,Goll20,Tarasevych21,
Rueckriegel22,Tarasevych22}.
We will explicitly do this 
in the following subsection.

\end{enumerate}

\subsection{Flow equations for the $t$-$J$ model}

To derive the Wetterich equation 
for the $t$-$J$ model we start again from
Eq.~\eqref{eq:GXdef} where it is now understood 
that the trace over the
projected Hilbert space excludes 
states with doubly occupied sites and 
the sum $\sum_{ p } J_i^p (\tau ) X^p_i ( \tau )$
in the source term is restricted to 
the nine X-operators acting on 
the projected Hilbert space.
Instead of Eq.~\eqref{eq:GXhubflow} 
we then obtain
\begin{widetext}
\begin{align}
& 
\partial_{\Lambda} {\cal{G}}_{\Lambda} [ J ]
=     
- \int_0^{\beta} d \tau  \sum_{ij \sigma} 
\left( 
\partial_{\Lambda} t_{ ij, \Lambda} 
\right)
\biggl[
\frac{ \delta^2  {\cal{G}}_{\Lambda} [ J ]}{
\delta J_i^{ \sigma 0} ( \tau ) 
\delta  J_j^{0 \sigma} ( \tau )}
+   
\frac{ \delta  {\cal{G}}_{\Lambda} [ J ]}{
\delta J_i^{ \sigma 0} ( \tau )} 
\frac{ \delta  {\cal{G}}_{\Lambda} [ J ] }{ 
\delta  J_j^{0 \sigma} ( \tau )} 
\biggr] 
\nonumber\\
& 
- \frac{1}{4} \int_0^{\beta} d \tau  
\sum_{ij \sigma} 
\left( 
\partial_{\Lambda} J_{ ij, \Lambda} 
\right)
\biggl[
\frac{ \delta^2  {\cal{G}}_{\Lambda} [ J ]}{
\delta J_i^{ \sigma  \bar{\sigma} } ( \tau )
\delta  J_j^{\bar{\sigma} \sigma} ( \tau )}
+   
\frac{ \delta  {\cal{G}}_{\Lambda} [ J ]}{
\delta J_i^{ \sigma \bar{\sigma}} ( \tau )} 
\frac{ \delta  {\cal{G}}_{\Lambda} [ J ] }{ 
\delta  J_j^{\bar{\sigma} \sigma} ( \tau )} 
- 
\frac{ \delta^2  {\cal{G}}_{\Lambda} [ J ]}{
\delta J_i^{ \sigma  {\sigma} } ( \tau ) 
\delta  J_j^{\bar{\sigma} \bar{\sigma}} ( \tau )}
-    
\frac{ \delta  {\cal{G}}_{\Lambda} [ J ]}{
\delta J_i^{ \sigma {\sigma}} ( \tau )} 
\frac{ \delta  {\cal{G}}_{\Lambda} [ J ] }{ 
\delta  J_j^{\bar{\sigma} \bar{\sigma}} ( \tau )} 
\biggr] .
\label{eq:GXtJflow}
\end{align}
The subtracted Legendre transform of  
${\cal{G}}_{\Lambda} [ J ]$
is now defined by 
\begin{align}
\Gamma_{\Lambda} [  \bar{X}  ]  
= {} &  
\int_0^{\beta} d \tau \sum_{ i, p} 
J_i^p ( \tau ) \bar{X}_i^p ( \tau ) 
-  {\cal{G}}_{\Lambda} [ J ]
\nonumber\\
& 
- \int_0^{\beta} d \tau \sum_{ ij \sigma} 
\left\{
R^{t}_{ ij, \Lambda}
\bar{X}_i^{ \sigma 0} ( \tau )  
\bar{X}_j^{0 \sigma} ( \tau )
+ \frac{1}{4} R^J_{ ij, \Lambda}
\left[ 
\bar{X}_i^{{\sigma} \bar{\sigma}} ( \tau ) 
\bar{X}_j^{\bar{\sigma} \sigma } (\tau ) 
-  
\bar{X}_{ i }^{ \sigma {\sigma}} ( \tau )
\bar{X}_j^{\bar{\sigma} \bar{\sigma}} ( \tau ) 
\right] 
\right\} ,
\label{eq:GammaXtJdef}
\end{align}
\end{widetext}
which satisfies the flow equation
\begin{align}
&  
\partial_{\Lambda} 
\Gamma_{\Lambda} [  \bar{X}  ]  
=  
\int_0^{\beta} d \tau  \sum_{ij \sigma} 
\Biggl\{ 
\left( 
\partial_{\Lambda} R^t_{ ij, \Lambda} 
\right)
\frac{ \delta^2  {\cal{G}}_{\Lambda} [ J ]}{
\delta J_i^{ \sigma 0} ( \tau ) 
\delta  J_j^{0 \sigma} ( \tau )}
\nonumber\\
& 
+ \frac{1}{4}  
\left( 
\partial_{\Lambda} R^J_{ ij, \Lambda} 
\right)
\biggl[ 
\frac{ \delta^2  {\cal{G}}_{\Lambda} [ J ]}{
\delta J_i^{ \sigma  \bar{\sigma} } ( \tau ) 
\delta  J_j^{ \bar{\sigma} \sigma} ( \tau )} 
-   
\frac{ \delta^2  {\cal{G}}_{\Lambda} [ J ]}{
\delta J_i^{  {\sigma} \sigma  } ( \tau ) 
\delta J_j^{ \bar{\sigma} \bar{\sigma}} ( \tau )} 
\biggr]
\Biggr\} .
\nonumber\\
\end{align}
Finally, 
we introduce a nine-component superfield
of the form \eqref{eq:superfield}; 
i.e., 
\begin{equation}
( \Phi_{\alpha} ) = 
\begin{pmatrix}
{\Phi}_{\uparrow}  \\  
{\Phi}_{\downarrow} \\
{\Phi}_0 \\ 
{\Phi}_2  
\end{pmatrix} ,
\end{equation}
where now
\begin{equation}
{\Phi}_{\sigma} = 
\begin{pmatrix} 
\bar{X}^{0 \sigma} \\ 
\bar{X}^{\sigma 0}
\end{pmatrix} , 
\; \; \;  \; \; \; 
\sigma = \uparrow , \downarrow ,
\end{equation}
and
\begin{equation}
{\Phi}_0 = 
\begin{pmatrix} 
\bar{X}^{ \uparrow \uparrow} \\ 
\bar{X}^{\downarrow \downarrow} \\
\bar{X}^{\uparrow \downarrow} \\ 
\bar{X}^{ \downarrow \uparrow} 
\end{pmatrix} , 
\; \; \;  \; \; \;
{\Phi}_2 = \bar{X}^{ 00 } .
\end{equation}
Defining the regulator matrix 
in superfield notation via
\begin{align}
& 
\int_0^{\beta} d \tau \sum_{ ij \sigma} 
\Bigl\{
R^{t}_{ ij, \Lambda }
\bar{X}_i^{ \sigma 0} ( \tau )  
\bar{X}_j^{0 \sigma} ( \tau )
\nonumber\\
& \hspace{7mm}  
+ \frac{1}{4} R^J_{ ij, \Lambda }
\left[ 
\bar{X}_i^{{\sigma} \bar{\sigma}} ( \tau ) 
\bar{X}_j^{\bar{\sigma} \sigma } (\tau ) 
-  
\bar{X}_{ i }^{ \sigma {\sigma}} ( \tau )
\bar{X}_j^{\bar{\sigma} \bar{\sigma}} ( \tau )
\right] 
\Bigr\}
\nonumber\\
= {} &  
\frac{1}{2} \int_\alpha \int_{\alpha'} 
\Phi_{\alpha}
[ \mathbf{R}_{\Lambda} ]_{\alpha \alpha' } 
\Phi_{\alpha'} ,
\label{eq:regulatorsym3}
\end{align}
the resulting  Wetterich equation for 
the $t$-$J$ model is
formally identical to the Wetterich equation  
for the Hubbard model
given in Eq.~\eqref{eq:Wetterich2}. 
One should keep in mind, 
however, 
that for the $t$-$J$ model 
the superfield $\Phi$ has only 
nine components and 
that the trace in Eq.~\eqref{eq:Wetterich2} 
is over the projected Hilbert space 
of the $t$-$J$ model.

As discussed at the end
of Sec.~\ref{subsec:HubbardFRG}, 
if we impose the initial condition
$J_{ ij , \Lambda=0} =0$
the Legendre transform 
$\Gamma_{\Lambda =0} [ \bar{X} ]$ 
defined in Eq.~\eqref{eq:GammaXtJdef}
does not exist because
all X-operators of the Bose type 
are conserved for $\Lambda =0$. 
Let us now show how to solve this problem 
using the second method mentioned 
at the end of Sec.~\ref{subsec:HubbardFRG}:
therefore we introduce 
a hybrid functional
$\Phi_{\Lambda} [ \bar{\psi} , \psi  , 
\bd{\eta} ]$ 
which generates vertices that are 
propagator-irreducible with 
respect to the Fermi type X-operators, 
and in addition 
exchange interaction-irreducible with 
respect to the Bose type X-operators.  
For simplicity, 
we drop the particle-number term 
$ \tilde{n}_j \tilde{n}_j / 4$
in the $t$-$J$ Hamiltonian and 
focus on the generating functional of 
connected correlation functions involving the 
four Fermi type X-operators 
$\tilde{c}_{i \sigma}^{\dagger} = 
X_i^{\sigma 0}$ and 
$\tilde{c}_{i \sigma} = 
X_i^{0 \sigma}$ and on 
the three Bose type X-operators 
representing the spin components,
\begin{subequations}
\begin{align}
S^{x}_i & = 
\frac{1}{2} \left( 
S^{+}_i + S^{-}_i 
\right)
= 
\frac{1}{2} \left( 
X_i^{ +-} + X_i^{-+} 
\right) ,
\\
S^{y}_i & = 
\frac{1}{2i} \left( 
S^{+}_i - S^{-}_i 
\right)
= \frac{1}{2i} \left( 
X_i^{ +-} - X_i^{-+} 
\right) ,
\\
S^z_i & = 
\frac{1}{2} \left( 
X^{++}_i - X_i^{--} \right) .
\end{align}
\end{subequations}
The generating functional of the 
connected correlation functions of 
this set of X-operators can then 
be written as
\begin{widetext}
\begin{equation}
e^{
{\cal{G}}_{\Lambda} [ \bar{j} , j , \bd{h} ] 
}
= {\rm Tr} \left\{ 
e^{ - \beta {\cal{H}}_1 } {\cal{T}} 
e^{ 
- \int_0^{\beta} d \tau
{\cal{H}}_{2,\Lambda} ( \tau ) 
}
e^{ 
\int_0^{\beta} d \tau \sum_{ i \sigma} 
\left[ 
\bar{j}_{i \sigma} ( \tau )  
\tilde{c}_{i \sigma} ( \tau ) 
+ 
\tilde{c}^{\dagger}_{ i \sigma} ( \tau )  
j_{ i \sigma} (\tau) 
\right]
}
e^{ 
\int_0^{\beta} d \tau  \sum_i 
\bd{h}_i ( \tau ) \cdot \bd{S}_i ( \tau ) }
\right\} ,
\label{eq:GXtJdef}
\end{equation}
where 
${\cal{H}}_1 = 
- \mu \sum_{ i \sigma} X^{\sigma \sigma}_i$ 
and
\begin{equation}
{\cal{H}}_{ 2 , \Lambda} =  
\sum_{ ij \sigma } t_{ij, \Lambda} 
\tilde{c}^{\dagger}_{i \sigma}
\tilde{c}_{j \sigma}
+ \frac{1}{2} \sum_{ ij } J_{ij, \Lambda}  
\bd{S}_i \cdot \bd{S}_j  .
\end{equation}
The generating functional of 
connected correlation functions of 
Fermi type X-operators and
interaction-amputated spin correlation functions 
is \cite{Krieg19,Goll19}
\begin{equation}
{\cal{F}}_{\Lambda} [ \bar{j} , j , \bd{m} ] = 
{\cal{G}}_{\Lambda} \bigl[ \bar{j} , j , 
\bd{h} \rightarrow - 
\sum_{j} {J}_{ij, \Lambda} \bd{m}_j 
\bigr]
- \frac{1}{2} \int_0^{\beta}  d \tau  
\sum_{ij} {J}_{ij, \Lambda}  
\bd{m}_{i} ( \tau) \cdot \bd{m}_j (\tau) .
\label{eq:amputated_functional}
\end{equation}
This functional satisfies 
the flow equation \cite{Krieg19},
\begin{align}
\partial_{\Lambda}
{{\cal{F}}}_{\Lambda} [ \bar{j} , j ,  \bd{m} ]  
= {} & 
- \zeta
\int_0^{\beta} d \tau  \sum_{ij \sigma}  
\left( 
\partial_{\Lambda} t_{ ij, \Lambda} 
\right)
\left[
\frac{ 
\delta^2 
{\cal{F}}_{\Lambda} [ \bar{j} , j , \bd{m} ]
}{
\delta j_{i \sigma} ( \tau ) 
\delta  \bar{j}_{j \sigma} ( \tau )
}
+ \frac{ 
\delta 
{\cal{F}}_{\Lambda} [ \bar{j} , j , \bd{m} ]
}{
\delta j_{i \sigma} ( \tau )
}
\frac{ 
\delta  
{\cal{F}}_{\Lambda} [ \bar{j} , j , \bd{m} ]
}{
\delta \bar{j}_{j \sigma} ( \tau )
} 
\right]
\nonumber\\
& + 
\frac{1}{2} \int_0^{\beta} d \tau 
\sum_{ij \alpha} 
\left( 
\partial_{\Lambda} 
{\mathbbm{{J}}}^{-1}_{\Lambda} 
\right)_{ij} 
\left[
\frac{ 
\delta^2 
{\cal{F}}_{\Lambda} [ \bar{j}, j , \bd{m} ] 
}{
\delta m_i^{\alpha} ( \tau )   
\delta m_j^{\alpha} ( \tau ) 
}
+
\frac{ 
\delta 
{\cal{F}}_{\Lambda} [ \bar{j} , j ,  \bd{m} ] 
}{
\delta m_i^{\alpha} ( \tau ) 
}
\frac{ 
\delta 
{\cal{F}}_{\Lambda} [ \bar{j} , j , \bd{m} ] 
}{
\delta m_j^{\alpha} ( \tau ) 
}
\right]
+ \frac{1}{2} {\rm Tr} \left[ 
\mathbf{{J}}_{\Lambda} 
\partial_\Lambda \mathbf{{J}}^{-1}_\Lambda 
\right] ,
\label{eq:flowGm}
\end{align}
where
 $\mathbbm{{J}}_{\Lambda}$ is a matrix 
 in the site labels $i$, $j$ 
 with matrix elements
 $ [\mathbbm{{J}}_{\Lambda}]_{ij} = 
 {J}_{ij, \Lambda}$ and
$\mathbf{J}$ is a matrix in the site ($i$),
imaginary time ($\tau$), 
and spin component ($\alpha$) labels with
matrix elements
\begin{equation}
[ \mathbf{J}_{\Lambda} ]_{ 
i \tau \alpha , 
j \tau^{\prime} \alpha^{\prime}
} 
= J_{ ij, \Lambda} 
\delta_{ \alpha \alpha^{\prime}} 
\delta ( \tau - \tau^{\prime} ) .
\end{equation}
The generating functional of vertices 
which are propagator-irreducible 
for the Fermi type fields and 
exchange-interaction irreducible 
for the Bose type fields 
is now defined via the following 
subtracted Legendre transform,
\begin{align}
\Gamma_{\Lambda} 
[ \bar{\psi} , \psi , \bd{\eta} ] 
= {} &
\int_0^{\beta} d \tau \sum_{ i \sigma} 
\left[
\bar{j}_{ i \sigma} ( \tau ) 
\psi_{ i \sigma} ( \tau ) + 
\bar{\psi}_{ i \sigma} ( \tau ) 
j_{ i \sigma} ( \tau ) 
\right]
+ \int_0^\beta d \tau \sum_i 
\bd{m}_i ( \tau ) \cdot \bd{\eta}_i ( \tau )
- {{\cal{F}}}_{\Lambda} [ \bar{j} , j ,  \bd{m} ]  
\nonumber\\
& 
- \int_0^{\beta} d \tau \sum_{ ij \sigma} 
t_{ ij, \Lambda}
\bar{\psi}_{i  \sigma } ( \tau )  
{\psi}_{j \sigma } ( \tau )
+ \frac{1}{2} \int_0^{\beta} d \tau \sum_{ij}  
[ {\mathbbm{J}}_{\Lambda}^{-1} ]_{ij}
\bd{\eta}_i ( \tau ) \cdot \bd{\eta}_j ( \tau ) ,
\label{eq:Ldef}
\end{align}
where
\begin{subequations}
\begin{align}
\psi_{ i \sigma} ( \tau ) 
& = 
\braket{ 
{\cal{T}} \tilde{c}_{ i \sigma } ( \tau ) 
}
= 
\frac{ 
\delta {{\cal{F}}}_{\Lambda} 
[ \bar{j} , j ,  \bd{m} ]  
}{
\delta \bar{j}_{ i \sigma} ( \tau ) 
},
\\
\bar{\psi}_{ i \sigma} ( \tau ) 
& = 
\braket{ 
{\cal{T}} 
\tilde{c}^{\dagger}_{ i \sigma } ( \tau ) 
}
= \zeta 
\frac{ 
\delta {{\cal{F}}}_{\Lambda} 
[ \bar{j} , j ,  \bd{m} ]  
}{
\delta {j}_{ i \sigma} ( \tau ) 
} ,
\\
\bd{\eta}_i ( \tau ) 
& =   
\frac{ 
\delta {{\cal{F}}}_{\Lambda} 
[ \bar{j} , j ,  \bd{m} ]  
}{
\delta \bd{m}_{ i } ( \tau ) 
} .
\end{align}
\end{subequations}
By construction, 
the functional  
$\Gamma_{\Lambda} 
[ \bar{\psi} , \psi , \bd{\eta} ]$
satisfies the flow equation
\begin{equation}
\partial_{\Lambda} 
\Gamma_{\Lambda} 
[ \bar{\psi} , \psi , \bd{\eta} ] 
= 
\zeta
\int_0^{\beta} d \tau  \sum_{ij \sigma} 
\left( 
\partial_{\Lambda} t_{ ij, \Lambda} 
\right)
\frac{ 
\delta^2 {\cal{F}}_{\Lambda} 
[ \bar{j} , j , \bd{m} ]}{
\delta j_{i \sigma} ( \tau ) 
\delta \bar{j}_{j \sigma} ( \tau )
}
- \frac{1}{2} \int_0^{\beta} d \tau 
\sum_{ij \alpha} 
\left( 
\partial_{\Lambda} 
{\mathbbm{{J}}}^{-1}_{\Lambda} 
\right)_{ij} 
\frac{ 
\delta^2 {\cal{F}}_{\Lambda} 
[ \bar{j}, j , \bd{m} ] }{
\delta m_i^{\alpha} ( \tau )  
\delta m_j^{\alpha} ( \tau ) 
}
- \frac{1}{2} {\rm Tr} \left[ 
\mathbf{{J}}_{\Lambda} 
\partial_\Lambda \mathbf{{J}}^{-1}_\Lambda 
\right] .
\label{eq:Phiflow}
\end{equation}
The right-hand side of
this flow equation has a finite limit 
for vanishing exchange couplings.
To see this more clearly, 
let us for simplicity set $t_{ ij, \Lambda} =0$ 
and focus only on the flow in the spin sector, 
which can be derived from the functional 
$\Phi_{\Lambda} [ \bd{\eta} ] = 
\Gamma_{\Lambda} [ 0 , 0 , \bd{\eta} ]$. 
It satisfies the flow equation
\begin{equation}
\partial_{\Lambda} 
{\Phi}_{\Lambda} [ \bd{\eta} ] 
=  
- \frac{1}{2} {\rm Tr} \left\{
\left[ 
\left(
\mathbf{{\Phi}}^{\prime \prime }_{\Lambda} 
[ \bd{\eta} ] -  
\mathbf{{J}}^{-1}_{\Lambda} 
\right)^{-1} +
\mathbf{{J}}_{\Lambda} 
\right]
\partial_{\Lambda} \mathbf{{J}}^{-1}_{\Lambda} 
\right\} 
= 
- \frac{1}{2} {\rm Tr} \left\{ 
\mathbf{{\Phi}}^{\prime \prime }_{\Lambda} 
[ \bd{\eta} ]
\left( 
\mathbf{1} -     
\mathbf{{J}}_{\Lambda}
\mathbf{{\Phi}}^{\prime \prime }_{\Lambda} 
[ \bd{\eta} ]   
\right)^{-1} 
\partial_{\Lambda} \mathbf{{J}}_{\Lambda} 
\right\} ,
\label{eq:WetterichPhi}
\end{equation}
where 
$\mathbf{\Phi}^{ \prime \prime}_{\Lambda}$ 
denotes the matrix of second derivatives 
with respect to the components of 
the $\bd{\eta}$-field,
\begin{equation}
{\bigl[} 
\mathbf{\Phi}^{ \prime \prime }_{\Lambda} 
[ \bd{\eta} ]  
{\bigr]}_{ 
i \tau \alpha, 
j \tau^{\prime} \alpha^{\prime} 
} 
= 
\frac{ \delta^2 \Phi_{\Lambda} [ \bd{\eta} ] }{
\delta \eta_i^{\alpha} ( \tau ) 
\delta \eta_j^{\alpha^{\prime}} ( \tau^{\prime} )
} .
\end{equation}
From the right-hand side 
of Eq.~\eqref{eq:WetterichPhi}, it is obvious 
that the functional 
$\Phi_{\Lambda} [ \bd{\eta} ]$ 
is well-defined even for 
$\mathbf{J}_{\Lambda} =0$.
The second functional derivatives 
on the right-hand side 
of Eq.~\eqref{eq:Phiflow} can be expressed 
in terms of the second derivatives of 
$\Gamma_{\Lambda} 
[ \bar{\psi} , \psi , \bd{\eta} ]$.
At this point, 
it is convenient to use again a 
superfield notation with a 
seven-component superfield
\begin{equation}
( \Phi_{\alpha} ) =
\begin{pmatrix} 
\psi_{ \uparrow} \\ 
\bar{\psi}_{\uparrow} \\
\psi_{\downarrow} \\ 
\bar{\psi}_{\downarrow} \\ 
\bd{\eta} 
\end{pmatrix} .
\end{equation}
Then the flow equation \eqref{eq:Phiflow} 
can be cast into the form
\begin{equation}
\partial_{\Lambda} \Gamma_{\Lambda} [ \Phi] 
= 
\frac{1}{2}
{\rm Tr} \left\{  
\mathbf{Z} 
\left( 
\partial_{\Lambda} \mathbf{T}_{\Lambda} 
\right) 
\left[
\left( 
\frac{\delta}{\delta \Phi} \otimes 
\frac{\delta}{\delta \Phi} 
\right)^T 
\Gamma_{\Lambda}  [ \Phi ]
+ \mathbf{T}_{\Lambda} 
\right]^{-1} 
\right\}
- \frac{1}{2} {\rm Tr} \left[ 
\mathbf{{J}}_{\Lambda} 
\partial_\Lambda \mathbf{{J}}^{-1}_\Lambda 
\right] ,
\label{eq:Wetterich3}
\end{equation}
where the generalized hopping matrix $\mathbf{T}_{\Lambda}$ 
in superfield space is defined by writing
\begin{equation}
\int_0^{\beta} d \tau \sum_{ ij \sigma} 
{t}_{ ij, \Lambda}
\bar{\psi}_{i  \sigma } ( \tau )  
{\psi}_{j \sigma } ( \tau )
- \frac{1}{2} \int_0^{\beta} d \tau \sum_{ij}  
[ {\mathbbm{J}}_{\Lambda}^{-1} ]_{ij}
\bd{\eta}_i ( \tau ) \cdot \bd{\eta}_j ( \tau ) 
=
\frac{1}{2} \int_{\alpha} \int_{\alpha'} 
\Phi_{\alpha} 
[ \mathbf{T}_{\Lambda} ]_{ \alpha \alpha' }
\Phi_{\alpha'} .
\end{equation}
\end{widetext}
The functional $\Gamma_{\Lambda} [ \Phi ]$ 
is well-defined for all $\Lambda$, 
including $\Lambda =0$ where 
both hopping and the exchange interaction vanish.

\section{Simple applications}
\label{sec:applications}

\subsection{ 
Hubbard-I approximation from X-FRG}
\label{sec:Hubbard1}

If we are only interested in 
correlation functions of the 
Fermi type X-operators 
(which is sufficient for the calculation 
of the single-particle Green function 
of the Hubbard model), 
we may solve the initial value problem by 
including only the Fermi type X-operators 
in the source term 
$\int_0^{\beta} d \tau \sum_{ i , p}  
J^p_{i} ( \tau ) X_i^p ( \tau )$ 
in the definition \eqref{eq:GXdef} of the
generating functional $G_{\Lambda} [ J ]$;  
see the first strategy discussed in  
Sec.~{\ref{subsec:HubbardFRG} 
after Eq.~\eqref{eq:Xabdyn}. 
Up to quadratic order in the expectation values
$\bar{X}^{ab} = \braket{ X^{ab} }$ 
of the X-operators 
the generating functional 
$\Gamma_{\Lambda} [ \bar{X} ]$
of the irreducible X-operator vertices 
defined in Eq.~\eqref{eq:GammaXdef}
has the following expansion 
in momentum-frequency space:
\begin{widetext}
\begin{align}
\Gamma_{\Lambda} [ \bar{X} ] 
& = 
\Gamma_{\Lambda} [ 0 ] 
+ \int_K \sum_{\sigma} 
\Bigl\{
\left[ 
t_{\bd{k}} + 
\Sigma_{\Lambda}^{ \sigma 0, 0 \sigma} ( K ) 
\right]
\bar{X}^{\sigma 0}_{-K} 
\bar{X}^{0 \sigma}_K
+ 
\left[ 
t_{\bd{k}} + 
\Sigma_{\Lambda}^{ 
2 \bar{\sigma} , 
\bar{\sigma} 2 } ( K ) 
\right]
\bar{X}^{2 \bar{\sigma}}_{-K} 
\bar{X}^{\bar{\sigma} 2}_K
\nonumber\\
& \hspace{23mm} 
+ \bar{\sigma} 
\left[ 
t_{\bd{k}} + 
\Sigma_{\Lambda}^{ \sigma 0,  
\bar{\sigma} 2} ( K ) 
\right]
\bar{X}^{\sigma 0}_{-K} 
\bar{X}^{ \bar{\sigma} 2}_K
+ \bar{\sigma} 
\left[ t_{\bd{k}} + 
\Sigma_{\Lambda}^{ 2 \bar{\sigma} ,  
0 {\sigma} } ( K ) 
\right]
\bar{X}^{2 \bar{\sigma}}_{-K} 
\bar{X}^{0{\sigma} }_K
\Bigr\} 
+ \ldots
\nonumber\\
& = 
\Gamma_{\Lambda} [ 0 ] 
+ \int_K \sum_{\sigma} 
{\bd{\psi}}^{\dagger}_{ K \sigma}
\begin{pmatrix}
t_{\bd{k}} + 
\Sigma_{\Lambda}^{ \sigma 0, 0 \sigma} ( K ) 
&
t_{\bd{k}} + \Sigma_{\Lambda}^{ \sigma 0, 
\bar{\sigma} 2} ( K ) 
\\
t_{\bd{k}} + 
\Sigma_{\Lambda}^{ 2 \bar{\sigma} ,  
0 {\sigma} } ( K ) 
&
t_{\bd{k}} + 
\Sigma_{\Lambda}^{ 2 \bar{\sigma} ,  
\bar{\sigma} 2} ( K )
\end{pmatrix}
{\bd{\psi}}_{ K \sigma} + \ldots \; ,
\label{eq:Gammaexp2}
\end{align}
\end{widetext}
where the collective label 
$K = ( \bd{k} ,  \omega)$ 
denotes momentum and 
fermionic Matsubara frequency, 
the integration symbol is defined as
$\int_K = 
\frac{1}{\beta N} \sum_{\bd{k}  \omega}$, 
where $N$ is the number of lattice sites, 
and the Fourier expansions of the
expectation values of the X-operators and 
the hopping are defined by
\begin{align}
\bar{X}_{i}^{ab} ( \tau ) 
& = 
\frac{1}{\beta N} \sum_{\bd{k}, \omega } 
e^{ i ( \bd{k} \cdot \bd{r}_i - \omega \tau ) } 
\bar{X}^{ab}_K ,
\\
t_{ ij} 
& = 
\frac{1}{N} \sum_{\bd{k}} 
e^{ i \bd{k} \cdot ( \bd{r}_i - \bd{r}_j ) } 
t_{\bd{k}} .
\end{align}
In the second line of Eq.~\eqref{eq:Gammaexp2},
we have introduced
two-component vectors analogous 
to Eq.~\eqref{eq:psi2flavor},
\begin{equation}
{\bd{\psi}}_{ K \sigma} = 
\begin{pmatrix} 
\bar{X}_K^{0 \sigma} \\
\bar{\sigma} \bar{X}_K^{ \bar{\sigma} 2} 
\end{pmatrix} , 
\; \; \;
{\bd{\psi}}^{\dagger}_{ K \sigma} = 
\left(  
\bar{X}_{-K}^{\sigma 0 } , 
\bar{\sigma} \bar{X}_{-K}^{2 \bar{\sigma}} 
\right) .
\label{eq:psi2flavorbar}
\end{equation}
For a given value of 
the deformation parameter $\Lambda$ 
the deformed two-point functions
of the Fermi type X-operators are
\begin{align}
&  
\begin{pmatrix} 
G_\Lambda^{ \sigma 0 , 0 \sigma} ( K ) 
& 
\bar{\sigma} 
G_{\Lambda}^{ \sigma 0, \bar{\sigma} 2} ( K ) 
\\
\bar{\sigma} 
G_\Lambda^{ 2 \bar{\sigma}  , 0 \sigma} ( K ) 
& 
G_{\Lambda}^{ 2 \bar{\sigma} , 
\bar{\sigma} 2} ( K )
\end{pmatrix}
=
\nonumber\\
& - \begin{pmatrix} 
t_{\bd{k}, \Lambda} +  
\Sigma_\Lambda^{ \sigma 0 , 0 \sigma} ( K ) 
& 
t_{\bd{k} , \Lambda} + 
\Sigma_{\Lambda}^{ 
\sigma 0, \bar{\sigma} 2} ( K ) 
\\
t_{\bd{k} , \Lambda} + 
\Sigma_\Lambda^{ 2 \bar{\sigma}, 0 \sigma} ( K )
& 
t_{\bd{k} , \Lambda} +  
\Sigma _{\Lambda}^{ 
2 \bar{\sigma} , \bar{\sigma} 2} ( K )
\end{pmatrix}^{-1} =
\nonumber\\
& 
\frac{- 1 }{ D_{\Lambda} ( K ) }
\begin{pmatrix}    
t_{\bd{k} , \Lambda} +  
\Sigma _{\Lambda}^{ 
2 \bar{\sigma} , \bar{\sigma} 2} ( K ) 
& 
- t_{\bd{k} , \Lambda} -  
\Sigma_{\Lambda}^{ 
\sigma 0, \bar{\sigma} 2} ( K ) 
\\
- t_{\bd{k} , \Lambda} - 
\Sigma_\Lambda^{ 2 \bar{\sigma} , 0 \sigma} ( K )
& 
t_{\bd{k}, \Lambda} +  
\Sigma_\Lambda^{ \sigma 0 , 0 \sigma} ( K )
\end{pmatrix} ,
\label{eq:correlatorvertices}
\end{align}
where 
$t_{\bd{k} , \Lambda} = 
t_{\bd{k}} + R_{ \bd{k} , \Lambda}^t $ 
is the Fourier transform of the 
deformed hopping and
\begin{align}
D_{\Lambda} ( K ) 
= {} &
\left[ 
t_{\bd{k}, \Lambda} +  
\Sigma_\Lambda^{ \sigma 0 , 0 \sigma} ( K ) 
\right]
\left[ 
t_{\bd{k} , \Lambda} +  
\Sigma _{\Lambda}^{ 
2 \bar{\sigma} , \bar{\sigma} 2} ( K ) 
\right]
\nonumber\\
& -  
\left[ 
t_{\bd{k} , \Lambda} + 
\Sigma_{\Lambda}^{ 
\sigma 0, \bar{\sigma} 2} ( K ) 
\right]
\left[ 
t_{\bd{k} , \Lambda} + 
\Sigma_\Lambda^{ 
2 \bar{\sigma}  , 0 \sigma} ( K ) 
\right] .
\end{align}
With the initial condition  
$t_{\bd{k} , \Lambda =0}=0$,
the initial self-energies are given by
the inverse two-point functions 
of the X-operators in the atomic limit.
A general method for calculating
arbitrary correlation functions of X-operators
in the atomic limit is presented in Appendix~\ref{app:spectral}.
In Appendix~\ref{app:static}, we summarize the
result for the two-point function.  
Using Eq.~\eqref{eq:Gabcd}, 
we obtain
\begin{equation}
\begin{pmatrix}  
\Sigma_0^{ \sigma 0 , 0 \sigma} ( K ) 
& 
\Sigma_{0}^{ \sigma 0, \bar{\sigma} 2} ( K ) 
\\
\Sigma_0^{ 2 \bar{\sigma}  , 0 \sigma} ( K ) 
& 
\Sigma _{0}^{ 
2 \bar{\sigma} , \bar{\sigma} 2} ( K )
\end{pmatrix} 
=
\begin{pmatrix}  
- \frac{ i \omega + \mu }{ x_0 + x_{\sigma} } 
& 0 \\
0 & - \frac{ i \omega + \mu - U 
}{ x_2 + x_{ \bar{\sigma}} } 
\end{pmatrix} ,
\label{eq:sigmainitial}
\end{equation}
where
\begin{equation}
x_a = \braket{ X^{aa}_i } , 
\; \; \;  \; \; \; 
a = 0, \uparrow, \downarrow, 2,
\end{equation}
are the expectation values of the 
diagonal X-operators in the atomic limit. 
The Hubbard-I approximation amounts 
to replacing the flowing self-energies 
of the X-operators by 
their initial values 
given in Eq.~\eqref{eq:sigmainitial}. 
In this approximation, 
the X-operator two-point functions are 
for $\Lambda \rightarrow 1$ 
(where $t_{\bd{k} , \Lambda} 
\rightarrow t_{\bd{k}}$) 
given by
\begin{align}
& 
\begin{pmatrix} 
G^{ \sigma 0 , 0 \sigma} ( K ) 
& 
\bar{\sigma} 
G^{ \sigma 0, \bar{\sigma} 2} ( K ) 
\\
\bar{\sigma} 
G^{ 2 \bar{\sigma}  , 0 \sigma} ( K ) 
& 
G^{ 2 \bar{\sigma} , \bar{\sigma} 2} ( K )
\end{pmatrix} 
\nonumber\\
= {} & 
\frac{1}{ D ( K ) }
\begin{pmatrix}      
\frac{   i \omega + \mu - U  
}{x_2 + x_{\bar{\sigma}}}  -  t_{\bd{k}} 
& 
t_{\bd{k} }  
\\
t_{\bd{k} } 
& 
\frac{  i \omega + \mu}{ x_0 + x_{\sigma}}  
-  t_{\bd{k}}
\end{pmatrix} ,
\end{align}
with
\begin{align}
D ( K ) 
= {} &  
\left[  
\frac{  i \omega + \mu}{ x_0 + x_{\sigma}}   
- t_{\bd{k}}   
\right]
\left[    
\frac{ 
i \omega + \mu - U}{x_2 + x_{\bar{\sigma}}
}   
-  t_{\bd{k}}   
\right] 
- t_{\bd{k}}^2
\nonumber\\
= {} & 
\left[ 
\frac{  i \omega + \mu}{ x_0 + x_{\sigma}} 
\right]
\left[ 
\frac{  i \omega + \mu - U}{
x_2 + x_{\bar{\sigma}}} 
\right]
\nonumber\\
&
- t_{\bd{k}} 
\left[ 
\frac{  i \omega + \mu}{ x_0 + x_{\sigma}} + 
\frac{  i \omega + \mu - U}{
x_2 + x_{\bar{\sigma}}} 
\right] .
\end{align}
The corresponding electronic Green function 
for spin $\sigma$ electrons is
\begin{align}
G_{\sigma} ( K ) 
= {} &  
G^{ 0 \sigma , \sigma 0} ( K )  +  
G^{ \bar{\sigma} 2  , 2 \bar{\sigma}} ( K )
\nonumber
\\
& + 
\bar{\sigma} 
G^{ 0 \sigma , 2 \bar{\sigma}} ( K ) + 
\bar{\sigma} 
G^{\bar{\sigma} 2  , \sigma 0} ( K )
\nonumber
\\
= {} & 
\frac{
\frac{ i \omega + \mu}{ x_0 + x_{\sigma}} + 
\frac{  i \omega + \mu - U}{
x_2 + x_{\bar{\sigma}}}
}{ 
\left[ 
\frac{ i \omega + \mu}{ x_0 + x_{\sigma}} 
\right]
\left[
\frac{  i \omega + \mu - U}{
x_2 + x_{\bar{\sigma}}} 
\right]
- t_{\bd{k}} 
\left[ 
\frac{  i \omega + \mu}{ x_0 + x_{\sigma}} + 
\frac{  i \omega + \mu - U}{
x_2 + x_{\bar{\sigma}}} 
\right]
}
\nonumber\\
= & {} 
\frac{   
\frac{ x_0 + x_{\sigma} }{  i \omega + \mu} + 
\frac{x_2 + x_{\bar{\sigma}}}{ 
i \omega + \mu - U}   
}{
1  - t_{\bd{k}} 
\left[ 
\frac{ x_0 + x_{\sigma}}{  i \omega + \mu}  + 
\frac{x_2 + x_{\bar{\sigma}}}{  
i \omega + \mu - U}     
\right]
} .
\label{eq:hub1}
\end{align}
Keeping in mind that  
in the atomic limit 
the probability of  
finding an electron with spin $\sigma$ is
$n_{\sigma} =  x_2 + x_{\sigma}$ 
and using 
\begin{align}
x_0 + x_{\sigma} 
& =  
1 - ( x_{\uparrow} + x_{\downarrow} + x_2 ) 
+ x_{\sigma} 
\nonumber\\
& =  
1 - x_2 - x_{ \bar{\sigma}}
= 1 - n_{\bar{\sigma}} ,
\end{align}
we see that $G_\sigma (K )$ 
in Eq.~\eqref{eq:hub1} can also 
be written as
\begin{align}
& G_{\sigma} (K) 
\nonumber\\
& =
\frac{ 
i \omega + \mu - U ( 1 - n_{ \bar{\sigma}} )
}{
( i \omega + \mu )(i \omega + \mu - U ) - 
t_{\bd{k}} 
\left[ i \omega + \mu - 
U ( 1 - n_{\bar{\sigma} } ) 
\right]
}
\nonumber
\\
& = \frac{ 
i \omega + \mu - U ( 1 - n_{ \bar{\sigma}} )
}{ 
( i \omega + \mu  - \epsilon^{+}_{\bd{k}} )
( i \omega + \mu - \epsilon^-_{\bd{k}} ) } ,
\label{eq:hub1b}
\end{align}
where
\begin{equation}
\epsilon_{\bd{k}}^{\pm} = 
\frac{ t_{\bd{k}} + U }{2} 
\pm \sqrt{  
\left( \frac{ t_{\bd{k}} - U }{2} \right)^2 
+ t_{\bd{k}} U n_{\bar{\sigma}} 
} .
\end{equation}
Equation~\eqref{eq:hub1b} is the 
well-known Hubbard-I approximation for the
electronic single-particle Green functions of 
the Hubbard model \cite{Fulde95,Ovchinnikov04},
which is exact both in the non-interacting limit
$U \rightarrow 0$ and in the 
atomic limit $t_{\bd{k}} \rightarrow 0$.
Note that for vanishing magnetic field 
$n_{\uparrow} = n_{\downarrow } $ is independent 
of the spin projection~$\sigma$. From
Eq.~\eqref{eq:hub1b} it is clear, that the X-operator self-energies
defined in Eq.~\eqref{eq:sigmainitial} are non-trivially related 
to the usual self-energy  $\Sigma(K)$ defined via $ G_\sigma^{-1}(K) = i\omega+\mu-t_{\bm{k}}-\Sigma(K)$.
Therefore even the initial condition of our X-FRG approach amounts to a
non-trivial resummation in perturbation theory.

\subsection{Quasi-particles in the hidden Fermi liquid}
\label{sec:hidden}

In this section, we focus on 
the Hubbard model for 
infinite on-site repulsion, 
i.e.~the $t$ model 
introduced in Sec.~\ref{sec:tdef}.  
The Hamiltonian of the model consists only 
of the projected kinetic energy 
given in Eq.~\eqref{eq:Htdef}.  
A slight complication arises from the fact 
that for fixed electronic filling $n$ 
the chemical potential $\mu_{\Lambda}$ 
depends on the flow-parameter $\Lambda$ 
when the hopping is switched on.  
In a conventional Fermi liquid 
one usually eliminates
the associated shift in the
self-energy via a suitable 
counter term \cite{Anderson93}.
To implement a similar procedure 
in our FRG approach to the $t$ model,
we use the operator identity 
$ X_i^{\sigma 0} X_i^{0 \sigma} = 
X_i^{\sigma \sigma}$ to rewrite the
deformed grand canonical Hamiltonian 
of the $t$ model as
\begin{equation}
	\label{eq: hred1}
\tilde{\cal{H}}_{ t, \Lambda} =  
{{\cal{H}}}_1 + 
{{\cal{H}}}_{2 , \Lambda} ,
\end{equation}
with
\begin{align}
	\label{eq: hred2}
{\cal{H}}_1 
& = 
- \mu_0 \sum_{ i \sigma} 
\tilde{c}^{\dagger}_{ i \sigma} 
\tilde{c}_{ i \sigma} 
= 
- \mu_0 \sum_{ i \sigma}  X^{\sigma \sigma}_i ,
\\
{{\cal{H}}}_{ 2 , \Lambda} 
& =  
\sum_{ i j \sigma} 
\left( 
t_{ ij, \Lambda} - \delta \mu_{\Lambda} 
\right) 
\tilde{c}^{\dagger}_{ i \sigma} 
\tilde{c}_{ j \sigma} ,
\end{align}
where 
\begin{equation}
\delta \mu_{\Lambda}  = 
\mu_{\Lambda} - \mu_0
\end{equation}
is the difference between 
the chemical potential $\mu_{\Lambda}$ 
at scale $\Lambda$ and 
the chemical potential $\mu_0$ 
in the atomic limit 
(i.e.,~for vanishing hopping),
which for a given electronic filling $n_0$ 
is determined by the atomic equation of state,
\begin{equation}
n_0 = \frac{ 2 e^{\beta \mu_0}}{ 
1 + 2 e^{\beta \mu_0 } } .
\label{eq:nnull}
\end{equation}
Solving for $\mu_0$ we obtain
\begin{equation}
\mu_0 = T \ln 
\left( \frac{n_0}{2 (1-n_0 ) } \right) .
\label{eq:munulldef}
\end{equation}
Assuming that at the initial value $\Lambda =0$ 
the deformed hopping $t_{ ij, \Lambda=0} $ 
vanishes, 
the initial generating functional 
of the connected correlation functions is
\begin{equation}
e^{{\cal{G}}_{0} [ J ] }
=  {\rm Tr} \left[ 
e^{ - \beta {\cal{H}}_1 } {\cal{T}}
e^{ \int_0^{\beta} d \tau \sum_{i, p} 
J^p_i ( \tau ) X^p_i ( \tau ) }   
\right] .
\label{eq:G0tdef}
\end{equation}
Since the sites are decoupled in this limit, 
the generating functional is 
the sum of single-site generating functionals,
\begin{equation}
{\cal{G}}_{0} [ J ] = \sum_i 
{\cal{G}}_{\rm site} [ J_i ] ,
\end{equation}
where the generating functional of 
the single-site (atomic) correlation functions is
\begin{equation}
{\cal{G}}_{\rm site} [ J ] =
\ln {\rm tr} \left[
e^{
\beta \mu_0 \sum_{\sigma} X^{\sigma \sigma}   
}
{\cal{T}}
e^{ \int_0^{\beta} d \tau \sum_{ p} 
J^p ( \tau ) X^p ( \tau ) } \right] .
\end{equation}
Here, 
the symbol  ${\rm tr} [ \ldots ]$ denotes 
the trace over the 
three-state projected Hilbert space
associated with a single lattice site.
Since we are only interested in 
the correlation functions of the
Fermi type X-operators, 
it is sufficient to
introduce only four Grassmann sources 
$J^{ 0 \uparrow}$, 
$J^{\uparrow 0}$, 
$J^{0 \downarrow}$ and 
$J^{ \downarrow 0}$.
The source term then reduces to
\begin{align}
& 
\int_0^{\beta}  d \tau \sum_{ p} 
J^p ( \tau ) X^p ( \tau ) 
\nonumber\\
= {} & 
\int_0^{\beta} \sum_{\sigma}  
\left[ 
J^{ 0 \sigma} ( \tau ) 
X^{0 \sigma} ( \tau ) + 
J^{\sigma 0} ( \tau )  
X^{ \sigma 0} ( \tau )  
\right]
\nonumber\\
= {} & 
\int_0^{\beta} d \tau \sum_{\sigma} 
\left[ 
J^{ 0 \sigma} ( \tau ) 
\tilde{c}_{ \sigma} ( \tau ) + 
J^{\sigma 0} ( \tau ) 
\tilde{c}^{\dagger}_{ \sigma } ( \tau ) 
\right]
\nonumber\\
= {} & 
\int_0^{\beta} d \tau \sum_{\sigma} 
\left[ 
\bar{j}_{  \sigma} ( \tau ) 
\tilde{c}_{ \sigma} ( \tau ) +  
\tilde{c}^{\dagger}_{ \sigma } ( \tau )  
j_{\sigma} ( \tau )
\right] ,
\end{align}
where in the last line we have set 
$J^{0 \sigma} ( \tau ) = 
\bar{j}_{\sigma} ( \tau )$ and
$J^{\sigma 0} ( \tau ) = 
\zeta j_{\sigma} ( \tau ) = 
- j_{\sigma} ( \tau )$.
With the notation
\begin{equation}
\psi_{i \sigma} = 
\bar{X}_i^{ 0 \sigma} = 
\braket{ \tilde{c}_{ i \sigma} } , 
\; \; \;  \; \; \;
\bar{\psi}_{i \sigma} = 
\bar{X}_i^{  \sigma 0} = 
\braket{ \tilde{c}^{\dagger}_{ i \sigma} } ,
\end{equation}
the generating functional 
$ \Gamma_{\Lambda} [ \bar{\psi} , \psi ] =
\Gamma_{\Lambda} [ \bar{X} ] $
defined in Eq.~\eqref{eq:GammaXtJdef} 
has for the $t$ model the 
following vertex expansion 
in momentum-frequency space:
\begin{align}
&  
\Gamma_{\Lambda} [ \bar{\psi} , \psi ] =   
\Gamma_{\Lambda} [0,0] +
\int_K \sum_{\sigma}  
\left[ 
t_{\bd{k}} + \Sigma_{\Lambda} ( K ) 
\right] 
\bar{\psi}_{ K \sigma} 
\psi_{ K \sigma}
\nonumber\\
& 
+ \frac{1}{(2!)^2} 
\int_{ K_1^{\prime} } 
\int_{ K_2^{\prime} } 
\int_{ K_2 } 
\int_{ K_1 } 
\sum_{ \sigma_1^{\prime} \sigma_2^{\prime}
\sigma_2 \sigma_1}
\delta_{ K_1^{\prime} + K_2^{\prime} , 
K_2 + K_1 }
\nonumber\\
& \hspace{10mm}
\times
\Gamma^{\bar{\psi} \bar{\psi} \psi \psi}_{\Lambda} 
( 
K_1^{\prime} \sigma_1^{\prime}, 
K_2^{\prime} \sigma_2^{\prime} ; 
K_2 \sigma_2,  K_1 \sigma_1
) 
\nonumber\\
& \hspace{10mm} 
\times
\bar{\psi}_{ K_1^{\prime} \sigma_1^{\prime} }
\bar{\psi}_{ K_2^\prime \sigma_2^{\prime} }
\psi_{ K_2 \sigma_2} \psi_{ K_1 \sigma_1} 
+   \ldots \; ,
\label{eq:vertexexpt}
\end{align}
where $\delta_{K,0}=\beta N \delta_{\bm{k},0}\delta_{\omega,0}$ and the ellipsis denotes 
higher-order vertices. 
The Fourier components 
are defined as 
\begin{subequations}
\begin{align}
\psi_{ K \sigma} & = 
\bar{X}^{ 0 \sigma}_{K} = 
\sum_i \int_0^{\beta} d \tau 
e^{ - i ( \bd{k} \cdot \bd{r}_i - \omega \tau )}
\psi_{ i \sigma} ( \tau ) ,
\\
\bar{\psi}_{ K \sigma} & = 
\bar{X}^{  \sigma 0 }_{- K} = 
\sum_i \int_0^{\beta} d \tau 
e^{  i ( \bd{k} \cdot \bd{r}_i - \omega \tau )}
\bar{\psi}_{ i \sigma} ( \tau ) .
\end{align}
\end{subequations}
The scale-dependent irreducible four-point vertex
in Eq.~\eqref{eq:vertexexpt}
is antisymmetric with respect to 
the exchange of the first two outgoing labels
$K_1^{\prime} \sigma_1^{\prime}
\leftrightarrow K_2^{\prime} \sigma_2^{\prime}$ 
and with respect to the exchange of 
the two incoming labels
$K_1 \sigma_1 \leftrightarrow K_2 \sigma_2$.
Note that the SU$(2)$ spin-rotational invariance 
of the Hubbard model implies that 
the spin-dependence of the 
antisymmetrized effective interaction
is of the form \cite{Kopietz10}
\begin{align}
&  
\Gamma^{\bar{\psi} \bar{\psi} \psi \psi }_{\Lambda} ( K_1^{\prime} \sigma_1^{\prime}, 
K_2^{\prime} \sigma_2^{\prime}; 
 K_2 \sigma_2 , K_1 \sigma_1 ) 
\nonumber\\
= {} &  
\delta_{ \sigma_1^{\prime} \sigma_1}
\delta_{\sigma_2^{\prime} \sigma_2}  
U_{\Lambda} ( K_1^{\prime},
K_2^{\prime}; K_2 , K_1 )
\nonumber\\
& - 
\delta_{ \sigma_1^{\prime} \sigma_2}
\delta_{\sigma_2^{\prime} \sigma_1}  
U_{\Lambda} ( K_1^{\prime},
K_2^{\prime}; K_1 , K_2 ),
\end{align}
where the function 
$U_{\Lambda} ( K_1^{\prime},
 K_2^{\prime}; K_1 , K_2 )$ 
is symmetric under the simultaneous exchange 
of its outgoing and incoming labels,
\begin{equation}
U_{\Lambda} 
( K_1^{\prime}, K_2^{\prime}; K_2 , K_1 ) = 
U_{\Lambda} 
( K_2^{\prime}, K_1^{\prime}; K_1 , K_2 ) .
\end{equation}
An alternative parametrization of the 
vertex expansion \eqref{eq:vertexexpt} 
is therefore
\begin{align}
& 
\Gamma_{\Lambda} [ \bar{\psi} , \psi ]  =  
\Gamma_{\Lambda} [0,0] +
\int_K \sum_{\sigma}  
\left[ 
t_{\bd{k}} + \Sigma_{\Lambda} ( K ) 
\right] 
\bar{\psi}_{ K \sigma} \psi_{ K \sigma}
\nonumber\\
& 
+ \frac{1}{2} 
\int_{ K_1^{\prime} } 
\int_{ K_2^{\prime}} 
\int_{ K_2} 
\int_{ K_1} 
\sum_{ \sigma_1 \sigma_2}
\delta_{ K_1^{\prime} + K_2^{\prime} , 
K_2 + K_1 }
\nonumber\\
& \; \; \; \; \times
U_{\Lambda} ( K_1^{\prime}, K_2^{\prime} ; 
K_2 , K_1 ) 
\bar{\psi}_{ K_1^{\prime} \sigma_1 } 
\bar{\psi}_{ K_2^\prime \sigma_2 }
\psi_{ K_2 \sigma_2} \psi_{ K_1 \sigma_1} 
\nonumber\\
& +  \ldots \; .
\label{eq:vertexexpt2}
\end{align}
It is important to emphasize that 
the four-point vertex and the 
higher-order vertices
represented by the ellipsis in 
the vertex expansions~\eqref{eq:vertexexpt} and
\eqref{eq:vertexexpt2} arise from 
the kinematic correlations 
due to the Hilbert space projection 
in the $t$ model model. 
If we replace the projected Fermi-like operators
$ \tilde{c}_{\sigma}$ and 
$\tilde{c}^{\dagger}_{\sigma}$ by 
canonical fermions and thus 
neglect the Hilbert space projection,
the generating functional 
${\cal{G}}_{\rm site} [ J ]$ reduces to  
a trivial quadratic functional 
of the sources, 
so that all interaction vertices 
in the vertex expansion \eqref{eq:vertexexpt}  
vanish and the self-energy
$\Sigma_{\Lambda} ( K )$ reduces to 
$ - i \omega - \mu$.

For a given value of $\Lambda$, 
the deformed two-point function 
of the projected fermions is
\begin{equation}
G_\Lambda ( \bd{k} , \omega ) 
= \frac{ - 1 }{ 
t_{\bd{k} , \Lambda}  - \delta \mu_{\Lambda} 
+ \Sigma_{\Lambda} (  \bd{k} , \omega ) } ,
\label{eq:G2pt}
\end{equation}
where the initial value of the self-energy 
is given by its atomic limit
derived in Appendix~\ref{app:static},
\begin{equation}
\Sigma_0 (  \omega ) = 
- \frac{ i \omega + \mu_0 }{ 
1 - \frac{n_0}{2} } .
\label{eq:sigmainitialtJ}
\end{equation}

Following the usual procedure \cite{Anderson93}, 
we assume that the counter-term 
$\delta \mu_{\Lambda}$ takes the shift 
in the chemical potential due to the
self-energy at zero frequency into account. 
Thus, 
we demand that 
\begin{equation}
\Sigma_{\Lambda} (  \bd{k}_F , 0) = 
\Sigma_0 ( \omega = 0 ) 
=
- \frac{ \mu_0 }{ 1 - \frac{ n_0 }{ 2 } } .
\label{eq:Sigmacounter}
\end{equation}
for wavevectors $\bd{k}_F$ 
on the Fermi surface.
As $ \mu_0 \to 0 $ for $ T \to 0 $
according to Eq.~\eqref{eq:munulldef},
the flowing Fermi surface is then determined by
\begin{equation}
t_{\bd{k}_F , \Lambda} = \delta \mu_{\Lambda} = \mu_{\Lambda}
\label{eq:FSdef}
\end{equation}
at zero temperature.
Actually, 
as pointed out by Anderson \cite{Anderson93}, 
in general the counter-term $\delta \mu$ 
depends on the momentum $\bd{k}_F$ 
on the Fermi surface.  
Here we can ignore this subtlety
because within our truncation 
the self-energy $\Sigma_{\Lambda} ( \omega )$
is momentum-independent. 
Once we have determined the 
two-point function \eqref{eq:G2pt}
within some approximation, 
we can obtain the equation of state
at a given $\Lambda$ in the implicit form
\begin{equation}
n_\Lambda = 
\frac{2}{ \beta N} \sum_{ \bd{k} , \omega}
G_{\Lambda} ( \bd{k} , \omega ) 
e^{ i \omega 0^+ } ,
\label{eq:eos}
\end{equation}
where the right-hand side depends 
on $\mu_\Lambda$ and $n_\Lambda$.

Before investigating the actual flow equations,
it is instructive to consider
the approximation where
the self-energy $ \Sigma_{\Lambda} ( \omega )$
is replaced by its atomic limit
\eqref{eq:sigmainitialtJ}.
In this case the above counter-term procedure
is not directly applicable because 
the condition \eqref{eq:Sigmacounter}
is trivially fulfilled.
Therefore
we fix the counter-term $ \delta \mu_\Lambda $
instead by demanding that the filling
does not change during the flow;
i.e.,~$ n_\Lambda = n_0 = n $.
For a given filling $n$,
the bare chemical potential 
$ \mu_0 = \mu_0 ( n , T)$ is then 
given by  Eq.~\eqref{eq:munulldef}.
At the end of the RG flow 
(i.e.,~for $\Lambda  \rightarrow 1$ where
$t_{\bd{k} , \Lambda} \rightarrow t_{\bd{k}}$ and
$\delta \mu _{\Lambda} \rightarrow \mu - \mu_0$) 
the two-point function is in this approximation 
given by
\begin{equation}
{G}_1 ( \bd{k} , \omega ) = 
\frac{Z_0}{ 
i \omega  - Z_0 ( t_{\bd{k}} - \delta \mu ) 
+ \mu_0 } ,
\label{eq:G1def}
\end{equation}
with quasi-particle residue
\begin{equation}
Z_0 = 1 - \frac{n}{2} .
\end{equation}
Note that if we remove 
the counter-term ($\delta \mu =0$),
Eq.~\eqref{eq:G1def} can also be obtained 
from the Hubbard-I approximation \eqref{eq:hub1b}
for the electronic single-particle Green function
of the Hubbard model by 
taking the limit $U \rightarrow \infty$.
At zero temperature where $\mu_0$ vanishes 
the equation of state \eqref{eq:eos} 
then reduces to
\begin{equation}
n =  
\frac{ 2 Z_0 }{N} \sum_{\bd{k}} 
\Theta ( \mu - t_{\bd{k}} ) .
\label{eq:eos1a}
\end{equation} 
In terms of the hole-doping $ x =1 - n$ 
this can be written as
\begin{equation}
\frac{1 - x}{1+x} = 
\frac{1}{N} \sum_{\bd{k}} 
\Theta ( \mu - t_{\bd{k}} ) .
\label{eq:eos1}
\end{equation}
In Fig.~\ref{fig:doping}, 
we show the resulting filling $n$ and 
hole-doping $x$ as a function of 
the chemical potential $\mu$ for 
the $t$ model with nearest neighbor hopping 
in a two-dimensional square lattice.
\begin{figure}
\centering
\includegraphics[width=1\linewidth]{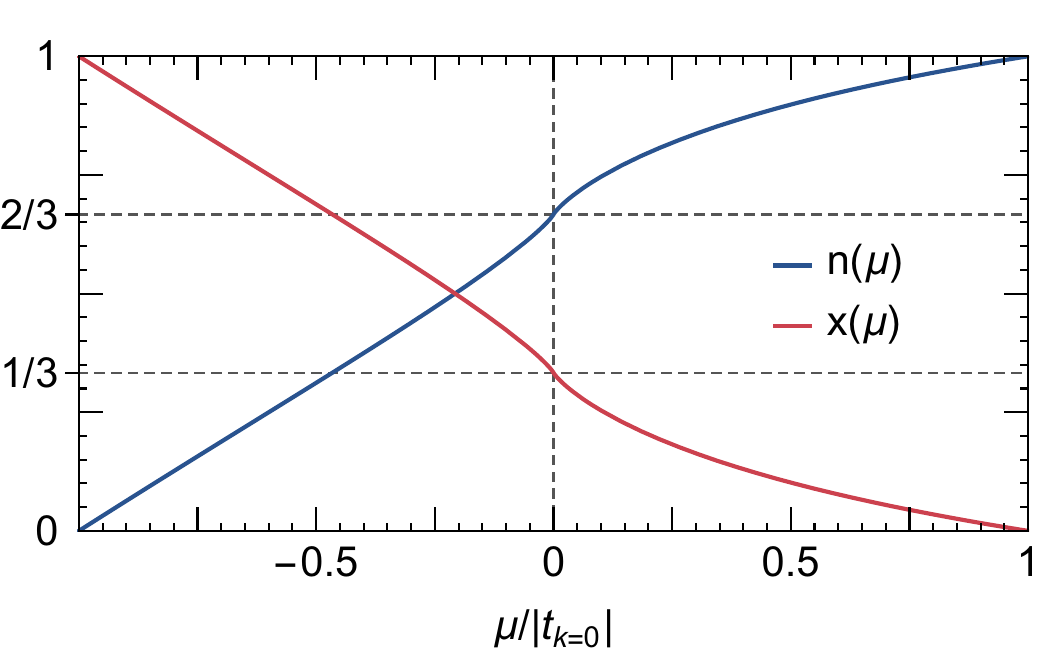}
\caption{Lattice filling $n$ and corresponding hole doping $x = 1-n$
of the $t$ model for nearest neighbor hopping on a square lattice as a function
of the chemical potential $\mu$.
The graph has been obtained from the numerical solution of the
approximate equation of state \eqref{eq:eos1} with $t_{\bd{k}} = - 2 t [ \cos ( k_x a )
 + \cos ( k_y a ) ]$ where $t > 0$ and $a$ is the lattice spacing.
}
 \label{fig:doping}
 \end{figure}
Note that according to Eq.~\eqref{eq:FSdef} 
the Fermi surface is for $T \rightarrow 0$ 
defined by
\begin{equation}
t_{\bd{k}_F} = \mu,
\label{eq:FSzero}
\end{equation}
so that for $\mu =0$ 
where the Fermi surface 
covers half of the first Brillouin zone 
(see Fig.~\ref{fig:FShalf})
we obtain $x =1/3$,  
corresponding to $ n = 2/3$.
This clearly shows that 
Luttinger's theorem \cite{Luttinger61} 
(which states that in a Fermi liquid the
volume of the Fermi surface is proportional 
to the electronic density) 
is violated in the
$t$ model \cite{Putikka98}. 
The violation of Luttinger's theorem 
in the $t$ model is a consequence of 
the projected Hilbert space. 
 \begin{figure}
 \centering
 \includegraphics[width=1\linewidth]{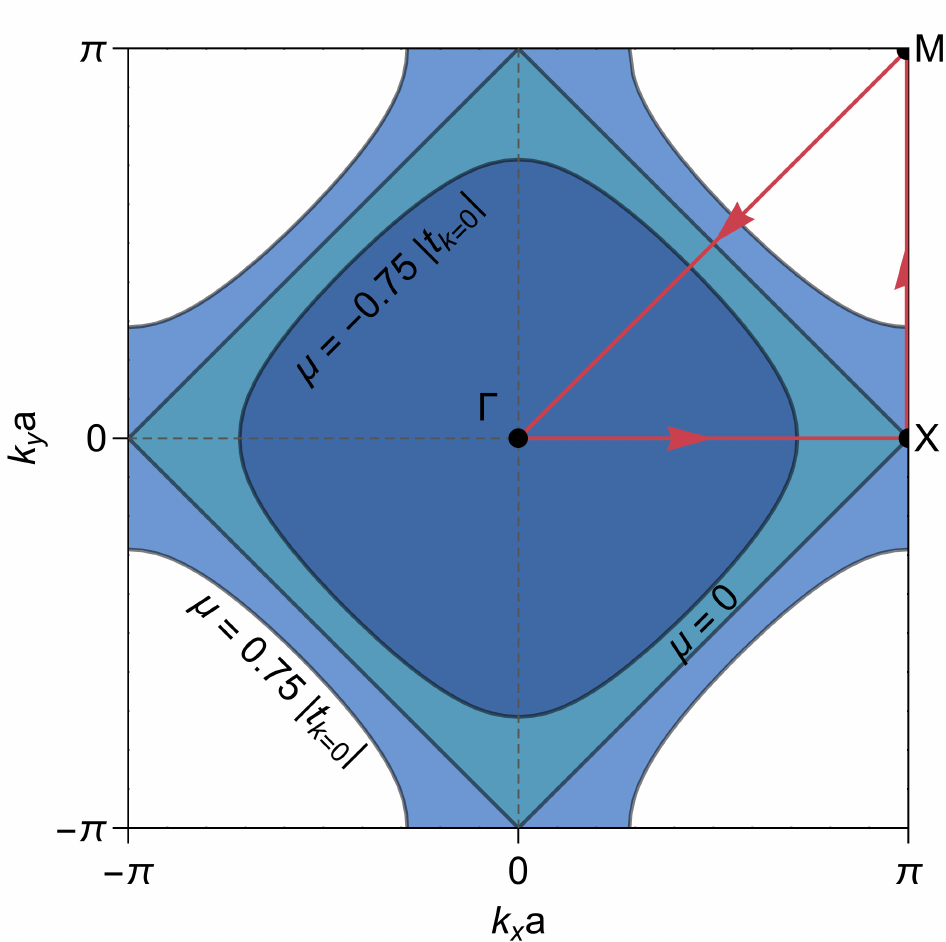}
 \caption{Fermi surface of the $t$ model with nearest-neighbor
 	 hopping on a square lattice for 
 	 $\mu =0$, $0.75\,|t_{\bm{k}=0}|$ and $-0.75\,|t_{\bm{k}=0}|$. 
	Although the Fermi surface at $\mu = 0$ covers half 
	of the first Brillouin zone, the corresponding filling is $n < 1$
	due to the projected Hilbert space. The black dots 
	denote the high symmetry points 
	$\bm{\Gamma} = (0,0)$, $\textmd{\textbf{X}} = (\pi,0)$
	 and $\textmd{\textbf{M}}= (\pi,\pi)$.
	 The red arrows show the path through the Brillouin zone taken in
	  Fig.~\ref{fig:damping}.
	}
 \label{fig:FShalf}
 \end{figure}

The equation of state  \eqref{eq:eos1a} 
has been obtained by approximating 
the self-energy $\Sigma_{\Lambda} ( K )$ 
by its atomic limit given 
in Eq.~\eqref{eq:sigmainitialtJ}.
To go beyond this approximation,
we now use the 
Wetterich equation \eqref{eq:Wetterich3} 
for the $t$-$J$ model in the limit
$J=0$ to derive a formally exact flow equation
for the self-energy $\Sigma_{\Lambda} ( K )$ 
of the $t$ model.
For vanishing $J$ the flow 
in the fermionic sector 
decouples from the bosonic sector, 
so that the flow equation of 
the fermionic self-energy
of the $t$ model is formally identical 
to the flow equation of the self-energy
of a system of 
canonical fermions~\cite{Kopietz10,Kopietz01},
\begin{align}
\partial_{\Lambda} \Sigma_{\Lambda} ( K )
= {} &   
\int_{ K^{\prime}}
\sum_{ \sigma^{\prime} } 
\dot{G}_{\Lambda} ( K^{\prime} )
\nonumber\\
& \times
\Gamma^{\bar{\psi} \bar{\psi} \psi \psi }_{\Lambda} 
( K \sigma, K^{\prime} \sigma^{\prime}; 
 K^{\prime} \sigma^{\prime} , K \sigma ) ,
\label{eq:sigmaflowt}
\end{align}
where the single-scale propagator is
\begin{equation}
\dot{G}_{\Lambda} ( K ) = 
G^2_{\Lambda} ( K )  
\partial_{\Lambda} 
\left(
t_{ \bd{k} , \Lambda} - \delta \mu_{\Lambda}
\right) .
\end{equation}
The exact flow equation of the four-point vertex 
$\Gamma^{\bar{\psi} \bar{\psi} \psi \psi }_{\Lambda} ( K \sigma, K^{\prime} \sigma^{\prime}; 
 K^{\prime} \sigma^{\prime} , K \sigma )$ involves the six-point vertex and can be found
in Refs.~[\onlinecite{Kopietz10,Kopietz01}].
These vertices are
related via skeleton 
equations to the corresponding 
connected correlation functions \cite{Kopietz10}.
In a simple level-1 truncation, 
we now approximate
the flowing four-point vertex 
on the right-hand side
of the flow equation \eqref{eq:sigmaflowt} by its
initial value for vanishing hopping.
In principle the corresponding four-point correlation function
can be obtained
using the recursive algorithm 
for the Matsubara kernel functions 
recently derived by 
Halbinger {\it{et al.}} \cite{Halbinger23}.
A more direct method 
which we have previously used to 
calculate the initial vertices in our 
recently developed spin-FRG approach 
to dimerized spin systems \cite{Rueckriegel22}  
is presented in Appendix~\ref{app:spectral} and
applied in Appendix~\ref{app:4point}.
For our purpose we need the four-point vertex
only for the special combinations of spin labels
$\sigma_1^{\prime} = 
\sigma_2^{\prime} = 
\sigma_2 = 
\sigma_1 = 
\sigma$ and 
$\sigma_1^{\prime} = 
\sigma_1 = 
\sigma = 
- \sigma_2^{\prime} = 
- \sigma_2$.
When all spin projections are equal we obtain
\begin{align}
& 
\Gamma_0^{ \bar{\psi} \bar{\psi} \psi \psi }
(  \omega \sigma , \omega^{\prime} \sigma ; \omega^{\prime} \sigma , \omega \sigma)
\nonumber\\
& =  
- \frac{ \beta  }{ Z_0^3 } 
\frac{n_0}{2} 
( 1 - \delta_{\omega , \omega^{\prime}} ) 
( i \omega + \mu_0 ) 
( i \omega^{\prime} + \mu_0 ) ,
\label{eq:Gamma4a}
\end{align}
while for the interaction between 
two projected fermions with opposite spin we find
\begin{align}
& 
\Gamma_0^{ \bar{\psi} \bar{\psi} \psi \psi }
(  \omega \sigma , \omega^{\prime} \bar{\sigma} ; \omega^{\prime} \bar{\sigma} , \omega \sigma)
=   
- \frac{ 
i \omega + i \omega^{\prime} + 2 \mu_0 
}{ Z_0^3 }
\nonumber\\
& + \frac{ \beta  }{  Z_0^4 }  
\frac{n_0}{2} 
\left( 
\frac{n_0}{2} + 
\delta_{ \omega , \omega^{\prime} } 
\right)  
(i \omega + \mu_0 ) 
( i \omega^{\prime} + \mu_0 ) .
\label{eq:Gamma4b}
\end{align}
Here, 
$\bar{\sigma} = - \sigma$ denotes the 
inverted spin projection and 
we have omitted the momentum labels 
because in the atomic limit 
the vertices depended only on the
frequency part $\omega$ 
of the collective labels 
$K = ( \bd{k} ,  \omega )$.
Note also that now
$ Z_0 = 1 - \frac{ n_0 }{ 2 } $,
since we no longer require 
the filling to remain constant 
during the flow.
Substituting the 
initial values~\eqref{eq:Gamma4a} and
\eqref{eq:Gamma4b} for the four-point vertex 
into the flow equation \eqref{eq:sigmaflowt}, 
we obtain
\begin{align}
& 
\partial_{\Lambda} \Sigma_{\Lambda} ( \omega ) =
\frac{( i \omega + \mu_0 )^2}{Z_0^4} 
\frac{n_0}{2} \left( 2- \frac{n_0}{2} \right)  
A_{\Lambda} ( \omega )
\nonumber\\
& - \frac{( i \omega + \mu_0 )}{Z_0^3} 
\left[ 
B_{\Lambda} + \frac{n_0}{2} \left(1 - n_0 \right) 
\frac{\beta}{Z_0} C_{\Lambda} 
\right] 
- \frac{C_{\Lambda}}{Z_0^3}, 
\label{eq:Sigmaflowt1}
\end{align}
where
\begin{subequations}
\begin{align}
A_{\Lambda} (\omega ) & = 
\frac{1}{N} \sum_{ \bd{k} }
\dot{G}_{\Lambda} ( \bd{k} , \omega ) ,
\\
B_{\Lambda} & = 
\frac{1}{\beta N} \sum_{ \bd{k} \omega^{\prime} }
\dot{G}_{\Lambda} ( \bd{k} , \omega^{\prime} ) ,
\\
C_{\Lambda} & = 
\frac{1}{\beta N} \sum_{ \bd{k} \omega^{\prime} }
\dot{G}_{\Lambda} ( \bd{k} , \omega^{\prime} ) 
( i \omega^{\prime} + \mu_0 ) .
\end{align}
\end{subequations}
A comprehensive analysis of the 
flow equation \eqref{eq:Sigmaflowt1} 
will be presented elsewhere \cite{Rueckriegel23}.
Here, 
we focus for simplicity 
on the limit $T \rightarrow 0$ 
where $\mu_0 \rightarrow 0$.
To satisfy the condition \eqref{eq:Sigmacounter} 
we should then choose the 
counter-term $\mu_{\Lambda}$ such that
\begin{align}
\partial_{\Lambda} \Sigma_{\Lambda} ( 0 ) 
& = 
 \frac{\mu_0^2}{Z_0^4}\frac{n_0}{2}\left(2-\frac{n_0}{2}\right)A_\Lambda(0)-\frac{\mu_0}{Z_0^3}B_\Lambda
\nonumber\\
&- \left[  
\frac{ n_0 (1 - n_0 )}{ 2 Z^4_0 } 
\tilde{\mu}_0
+ \frac{1}{Z_0^3} 
\right]
C_{\Lambda}
\nonumber\\
& = 0 ,
\label{eq:sigmacounter}
\end{align}
where $\tilde{\mu}_0 = \beta \mu_0$ is independent of temperature; see Eq.~\eqref{eq:munulldef}.
Equation~\eqref{eq:sigmacounter} can be satisfied 
by choosing the counter-term 
$\mu_{\Lambda}$ such that
\begin{equation}
C_{\Lambda} = \frac{\mu_0}{1+\frac{n_0}{2}\frac{1-n_0}{Z_0}\tilde{\mu}_0}\left[\frac{n_0}{2}\frac{\left(2-\frac{n_0}{2}\right)}{Z_0}\mu_0 A_\Lambda(0)-B_\Lambda\right],
\label{eq:Cnull}
\end{equation}
which 
together with the initial condition 
$\Sigma_0 ( 0 ) = - \mu_0 / Z_0  \rightarrow 0$ 
at $T \rightarrow 0$
guarantees that $\Sigma_{\Lambda} ( 0 )$ 
vanishes at zero temperature for all values
of the deformation parameter $\Lambda$.
At $ T = 0 $ the condition \eqref{eq:Cnull} reduces
to $ C_\Lambda = 0$.
Solving
for $\partial_\Lambda \mu_{\Lambda}$, 
we conclude that at $T=0$ the 
scale-dependent chemical potential 
$\mu_{\Lambda} $ is determined by the
flow equation
\begin{equation}
\partial_{\Lambda} \mu_{\Lambda}  =
\frac{ 
\frac{1}{\beta N} \sum_{ \bd{k} \omega^{\prime} }
( \partial_{\Lambda} t_{\bd{k} , \Lambda} )
{G}^2_{\Lambda} ( \bd{k} , \omega^{\prime} )  
i \omega^{\prime}  
}{
\frac{1}{\beta N} \sum_{ \bd{k} \omega^{\prime} }
{G}^2_{\Lambda} ( \bd{k} , \omega^{\prime} )  i \omega^{\prime} 
} .
\label{eq:muflow}
\end{equation}
Replacing $C_\Lambda$ according to Eq.~\eqref{eq:Cnull}, the flow equation~\eqref{eq:Sigmaflowt1} 
for the self-energy becomes
\begin{widetext}
\begin{align}
\partial_{\Lambda} \Sigma_{\Lambda} ( \omega )  
=
-  \frac{i\omega}{1+\frac{n_0}{2}\frac{1-n_0}{Z_0}\tilde{\mu}_0}   \frac{B_{\Lambda}}{Z_0^3}
+\frac{ \frac{n_0}{2} ( 2 - \frac{n_0}{2} )
}{ Z_0^4} 
&\left[\frac{ ( i \omega + \mu_0)^2}{N} \sum_{\bd{k}} 
\frac{ \partial_{\Lambda} 
( t_{\bd{k} , \Lambda} - \mu_{\Lambda} ) }{  
\left[ 
t_{\bd{k} , \Lambda} -  \mu_{\Lambda}
+ \Sigma_{\Lambda} ( \omega )  \right]^2 }\right.\nonumber\\
&\left.-\biggl(\mu_0+ \frac{i\omega}{1+\frac{2}{n_0}\frac{Z_0}{(1-n_0)\tilde{\mu}_0}}\biggr)\frac{\mu_0}{N}\sum_{\bd{k}} 
\frac{ \partial_{\Lambda} 
	( t_{\bd{k} , \Lambda} - \mu_{\Lambda} ) }{  
	\left[ 
	t_{\bd{k} , \Lambda} -  \mu_{\Lambda}
	+ \Sigma_{\Lambda} ( 0 )  \right]^2 } \right],
\label{eq:Sigmaflowret}
\end{align}
%
%
which should be integrated with 
the initial condition 
\begin{equation}
\Sigma_{\Lambda =0}  ( \omega ) = 
\Sigma_0 ( \omega ) = 
-  \frac{ i \omega+\mu_0}{Z_0},  
\; \; \;  \; \; \;
Z_0 = 1 - \frac{n_0}{2} .
\label{eq:SigmaRinit}
\end{equation} 
To make progress analytically, 
we next modify
the flow equation \eqref{eq:Sigmaflowret} 
using the so-called 
Katanin substitution~\cite{Katanin04}, 
which has been shown to 
reduce the errors due to 
the violation of Ward identities
by truncations of the FRG flow equations 
for canonical fermions.
For the flow equation \eqref{eq:Sigmaflowret},
the Katanin substitution amounts to replacing
the single-scale propagator 
$\dot{G}_{\Lambda} ( K )$ 
by the total $\Lambda$-derivative of 
the scale-dependent propagator
$G_{\Lambda} ( K )$; 
i.e.,
\begin{equation}
\frac{ 
\partial_{\Lambda} ( t_{\bd{k} , \Lambda}  
- \mu_{\Lambda} ) 
}{  
\left[ 
t_{\bd{k} , \Lambda} - \mu_{\Lambda} 
+ \Sigma_{\Lambda} ( \omega )  \right]^2
}
\rightarrow  
\partial_{\Lambda} \left[  
\frac{-1}{ 
t_{\bd{k} , \Lambda} - \mu_{\Lambda} 
+ \Sigma_{\Lambda} ( \omega ) } 
\right] .
\end{equation}
Then both sides of the 
flow equation Eq.~\eqref{eq:Sigmaflowret} 
become total $\Lambda$-derivatives, 
so that we may explicitly integrate it 
over the flow parameter $\Lambda$.
With the Katanin substitution,
the contribution from the first term in Eq.~\eqref{eq:Sigmaflowret}
involving $B_{\Lambda}$ 
becomes proportional to the difference
between initial and final fillings,
\begin{align}
\int_0^1 d \Lambda B_{\Lambda} 
\rightarrow 
\frac{1}{ \beta N} \sum_{\bd{k} \omega} 
\left[ 
G ( \bd{k} , \omega ) + 
\frac{1}{\Sigma_0 ( \omega )} 
\right]
= 
\frac{n}{2} - \frac{1}{\beta} \sum_{\omega} \frac{Z_0}{ i \omega + \mu_0 } 
e^{ i \omega 0^+ }
=
\frac{n - n_0 }{2} .
\end{align}
Here, 
we have used Eqs.~\eqref{eq:nnull} 
and \eqref{eq:munulldef} to express 
the Matsubara sum in terms of the filling, 
\begin{equation}
\frac{1}{\beta} \sum_{\omega} 
\frac{Z_0}{ i \omega + \mu_0 } 
e^{ i \omega 0^+ }
= 
\frac{ 1 - \frac{n_0}{2}}{ 
e^{ - \beta \mu_0} +1 } 
= 
\frac{ e^{ \beta \mu_0}}{
1 + 2 e^{\beta \mu_0}} = \frac{n_0 }{2} .
\end{equation}
With the initial condition \eqref{eq:SigmaRinit},
we thus find that the self-energy
$\Sigma ( \omega )$ satisfies 
for $T \rightarrow 0$ 
the following implicit equation:
%
%
\begin{align}
\Sigma ( \omega ) 
= 
- \frac{ i \omega}{Z_0} 
- \frac{i \omega}{ Z_0^3} \frac{1}{1+\frac{n_0}{2}\frac{1-n_0}{Z_0}\tilde{\mu}_0} \frac{n - n_0 }{2 }
+\frac{i \omega}{ Z_0^3} \frac{n_0}{2}\left( 2- \frac{n_0}{2} \right)
\left[\frac{1 }{1+\frac{2}{n_0}\frac{Z_0}{(1-n_0)\tilde{\mu}_0}}-\frac{1}{N}
\sum_{\bd{k}} \frac{ 
t_{\bd{k}} - \mu + \Sigma ( \omega ) 
+ \frac{ i \omega}{Z_0}
}{
t_{\bd{k}} - \mu  + \Sigma ( \omega ) 
} \right].
\label{eq:Sigmaint}
\end{align}

\end{widetext}
For $\omega \rightarrow 0$ this yields  
to leading order
\begin{equation}
\Sigma ( \omega ) = - \frac{i \omega}{Z} + {\cal{O}} ( \omega^2 ),
\label{eq:Sigmalow}
\end{equation}
where
\begin{align}
\frac{1}{Z} 
& = \frac{1}{Z_0}+\frac{1}{Z_0^3}\frac{1}{1+\frac{n_0}{2}\frac{1-n_0}{Z_0}\tilde{\mu}_0} \left[
\frac{n_0}{2}\left(2-\frac{n_0}{2}\right) + \frac{n - n_0 }{ 2 } \right] 
\nonumber\\
& = 
\frac{1}{Z_0}+\frac{1}{Z_0^3}\frac{1}{1+\frac{n_0}{2}\frac{1-n_0}{Z_0}\tilde{\mu}_0} \left[
\frac{n + n_0 }{ 2 } - \frac{n_0^2}{4} \right] .
\end{align}
We conclude that the 
effective kinematic interaction 
due to the Hilbert space projection
reduces the quasi-particle residue to
\begin{equation}
Z = 
Z_0^3 \frac{1+\frac{x_0n_0}{2 Z_0}\tilde{\mu}_0}{1+\frac{x_0n_0}{2}\tilde{\mu}_0Z_0+\frac{n-n_0}{2}} , 
\label{eq:Zqp}
\end{equation}
where we have again introduced the 
hole doping $x_0 = 1 - n_0$.
To estimate the quasi-particle damping,
we substitute the 
low-energy expansion \eqref{eq:Sigmalow}
for the self-energy on the 
right-hand side of Eq.~\eqref{eq:Sigmaint}
and analytically continue both sides 
to real frequencies to obtain 
the retarded self-energy
$\Sigma^R ( \omega ) = 
\Sigma ( \omega ) |_{ i \omega \rightarrow \omega + i 0^+}$.
Taking the imaginary part yields
\begin{equation}
{\rm Im} \Sigma^R ( \omega ) = 
- \pi 
 \frac{n_0}{2} \left( 2- \frac{n_0}{2}\right)\frac{Z}{
Z_0^4} 
\frac{ \omega^2}{N} \sum_{\bd{k}} 
\delta ( \omega - Z [ t_{\bd{k}} - \mu] ) .
\end{equation} 
In terms of the (bare) density of states
\begin{equation}
\nu_0 ( \epsilon ) = \frac{ 1}{N}
\sum_{\bd{k}} \delta ( \epsilon -  t_{\bd{k}} )
\end{equation}
this can be written as
\begin{equation}
Z {\rm Im} \Sigma^R ( \omega ) = 
- \pi  
\frac{n_0}{2} \left( 2- \frac{n_0}{2}\right)\frac{Z}{
	Z_0^4} 
\omega^2 \nu_0 (  \mu + \omega /{Z} ) .
\end{equation}
We conclude that at low-energies the 
retarded propagator of the projected fermions
in the $t$ model has the quasi-particle form
\begin{equation}
G^R ( \omega ) = 
\frac{ Z }{ \omega - \xi_{\bd{k}} 
+ i \gamma_{\bd{k}}} ,
\label{eq:Gret}
\end{equation}
where the quasi-particle residue $Z$ 
is given in Eq.~\eqref{eq:Zqp},
the quasi-particle dispersion is
\begin{equation}
\xi_{\bd{k}} 
= Z ( t_{\bd{k}}  - \mu ) 
= Z ( t_{\bd{k}} - t_{ \bd{k}_F }) ,
\label{eq:dispersionres}
\end{equation}
and the quasi-particle damping is
\begin{align}
\gamma_{\bd{k}} 
& = 
- Z {\rm Im} \Sigma^R ( \omega = \xi_{\bd{k}} ) 
\nonumber\\
& = 
\pi \frac{n_0}{2} \left( 2- \frac{n_0}{2}\right)\frac{Z}{
	Z_0^4} 
\nu_0 ( t_{\bd{k}} ) \xi_{\bd{k}}^2 .
\label{eq:dampingres}
\end{align}
Recall that according to Eq.~\eqref{eq:FSzero},
we have constructed our counter-term such that
for vanishing temperature the Fermi surface 
is given by $t_{ \bd{k}_F} = \mu$.
To explicitly calculate the 
chemical potential $\mu$ for 
a given lattice filling we should
integrate the flow equation \eqref{eq:muflow} 
for the counter-term. 
The results of this calculation
will be reported elsewhere \cite{Rueckriegel23}. 
As an estimate we may use the approximate 
equation of state \eqref{eq:eos1a}
with $ n \approx n_0 $; see Fig.~\ref{fig:damping}.

\begin{figure}
	\centering
	\includegraphics[width=1\linewidth]{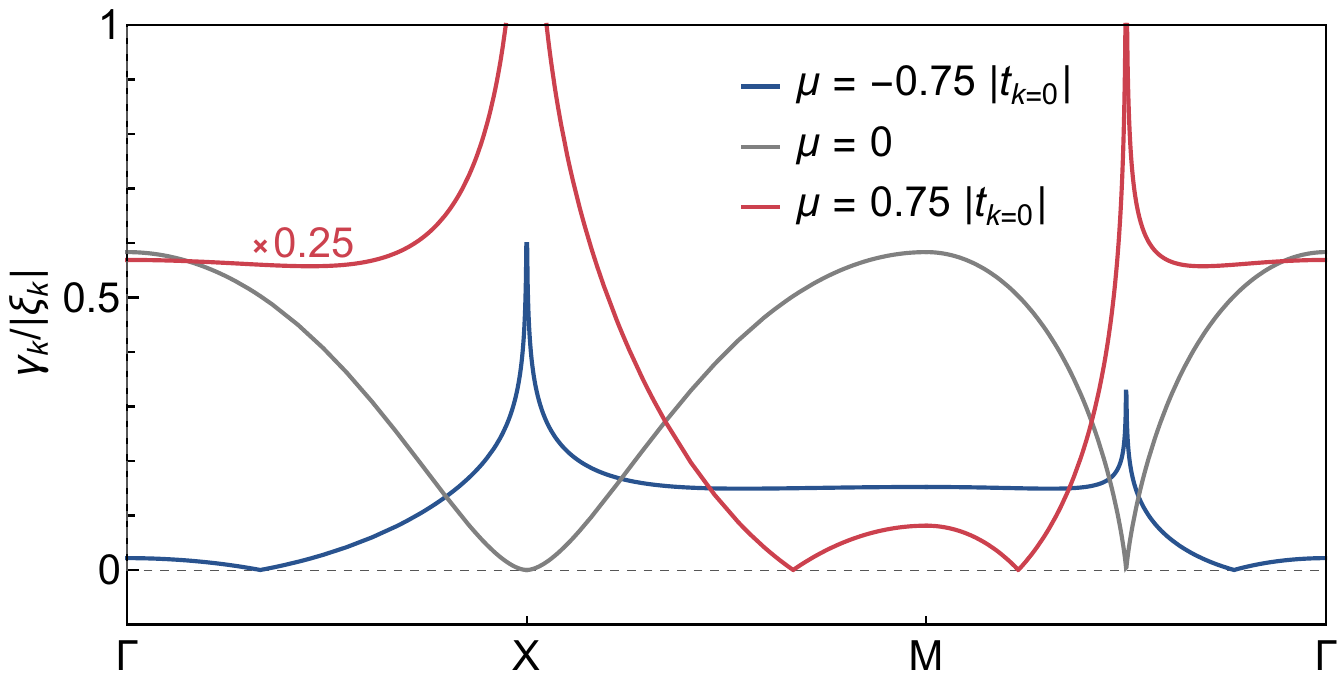}
	\caption{Momentum dependence of the  damping coefficient $\gamma_{\bd{k}}/|\xi_{\bd{k}}|$ on a square lattice, plotted along the path shown in Fig.~\ref{fig:FShalf} for chemical potential $\mu =0$, $0.75\,|t_{\bm{k}=0}|$ and $-0.75\,|t_{\bm{k}=0}|$. Note that the damping becomes arbitrarily small in the vicinity of the Fermi surface, indicating well-defined quasi-particles. The singularities in the damping for $ \mu \neq 0 $ reflect the van-Hove singularities of the two-dimensional density of states.
	}
	\label{fig:damping}
\end{figure}

Our results (\ref{eq:Gret}--\ref{eq:dampingres})
imply that at zero temperature
the $t$ model supports 
well-defined quasi-particles 
in the sense that the 
quasi-particle residue is finite
and the ratio $\gamma_{\bd{k}} / \xi_{\bd{k}}$ 
of the quasi-particle damping to 
the excitation energy
becomes arbitrarily small  
for $\bd{k} \rightarrow \bd{k}_F$, 
provided that the density of states is finite.
Thus we find the ground state of the $t$ model to be
a hidden Fermi liquid~\cite{Anderson08,Anderson09,Casey11}, which in this respect
behaves just like an ordinary Fermi liquid.
However, 
due to the projected Hilbert space 
Luttinger's theorem is violated 
in a hidden Fermi liquid.

Let us conclude this section with a caveat.
Within  our level-1 truncation of 
the FRG flow equations 
the effective interaction
does not have any momentum dependence, 
so that our self-energy depends 
only on frequency.
While in reduced dimensions this approximation 
is probably not justified,
we expect it to be valid in 
three and higher dimensions 
because in the limit of infinite dimensions
(which can be realized 
by a proper rescaling of the hopping) 
the self-energy of the Hubbard model is known 
to be momentum-independent~\cite{Metzner89,
Georges96}. 
Unfortunately, 
the limit of infinite on-site repulsion 
is not easily accessible 
within dynamical mean-field theory, 
so that we cannot compare the results 
obtained in this section with 
numerical calculations based on 
dynamical mean-field theory.

\section{Summary and outlook}
\label{sec:summary}

In this work, we have developed a
new functional renormalization group approach 
to strongly correlated electronic lattice models
with projected Hilbert spaces using 
Hubbard X-operators. Our approach is
complementary to the usual truncated vertex expansion
for fermionic many-body systems~\cite{Metzner12,Dupuis21,Kopietz10,Andergassen2020}, which breaks down at
intermediate coupling.
The main advantage of our X-FRG approach is that
the kinematic constraints imposed by 
the Hilbert space projection can be
taken into account non-perturbatively.   
The X-FRG approach relies on 
the simple insight that
formally exact FRG flow equations 
for generating functionals can be derived 
even if the Hamiltonian cannot be expressed 
in terms of operators satisfying 
canonical commutation relations.
In previous works~\cite{Krieg19,Tarasevych18,
Goll19,Goll20,Tarasevych21,
Tarasevych22,Rueckriegel22,Tarasevych22b} 
we have shown 
that this insight can be used 
to derive exact flow equations for the
time-ordered spin correlation functions 
of quantum spin systems.
Although a diagrammatic representation 
of perturbation theory in terms of 
Hubbard X-operators
has been developed \cite{Ovchinnikov04,
Izyumov88,Izyumov05},
the diagrammatic structure is rather complicated.
On the other hand, 
the diagammatic representation of 
the flow equations for the irreducible 
vertices obtained within our X-FRG approach 
is identical to the 
familiar diagrammatics \cite{Kopietz10} 
generated by
the usual Wetterich equation \cite{Wetterich93}
for fermionic or bosonic many-body systems.

To demonstrate that the X-FRG can indeed be used
to obtain useful results for 
strongly correlated electrons, 
in Sec.~\ref{sec:applications}
we have presented two simple applications. 
On the one hand, 
we have shown that the so-called 
Hubbard-I approximation can be obtained 
simply by approximating the 
irreducible two-point vertices 
by their initial conditions corresponding 
to the atomic limit. 
More importantly, 
we have used our X-FRG approach 
to calculate the 
single-particle Green function 
in the hidden Fermi liquid state \cite{Anderson08,Anderson09,Casey11}
of the Hubbard model at 
infinite on-site repulsion 
(i.e., the $t$ model). 
In this case 
Luttinger's theorem \cite{Luttinger61}, 
stating that in a normal Fermi liquid 
the volume of the Fermi surface 
is proportional to the density, 
is violated due to the projected Hilbert space. 
Nevertheless, 
the quasi-particles for momenta
in the vicinity of the Fermi surface 
are well-defined in the sense that 
their damping is small compared 
with their excitation energy.
Our calculation of the quasi-particle damping 
of the $t$ model
presented in Sec.~\ref{sec:applications} 
is rather simple and elegant;
we are not aware of any other 
analytical method which 
is able to calculate the quasi-particle damping 
in strongly correlated systems with  
projected Hilbert spaces. 
Although the quasi-particle energies  
can also be estimated by means a
Gutzwiller projection \cite{Gutzwiller63,
Buenemann03,Fukushima05} 
of a suitable variational state, 
to the best of our knowledge 
the quasi-particle damping has so far 
not been obtained from Gutzwiller projections.

Our X-FRG approach can be extended 
in many directions 
which we are planning to 
explore in future publications.
Possible applications include 
the interplay between 
fermionic quasi-particle excitations
and spin- and charge fluctuations 
in the $t$-$J$ model \cite{Izyumov90,Izyumov02},
the development of systematic 
strong-coupling expansions 
in powers of the hopping, 
or the study of the dynamics of holes 
in an antiferromagnetic background \cite{Schmittrink88,Kane89,Kopietz90}.
It would also be interesting to explore 
the connection between our X-FRG approach 
and the dual fermion approach \cite{Rubtsov09}, 
which also relies on the 
local strong-coupling limit as a 
non-perturbative starting point.

\section*{Acknowledgements}
We thank Johannes Reuther and Björn Sbierski for
organizing a workshop on \textit{functional renormalization
for quantum spin systems}, which helped sharpen some of the
ideas presented in this work.
This work was financially supported by the  Deutsche Forschungsgemeinschaft (DFG, German
Research Foundation) through project number 431190042.

\appendix

\renewcommand{\appendixname}{APPENDIX}

\renewcommand{\thesection}{\Alph{section}}

\section{GENERALIZED SPECTRAL REPRESENTATION}

\label{app:spectral}

\renewcommand{\theequation}{A\arabic{equation}}

\setcounter{equation}{0}

In the X-FRG approach,
the initial conditions of the flow equations
generally involve the exact correlation functions
of a solvable quantum-mechanical system. 
In this appendix, 
we therefore give a scheme to 
efficiently construct 
imaginary-time ordered correlation functions 
for any diagonalizable Hamiltonian $\mathcal{H}$
with operators in 
an explicit matrix representation.
Note that a related algorithm 
for this purpose has recently been discussed by 
Halbinger {\it{et al.}}~[\onlinecite{Halbinger23}].
Our aim is to evaluate the imaginary-time ordered correlation function in Matsubara frequency space given by 
\begin{widetext}
	\begin{equation}\label{eq:fouriertransform}
		\beta\delta_{\omega_1+ \dotsc +\omega_n,0}C^{\alpha_1\dots \alpha_n}(\omega_1,\dots, \omega_n) = \int_{0}^{\beta}d\tau_1 \dotsc \int_{0}^{\beta}d\tau_n e^{i(\omega_1\tau_1+ \dotsc + \omega_n\tau_n)} \braket{\mathcal{T} \left\{A^{\alpha_1}(\tau_1)\dotsc A^{\alpha_n}(\tau_n)\right\}} 
		.
	\end{equation}
\end{widetext}
Here, 
$A^{\alpha}(\tau)$ denotes an arbitrary operator with flavor index $\alpha$ and 
imaginary-time dependence given by
\begin{equation}
	A^\alpha(\tau)= e^{\mathcal{H}\tau}A^\alpha e^{-\mathcal{H}\tau}.
\end{equation} The imaginary-time ordering operator $\mathcal{T}$ acts
as \begin{align}
	\mathcal{T} \left\{A^{\alpha_1}(\tau_1)A^{\alpha_2}(\tau_2)\right\} = {}& \Theta(\tau_1-\tau_2)A^{\alpha_1}(\tau_1)A^{\alpha_2}(\tau_2) \nonumber \\
	 & + \zeta \Theta(\tau_2-\tau_1)A^{\alpha_2}(\tau_2)A^{\alpha_1}(\tau_1),
\end{align}
where $\zeta = -1$ if both operators are of the Fermi type based on the classification introduced in Sec.~\ref{subsec:hubmodel}, otherwise $\zeta = 1$. 
The symbol 
\begin{equation}
	\langle \ldots \rangle = \frac{ {\rm tr} [ e^{ - \beta {\cal{H}} } \ldots ] }{
		{\rm tr} e^{ - \beta {\cal{H}} } }
\end{equation}
denotes the thermal expectation value. 
In order to evaluate it, 
we introduce the complete basis 
$\ket{m_1}, \dotsc ,\ket{m_d}$ 
of energy eigenstates with
\begin{equation}
	\mathcal{H} \ket{m_j} = \epsilon_{m_j} \ket{m_j}.
\end{equation}
Here, 
the $ \epsilon_{m_j} $
are the corresponding eigenenergies. 
We also introduce the partition function $\mathcal{Z} = {\rm tr} e^{ - \beta {\cal{H}} } = \sum_{m}e^{-\beta \epsilon_m}$ and $\epsilon_{ij} = \epsilon_{m_i}-\epsilon_{m_j}$. 
Performing the time-ordering and relabeling the integration variables appropriately allows us to separate the right-hand side of Eq.~\eqref{eq:fouriertransform} into a universal frequency dependent coefficient $\Omega^{(n)}$ and a matrix product of operators:
\begin{widetext}
	\begin{align}\label{eq:permutationsum}
		& \int_{0}^{\beta}d\tau_1 \int_0^{\tau_1}d\tau_2 \dotsc \int_{0}^{\tau_{n-1}}d\tau_n \sum_{\mathcal{P}}\sum_{\left\{m_i\right\}}\left[\zeta^{\mathcal{P}}\left(\frac{e^{-\beta \epsilon_{m_1}}}{ {\cal Z } }\prod_{i=1}^{n} e^{(i\omega_{P(i)}+\epsilon_{i(i+1)})\tau_i}\right)\left(\prod_{i=1}^{n}\bra{m_i} A^{\alpha_{P(i)}} \ket{m_{i+1}}\right)\right]
		\nonumber\\
		={}&\sum_{\mathcal{P}}\sum_{\left\{m_i\right\}}\zeta^{\mathcal{P}} \frac{e^{-\beta \epsilon_{m_1}}}{ {\cal Z } } \Omega^{(n)}(\omega_{P(1)},\dotsc,\omega_{P(n)}; \epsilon_{m_1},\dotsc, \epsilon_{m_n})\prod_{i=1}^{n}  \bra{m_i} A^{\alpha_{P(i)}} \ket{m_{i+1}}.
	\end{align}
\end{widetext}
Here, 
$\sum_{\left\{m_i\right\}}$ sums the eigenenergies and operator projections over all eigenstates. The $\sum_{\mathcal{P}}$ represents a summation over all permutations of the tuple $\left(\omega_i,\alpha_i\right)$
consisting of frequency and associated operator label that result from the time-ordering. 
The factor $\zeta^{\mathcal{P}}$ gives a minus sign whenever an uneven number of Fermi type operators has been commuted. The products are defined periodic such that $n+1 \equiv 1$, which is a consequence of the cyclic invariance of the trace. 

We now give a blueprint to construct the frequency dependent coefficients $\Omega^{(n)}$ to any given order $n$. The integration procedure is in itself quite straightforward. However due to the Matsubara character of the frequencies that fulfill 
\begin{equation}
	\label{eq:matsubaraf}
	e^{i \omega_i \beta} = \zeta = \pm 1,
\end{equation} 
 the expression becomes ill defined with respect to various limits. Those limits should be taken into account while still treating the frequencies as continuous, giving rise to terms with additional constraints on the Matsubara frequency conservation. These terms correspond to the anomalous contributions presented by Halbinger {\it{et al.}}~[\onlinecite{Halbinger23}].
 To systematically identify all possible divergencies and properly account for them, we modify the evaluation. 
For example, 
the innermost integral shall give either
\begin{align}\label{eq:integrationscheme}
	\int_{0}^{\tau_{n-1}} d\tau_n e^{(i\omega_n+\epsilon_{n1})\tau_n} =
	\begin{cases}
	&\delta_{\omega_n,0}\delta_{\epsilon_{n1},0}\tau_{n-1}  
\;\;\; \text{or}	\\
	\\
	&\dfrac{e^{(i\omega_n+\epsilon_{n1})\tau_{n-1}}-1}{i\omega_n+\epsilon_{n1}} .
	\end{cases}
\end{align}
This distinction is made within each integration, even though the limit of vanishing frequency and energy difference is for the moment well-defined in the second case of Eq.~\eqref{eq:integrationscheme}. The complete result to any order is given by the sum of all possible branches emerging from taking all possible combinations of these different cases. The relation~\eqref{eq:matsubaraf} can then be applied without the need to consider any more limits.
Additionally, 
due to energy conservation only terms proportional to $\delta_{\omega_1+\dotsc+\omega_n,0}$ will appear in the final expression and each frequency Kronecker delta will be accompanied by a factor of $\beta$.

A general formula for this construction procedure is given by
\begin{widetext}
	\begin{align}
			&\Omega^{(n)}(\omega_1,\dotsc,\omega_n; \epsilon_{m_1},\dotsc, \epsilon_{m_n})
			=  \sum_{l=1}^{n}\frac{\beta^l}{l!}\sum_{b_1=1}^{n-1} \sum_{b_2 > b_1}^{n-1} \dotsc \sum_{b_{l-1} > b_{l-2}}^{n-1} g_{1,b_1}(\omega_1,\dotsc,\omega_{b_1};\epsilon_{m_1},\dotsc,\epsilon_{m_{b_l+1}})
			\nonumber\\
			&\times g_{b_1+1,b_2}(\omega_{b_1+1},\dotsc,\omega_{b_2};\epsilon_{m_{b_1+1}},\dotsc,\epsilon_{m_{b_2+1}}) \dotsc  g_{b_{l-1}+1,n}(\omega_{b_{l-1}+1},\dotsc,\omega_{n};\epsilon_{m_{b_{l-1}+1}},\dotsc,\epsilon_{m_n},\epsilon_{m_1}).
\label{frequencyterm}
	\end{align}
\end{widetext}
Here, the nested sums over the variables $b_i$ construct all possible stars and bars partitions~\cite{Feller} using $l-1$ bars of an ordered list given by the frequency arguments. An illustrative example for four frequencies is shown in Fig.~\ref{fig:4pointbars}. 
\begin{figure}
	\centering
	\includegraphics[width=0.8\linewidth]{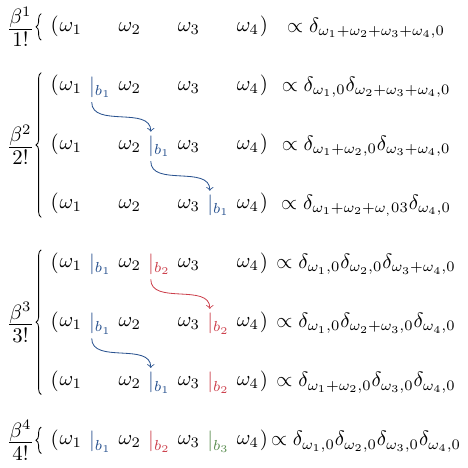}
	\caption{Construction scheme of all relevant summands of one permutation for the frequency dependent coefficient $\Omega^{(4)}$ of the four-point correlation function. The colored bars represent the summation indices of Eq.~\eqref{frequencyterm} that count the allowed partitions.
	}
	\label{fig:4pointbars}
\end{figure}
The contribution of a single partition is generated via the propagator function
\begin{widetext}
	\begin{equation}\label{eq:gdef}
		g_{x,y}(\omega_x,\dotsc,\omega_y;\epsilon_x,\dotsc \epsilon_{y+1})=
		\delta_{\sum_{j=x}^{y} \epsilon_{j(j+1)},0} \delta_{\sum_{j=x}^{y} \omega_{j},0} \prod_{j=1}^{y-x} \frac{1}{\sum_{i=1}^{j}\left[i\omega_{y-i+1}+\epsilon_{(y-i+1)(y-i+2)}\right]}.
	\end{equation}
\end{widetext}
This term can be understood by following the nested imaginary-time integrations using the scheme shown in Eq.~\eqref{eq:integrationscheme}, with the aim to create a specific frequency Kronecker delta containing $j$ neighboring frequencies. The propagator product in Eq.~\eqref{eq:gdef} is then a consequence of the $j-1$ integrations necessary to obtain an exponential containing the desired sum of frequencies.

To give an explicit example, we construct the coefficient $\Omega^{(4)}$, which determines the four-point correlation function. We first consider the order $\left\{\omega_1,\omega_2,\omega_3, \omega_4\right\}$ and note that the frequencies should afterwards be permuted in accordance with Eq.~\eqref{frequencyterm}.
Sorted by powers of $\beta$ the coefficients have the structure
\begin{align}
	&\nonumber\Omega^{(4)}(\omega_1,\omega_2,\omega_3,\omega_4; \epsilon_{m_1},\epsilon_{m_2},\epsilon_{m_3},\epsilon_{m_4}) = \\ &\frac{\beta^4}{4!}\Omega_4^{(4)}+\frac{\beta^3}{3!}\Omega_3^{(4)}+\frac{\beta^2}{2!}\Omega_2^{(4)}+\beta\Omega_1^{(4)}.
\end{align}
The term proportional to $\beta$ has only one possible partition accompanied by a non-trivial frequency dependence:
\begin{align}
	\nonumber \Omega^{(4)}_1 ={} &\frac{\delta_{\omega_1+\omega_2+\omega_3+\omega_4,0}}{i\omega_4+i\omega_3+i\omega_2+ \epsilon_{21}}\\
	&\times\frac{1}{i\omega_4+i\omega_3+\epsilon_{31}} \frac{1}{i\omega_4+\epsilon_{41}}.
\end{align}	
The higher-order coefficients are given by
\begin{align}
	\nonumber	\Omega_2^{(4)}={}&\delta_{\omega_1,0}\delta_{\epsilon_{12},0}\frac{\delta_{\omega_2+\omega_3+\omega_4,0}\delta_{\epsilon_{21},0}}{i\omega_4+i\omega_3+\epsilon_{31}} \frac{1}{i\omega_4+\epsilon_{41}}\\
	&\nonumber +\frac{\delta_{\omega_1+\omega_2,0}\delta_{\epsilon_{13},0}}{i\omega_2+\epsilon_{23}} \frac{\delta_{\omega_3+\omega_4,0}\delta_{\epsilon_{31},0}}{i\omega_4+\epsilon_{41}}\\
	&+\frac{\delta_{\omega_1+\omega_2+\omega_3,0}\delta_{\epsilon_{14},0}}{i\omega_3+i\omega_2+\epsilon_{24}}\frac{1}{i\omega_3+\epsilon_{34}} \delta_{\omega_4,0} \delta_{\epsilon_{41},0},
\end{align}
and
\begin{align}
	\nonumber \Omega^{(4)}_3 ={}& \delta_{\omega_1,0}\delta_{\epsilon_{12},0}\delta_{\omega_2,0}\delta_{\epsilon_{23},0}\frac{\delta_{\omega_3+\omega_4,0}\delta_{\epsilon_{31},0}}{i\omega_4+\epsilon_{41}}\\
	&\nonumber + \delta_{\omega_1,0}\delta_{\epsilon_{12},0} \frac{\delta_{\omega_2+\omega_3,0}\delta_{\epsilon_{24},0}}{i\omega_3+\epsilon_{34}}\delta_{\omega_4,0}\delta_{\epsilon_{41},0}\\
	& + \frac{\delta_{\omega_1+\omega_2,0}\delta_{\epsilon_{13},0}}{i\omega_2+\epsilon_{23}}\delta_{\omega_3,0}\delta_{\epsilon_{34},0}\delta_{\omega_4,0}\delta_{\epsilon_{41},0}.
\end{align}
Finally, 
the highest order of $\beta$ will then include $4-1$ bars, dividing this list in four single frequencies:
\begin{equation}
	\Omega^{(4)}_4 = \delta_{\omega_1,0}\delta_{\epsilon_{12},0} \delta_{\omega_2,0}\delta_{\epsilon_{23},0} \delta_{\omega_3,0}\delta_{\epsilon_{34},0} \delta_{\omega_4,0}\delta_{\epsilon_{41},0}.
\end{equation}
The structure of these terms is independent of the physical model, so that it is sufficient to construct them once.
The general recipe for order $n$ coefficients can be summarized as follows:
\begin{itemize}
	\item Begin by considering a single permutation given by the tuple list $\left\{(\omega_1,\alpha_1),\dotsc, (\omega_n,\alpha_n)\right\}$.
	\item For each order $\beta^k/k!$ with $k\in [1,n]$, find all possible Kronecker delta combinations. This is equivalent to finding all possible stars and bars partitions of the ordered tuple list with $k-1$ bars.
	\item Multiply each Kronecker delta with suitable frequency dependent propagator functions \eqref{eq:gdef} according to Eq.~\eqref{frequencyterm}. 
\end{itemize}

\section{TWO-POINT FUNCTIONS OF X-OPERATORS IN THE HUBBARD ATOM}

\label{app:static}

\renewcommand{\theequation}{B\arabic{equation}}

\setcounter{equation}{0}

In this appendix, 
we calculate the
two-point correlation functions of the X-operators 
of the Hubbard atom; i.e.,~the Hubbard model for vanishing hopping
with Hamiltonian
 \begin{equation}
  {\cal{H}}_1  =  \sum_i \left[  U  n_{ i \uparrow } n_{i \downarrow}   - \mu {n}_i - 
 h S^z_i   \right] = \sum_{ i  a} \epsilon_a X_i^{aa},
 \end{equation}
where for fixed chemical potential $\mu$ and magnetic field $h$ the
energies  $\epsilon_a$ are
 \begin{equation}
 \epsilon_0 =0, \; \; \; \epsilon_{\uparrow} = - \mu - \frac{h}{2},
 \; \; \; \epsilon_{\downarrow} = - \mu + \frac{h}{2}, \; \; \; 
 \epsilon_2 = U - 2 \mu.
 \end{equation}
The generating functional 
of imaginary-time ordered connected correlation functions of the X-operators
can then be written as
 \begin{equation}
 e^{{\cal{G}}_{0} [ J ] }
=  {\rm Tr} \left[ e^{ - \beta {\cal{H}}_1 } {\cal{T}}
e^{ \int_0^{\beta} 
 d \tau       \sum_{i, p} J^p_i ( \tau ) X^p_i ( \tau ) }   \right].
 \label{eq:G0def}
 \end{equation}
Since all sites are decoupled in this limit, the generating functional decouples into a sum of single-site generating functionals:
 \begin{equation}
 {\cal{G}}_{0} [ J ] = \sum_i {\cal{G}}_{\rm site} [ J_i ],
 \end{equation}
where the generating functional of the single-site (atomic)
correlation functions is
 \begin{equation}
{\cal{G}}_{\rm site} [ J ] =
\ln {\rm tr} \left[
 e^{ - \beta {\cal{H}}_{\rm site}  }
  {\cal{T}}
 e^{ \int_0^{\beta} 
 d \tau       \sum_{ p} J^p ( \tau ) X^p ( \tau ) }   \right],
 \hspace{7mm}
 \end{equation}
with single-site Hamiltonian
 \begin{equation}
  {\cal{H}}_{\rm site} =  
U n_{\uparrow} n_{\downarrow} - \mu n - h S^z = \sum_a \epsilon_a X^{aa}.
 \label{eq:Hsite}
 \end{equation}
The symbol  ${\rm tr} [ \ldots ]$ denotes the trace  over the four-state fermionic Fock space  associated with
a single lattice site.
The single-site 
functional ${\cal{G}}_{\rm site} [ J ] $ cannot be calculated in closed form, but can 
be expanded in powers of the sources,
 \begin{align}
&  {\cal{G}}_{\rm site} [ J ]  = {\cal{G}}_{\rm site} [ 0 ] 
  +  \int_0^{\beta } d \tau \sum_{ p }  \langle X^{p} ( \tau ) \rangle J^{p} ( \tau ) 
 \nonumber
 \\
   & + \frac{1}{2!}  \int_0^{\beta } d \tau_1 d \tau_2   \sum_{p_1 p_2 } G_0^{p_1 p_2} ( \tau_1 , \tau_2 )
 J^{p_1} ( \tau_1 ) J^{p_2} ( \tau_2 )  
 \nonumber
 \\
 &+  \ldots,
 \end{align}
where
 \begin{align}
  G_0^{p_1 p_2} ( \tau_1 , \tau_2 )  ={} &  \braket{ {\cal{T}} \left\{X^{p_2} ( \tau_2 ) X^{p_1} ( \tau_1 ) \right\}}
 \nonumber
 \\
  &-   \braket{X^{p_2} ( \tau_2 ) } \braket{X^{p_1} ( \tau_1 )} .
 \label{eq:G2def}
 \end{align}
Here, the thermal expectation value is evaluated with respect to the single-site Hamiltonian
${\cal{H}}_{\rm site}$.
The expectation values $ \braket{X^{p} ( \tau )} $ are only finite for the diagonal Bose type X-operators
$p = 00, \uparrow \uparrow , \downarrow \downarrow  , 22$. If we introduce only fermionic sources,
the linear term is absent.

Let us now explicitly calculate the connected correlation functions of the Fermi type X-operators up to fourth order.
Therefore
we work with the occupation number basis of the four-dimensional single-site Fock space consisting 
of the vacuum state $ \ket{ 0 }$, the two single-particle states
$ \ket{ \uparrow } = c^{\dagger}_{\uparrow} \ket{ 0 }$ and $\ket{ \downarrow } = c^{\dagger}_{\downarrow} \ket{ 0 }$, and the antisymmetrized two-particle states
$\ket{ 2 } = c^{\dagger}_{\downarrow} c^{\dagger}_{\uparrow} \ket{ 0 } = 
-  c^{\dagger}_{\uparrow} c^{\dagger}_{\downarrow} \ket{ 0 } $.
Representing these states by
 \begin{align}
 \ket{ 0 } \rightarrow \bd{e}_0  ={} &  \left( \begin{array}{c} 1 \\ 0 \\ 0 \\ 0 \end{array} \right),  \; \; \;  \; \; \; 
 \ket{ \uparrow } \rightarrow \bd{e}_\uparrow = \left( \begin{array}{c} 0 \\ 1 \\ 0 \\ 0 \end{array} \right), 
 \nonumber 
 \\ \ket{ \downarrow } \rightarrow \bd{e}_\downarrow ={} &  \left( \begin{array}{c} 0 \\ 0 \\ 1 \\ 0 \end{array} \right),  \; \; \;  \; \; \; 
 \ket{ 2 } \rightarrow \bd{e}_2 = \left( \begin{array}{c} 0 \\ 0 \\ 0 \\ 1 \end{array} \right),  
 \end{align}
the X-operators are represented by the matrices
 \begin{equation}
 X^{ab} \rightarrow \hat{X}^{ab} = \bd{e}_a \bd{e}_b^T, 
 \; \; \;  \; \; \; 
 a,b \in \{ 0, \uparrow,
 \downarrow , 2 \},
 \end{equation}
and from Eq.~\eqref{eq:cX} we conclude that
the canonical creation and annihilation operators are represented by the following
$4 \times 4$ matrices,
\begin{align}
\hat{c}_{\uparrow}  ={}&      \left( \begin{array}{cc|cc}
 0 & 1 & 0 & 0 \\
 0 & 0 & 0 & 0 \\
 \hline
 0 & 0 & 0 & - 1 \\
 0 & 0 & 0 & 0
 \end{array} \right),
  \; \; \;  
 \hat{c}^{\dagger}_{\uparrow} =    \left( \begin{array}{cc|cc}
 0 & 0 & 0 & 0 \\
 1 & 0 & 0 & 0 \\
 \hline
 0 & 0 & 0 & 0 \\
 0 & 0 & - 1 & 0
 \end{array} \right),
 \\
 \hat{c}_{\downarrow}   ={}&    \left( \begin{array}{cc|cc}
 0 & 0 & 1 & 0 \\
 0 & 0 & 0 & 1 \\
 \hline
 0 & 0 & 0 & 0 \\
 0 & 0 & 0 & 0
 \end{array} \right),
  \; \; \; \; ~
 \hat{c}^{\dagger}_{\downarrow} =  \left( \begin{array}{cc|cc}
 0 & 0 & 0 & 0 \\
 0 & 0 & 0 & 0 \\
 \hline
 1 & 0 & 0 & 0 \\
 0 & 1 & 0 & 0
 \end{array} \right),
 \end{align}
where the lines emphasize the block-diagonal structure of the matrices.
One easily verifies that these matrices satisfy the canonical anticommutation relations
 \begin{equation}
  \hat{c}_{\sigma}  \hat{c}^{\dagger}_{\sigma^{\prime}}  + \hat{c}^{\dagger}_{\sigma^{\prime}}   \hat{c}_{\sigma} =
 \delta_{\sigma \sigma^{\prime} }    \hat{1}
 \end{equation}
and
 \begin{equation}
  \hat{c}_{\sigma}  \hat{c}_{\sigma^{\prime}}  + \hat{c}_{\sigma^{\prime}}   \hat{c}_{\sigma} = \hat{0} =
 \hat{c}^{\dagger}_{\sigma}  \hat{c}^{\dagger}_{\sigma^{\prime}}  + \hat{c}^{\dagger}_{\sigma^{\prime}}   \hat{c}^{\dagger}_{\sigma}.
 \end{equation}
Here, 
$\hat{1}$ is the $4 \times 4$ unit matrix and $\hat{0}$ is the $4 \times 4$ null matrix.
The occupation number operators are represented by
 \begin{align}
 \hat{n}_{\uparrow} = 
\left( \begin{array}{cc|cc}
 0 & 0 & 0 & 0 \\
 0 & 1 & 0 & 0 \\
 \hline
 0 & 0 & 0 & 0 \\
 0 & 0 & 0 & 1
 \end{array} \right), 
 \; \; \; 
 \hat{n}_{\downarrow} = 
\left( \begin{array}{cc|cc}
 0 & 0 & 0 & 0 \\
 0 & 0  & 0 & 0 \\
 \hline
 0 & 0 & 1 & 0 \\
 0 & 0 & 0 & 1
 \end{array} \right), 
 \end{align}
and the single-site Hamiltonian ${\cal{H}}_{\rm site}$
in Eq.~\eqref{eq:Hsite}
is represented by the diagonal matrix
 \begin{equation}
\hat{H}  =   \left( \begin{array}{cc|cc}
 0 & 0 & 0 & 0 \\
 0 & - \mu - \frac{h}{2}   & 0 & 0 \\
 \hline
 0 & 0 &  - \mu + \frac{h}{2}  & 0 \\
 0 & 0 & 0 & U - 2 \mu
 \end{array} \right)
  =  \sum_a \epsilon_a \hat{X}^{aa}.
\end{equation}

Let us now calculate the two-point correlation functions of the X-operators,
 \begin{align}
& \left\langle {\cal{T}}  \left\{
 \hat{X}^{a_1 b_1} ( \tau_1 ) \hat{X}^{a_2 b_2} ( \tau_2 ) \right\} \right \rangle
-  
 \braket{\hat{X}^{a_1 b_1} ( \tau_1 )} \braket{\hat{X}^{a_2 b_2} ( \tau_2 ) }
 \nonumber
 \\
  ={}&   \frac{1}{\mathcal{Z} } {\rm tr} \left[ e^{ - \beta \hat{{H}} }   {\cal{T}}   \left\{
  \hat{X}^{a_1 b_1} ( \tau_1 ) \hat{X}^{a_2 b_2 } ( \tau_2 ) \right\}
 \right] 
 \nonumber
 \\
 & - \frac{1}{\mathcal{Z}^2 } {\rm tr} \left[ e^{ - \beta \hat{{H}} } 
  \hat{X}^{a_1 b_1} ( \tau_1 ) \right]
{\rm tr} \left[ e^{ - \beta \hat{{H}} } 
  \hat{X}^{a_2 b_2} ( \tau_2 ) \right],
 \end{align}
where  the single-site partition function is explicitly
 \begin{align}
 \mathcal{Z}  ={} & {\rm tr} e^{ - \beta \hat{{H}} } = \sum_a e^{ - \beta \epsilon_a } 
 \nonumber
 \\
  ={} & 
 1 +  e^{ \beta ( \mu +  \frac{h}{2} ) } +
  e^{ \beta ( \mu - \frac{h}{2} ) } +e^{\beta ( 2 \mu - U ) },
 \end{align}
and
 \begin{equation}
 \hat{X}^{ab} ( \tau ) = e^{ \hat{H} \tau } \hat{X}^{ab} e^{ - \hat{H} \tau } =
 e^{  (\epsilon_{a} - \epsilon_b )  \tau } \hat{X}^{ab}
 = e^{ -  \epsilon_{ab} \tau } \hat{X}^{ab}.
 \end{equation}
Here, the energy difference $\epsilon_{ab}$ is  defined by
 \begin{equation}
 \epsilon_{ab} = \epsilon_b - \epsilon_a,
 \end{equation}
where the inverted  order of  labels on the right-hand side is introduced 
to emphasize the analogy with canonical fermions \cite{footnotecanonical} and 
to facilitate the comparison with the notation introduced in the
textbook [\onlinecite{Izyumov88}].
Note that only the diagonal X-operators have a finite expectation value,
\begin{equation}
  \bar{X}^{ ab} =  \braket{\hat{X}^{ab}}  = \delta_{ab} x_a ,  \; \; \;  \; \; \; 
 x_a = \frac{ e^{ - \beta \epsilon_a}}{\mathcal{Z}},
 \end{equation}
where the 
$x_a$ are the normalized ($ \sum_a x_a =1$) occupation probabilities
of the states $|a \rangle$ in the atomic limit.
The time-ordered connected two-point function
for any pair of Fermi type X-operators is
 \begin{align}
&  \left\langle {\cal{T}}  \left\{
 \hat{X}^{a_1 b_1} ( \tau_1 ) \hat{X}^{a_2 b_2} ( \tau_2 ) \right\} \right \rangle
 \nonumber
 \\
  ={}& 
 -  G_{a_1 b_1}  ( \tau_1 - \tau_2 ) 
\langle [ X^{a_1 b_1} ( \tau_2 ) , X^{a_2  b_2} ( \tau_2 ) ]_{+} \rangle
 \nonumber
 \\
  ={}&  - G_{a_1 b_1} ( \tau_1 - \tau_2 ) 
 \left( 
 \delta_{ b_1 a_2 } \langle X^{a_1 b_2} ( \tau_2)  \rangle 
 + \delta_{a_1 b_2 } \langle X^{a_2 b_1}( \tau_2)  \rangle 
\right) 
 \nonumber
 \\
  ={}&  - G_{a_1 b_1} ( \tau_1 - \tau_2 ) 
 \left( \delta_{  b_1 a_2} \bar{X}^{a_1 b_2}  +   \delta_{a_1 b_2 } \bar{X}^{a_2 b_1}  
 \right)  
 \nonumber
 \\
  ={}&    - G_{a_1 b_1} ( \tau_1 - \tau_2 )  \delta_{a_1 b_2 } \delta_{ b_1 a_2}  
 ( x_{a_2} + x_{b_2} )    .
 \end{align}
Here, $ [ \; , \; ]_{+}$ denotes the anti-commutator and
$G_{ab} ( \tau )$ is formally identical to 
the imaginary-time propagator of non-interacting fermions at energy
$\epsilon_{ab} = \epsilon_{b} - \epsilon_a$,
 \begin{equation}
 G_{ab} ( \tau  ) = - e^{ - \epsilon_{ab} \tau } [ \Theta (  \tau ) ( 1- n_{ab} )  - \Theta (- \tau ) n_{ab}    ],
 \end{equation} 
with the Fermi function
 \begin{equation}
 n_{ab} = \frac{1}{ e^{\beta  \epsilon_{ab} } +1 }.
 \end{equation}
On the other hand, for two non-diagonal Bose type X-operators
we have
\begin{align}
&\left\langle {\cal{T}}  \left\{
 \hat{X}^{a_1 b_1} ( \tau_1 ) \hat{X}^{a_2 b_2} ( \tau_2 ) \right\} \right \rangle
 \nonumber
 \\
  ={}&   -  G_{a_1 b_1}  ( \tau_1 - \tau_2 ) 
 \langle [ X^{a_1 b_1} ( \tau_2 ) , X^{a_2  b_2} ( \tau_2 ) ] \rangle
 \nonumber
 \\
  ={}&   - G_{a_1 b_1} ( \tau_1 - \tau_2 ) 
 \left( 
 \delta_{ b_1 a_2 } \langle X^{a_1 b_2} ( \tau_1)  \rangle 
- \delta_{a_1 b_2 } \langle X^{a_2 b_1}( \tau_1)  \rangle 
 \right) 
 \nonumber
 \\
  ={}&  - G_{a_1 b_1} ( \tau_1 - \tau_2 ) 
 \left(  \delta_{  b_1 a_2}
 \bar{X}^{a_1 b_2}    - \delta_{a_1 b_2 } \bar{X}^{a_2 b_1}     \right)   
 \nonumber
 \\
  ={}&    G_{a_1 b_1} ( \tau_1 - \tau_2 )  \delta_{a_1 b_2 } \delta_{ b_1 a_2}  
( x_{a_2} -  x_{b_2} )    
 ,
 \end{align}
where $G_{ab} ( \tau )$ is now the usual imaginary-time propagator of non-interacting bosons,
 \begin{equation}
 G_{ab} ( \tau  ) = - e^{ - \epsilon_{ab} \tau } [ \Theta (  \tau ) ( 1+  n_{ab} )  +  \Theta (- \tau ) n_{ab}    ],
 \end{equation}
where $n_{ab}$ is now the Bose function,
 \begin{equation}
 n_{ab} = \frac{1}{ e^{\beta  \epsilon_{ab}  }  - 1 }.
 \end{equation}
Both cases can be compactly written as
 \begin{align}
& \left\langle {\cal{T}}  \left\{
 \hat{X}^{a_1 b_1} ( \tau_1 ) \hat{X}^{a_2 b_2} ( \tau_2 ) \right\} \right \rangle  
 \nonumber
 \\
 ={} &  \zeta   G_{a_1 b_1}^{\zeta} ( \tau_1 - \tau_2 ) \delta_{a_1 b_2 } \delta_{ b_1 a_2}  
 ( x_{a_2} - \zeta  x_{b_2} )  
 \nonumber
 \\
  ={} &  G_{a_1 b_1}^{\zeta} ( \tau_1 - \tau_2 ) \delta_{a_1 b_2 } \delta_{ b_1 a_2}  
 ( \zeta x_{a_2} -   x_{b_2} )   ,
\label{eq:XX2}
 \end{align}
where
\begin{equation}
 G_{ab}^{\zeta} ( \tau  ) = - e^{ - \epsilon_{ab} \tau } [ \Theta (  \tau ) ( 1+  \zeta n^{\zeta}_{ab} )  +  \zeta 
\Theta (- \tau ) n^{\zeta}_{ab}    ],
 \end{equation} 
and
\begin{equation}
 n^{\zeta}_{ab} = \frac{1}{ e^{\beta \epsilon_{ab}  }  - \zeta },
 \end{equation}
with $\zeta =1$ for Bose type X-operators and $\zeta =-1$ for Fermi type
X-operators.
The functions  $G^{\zeta}_{ab} ( \tau - \tau^{\prime} )$ satisfy the usual
Kubo-Martin-Schwinger boundary conditions, i.e., 
$G_{ab}^{\zeta} ( \tau - \tau^{\prime} )$
is for $\zeta = -1$ anti-periodic in both times $\tau$ and $\tau^{\prime}$ with period $\beta$, while for $\zeta =1$ it is periodic in both times.
As a consequence, $G^{\zeta}_{ab} ( \tau - \tau^{\prime} )$
can be represented as a Matsubara sum,
 \begin{equation}
  G^{\zeta}_{ab} ( \tau  )= \frac{1}{\beta} \sum_\omega e^{ - i \omega \tau } G_{ab} (  \omega ),
 \end{equation}
where $\omega = 2 \pi n T$ for $\zeta =1$ and
$\omega = 2 \pi ( n + \frac{1}{2} ) T$ for $\zeta = -1$.
The Fourier coefficients depend on $\zeta$ only via the Matsubara frequencies,
 \begin{equation}
  G_{ab} ( \omega ) = \int_0^{\beta} d \tau e^{ i \omega \tau } G^{\zeta}_{ab} ( \tau ) = \frac{1}{ i \omega - \epsilon_{ab}}.
 \end{equation}
Actually, the simplest way to derive the two-point correlation function of the
X-operators in frequency space is by direct calculation of the Fourier coefficients using
$X^{ab} ( \tau ) = e^{- \epsilon_{ab} \tau } X^{ab} (0)$ and Eq.~\eqref{eq:fouriertransform} and \eqref{eq:permutationsum}.
The correlation function of two non-diagonal X-operators is
 \begin{align}
   & \int_0^{\beta} d \tau 
e^{ i \omega \tau }  \braket{X^{a_1 b_1} ( \tau ) X^{a_2 b_2 } (0) }
\nonumber
 \\
  ={} &    \delta_{ a_1 b_2} \delta_{ b_1 a_2}  \frac{ 
  \zeta x_{a_2} -  x_{b_2}}{
 i \omega   - \epsilon_{a_1 b_1 }   } .
\end{align}
Note that according to the definition \eqref{eq:G2def} the two-point function
of non-diagonal X-operators has the inverted ordering of the labels,
 \begin{align}
 &G_0^{a_1 b_1,  a_2 b_2} ( \tau_1 , \tau_2 )     =
\left\langle {\cal{T}}  \left\{
 \hat{X}^{a_2 b_2} ( \tau_2 ) \hat{X}^{a_1 b_1} ( \tau_1 ) \right\} \right \rangle
 \nonumber
 \\
 ={}&
G_{a_2 b_2}^{\zeta} ( \tau_2 - \tau_1 ) \delta_{a_1 b_2 } \delta_{ b_1 a_2}  
 ( \zeta x_{a_1} -   x_{b_1} )   .
 \end{align}
In frequency space this becomes
 \begin{align}
  & \tilde{G}_0^{a_1 b_1 , a_2 b_2 } ( \omega_1 , \omega_2 ) 
 \nonumber
 \\
={}& 
 \int_0^{\beta} d \tau_1 \int_0^{\beta} d \tau_2 e^{ i ( \omega_1 \tau_1 + \omega_2 \tau_2 )}
 G_0^{a_1 b_1 , a_2 b_2 } ( \tau_1 , \tau_2 )
 \nonumber
 \\
={}& 
 \int_0^{\beta} d \tau_1 \int_0^{\beta} d \tau_2 e^{ i ( \omega_1 \tau_1 + \omega_2 \tau_2 )}
 G_{a_2 b_2 } ( \tau_2 - \tau_1 )
 \nonumber
 \\
 & \hspace{20mm} \times
  \delta_{ a_1 b_2} \delta_{b_1 a_2} ( \zeta x_{a_1} -  x_{b_1} )
 \nonumber
 \\
={}& \beta \delta_{\omega_1 + \omega_2,0} \delta_{a_1 b_2} \delta_{b_1 a_2}
 \frac{ \zeta x_{a_1} - x_{b_1} }{ i \omega_2 - \epsilon_{a_2 b_2} }
 \nonumber
 \\
 ={}& \beta \delta_{\omega_1 + \omega_2,0} \delta_{a_1 b_2} \delta_{b_1 a_2}
 \frac{ x_{b_1} - \zeta x_{a_1}  }{ i \omega_1 - \epsilon_{a_1 b_1} }
 \nonumber
 \\
   ={}& \beta \delta_{\omega_1 + \omega_2,0} \delta_{a_1 b_2} \delta_{b_1 a_2}
 \left[ \frac{ x_{b_1}}{ i \omega_1 - \epsilon_{a_1 b_1}}
 + \zeta \frac{  x_{b_2}}{ i \omega_2 -  \epsilon_{a_2 b_2}} \right],
 \nonumber
 \\
 \end{align}
which is manifestly (anti)-symmetric with respect to the exchange $1 \leftrightarrow 2$; i.e.,
 \begin{equation}
 \tilde{G}_0^{a_1 b_1 , a_2 b_2 } ( \omega_1 , \omega_2 ) 
 = \zeta \tilde{G}_0^{a_2 b_2 , a_1 b_1 } ( \omega_2 , \omega_1 ).
 \end{equation}
For completeness let us also give the 
propagator of two diagonal X-operators,
 \begin{equation}
 G_0^{a_1 a_1,  a_2 a_2} ( \tau_1 ,  \tau_2 )   = 
\delta_{ a_1 a_2 }  x_{a_1}  - x_{a_1} x_{a_2}.
 \end{equation}
It is independent of time because the diagonal X-operators do not have any dynamics in the atomic limit.
In frequency space it is convenient to factor our the frequency-conserving $\delta$-function, defining
 \begin{equation}
 \tilde{G}_0^{a_1 b_1 , a_2 b_2 } ( \omega_1 , \omega_2 )  = \beta \delta_{\omega_1 +
 \omega_2 , 0} {G}_0^{a_1 b_1 , a_2 b_2 } (  \omega_1  ),
 \end{equation}
with
 \begin{equation}
{G}_0^{a_1 b_1 , a_2 b_2 } (  \omega_1  ) 
 = \delta_{a_1 b_2} \delta_{b_1 a_2}
  \frac{ x_{b_1} - \zeta x_{a_1}}{ i \omega_1 - \epsilon_{a_1 b_1}}.
 \label{eq:Gabcd}
 \end{equation}
As a simple check, let us use the above results to calculate  the time-ordered electronic single-particle
Green function,
 \begin{align}
 G_{\sigma} ( \tau - \tau^{\prime} ) ={} &  - \braket{  {\cal{T}} 
 \hat{c}_{\sigma} ( \tau ) \hat{c}^{\dagger}_{\sigma} ( \tau^{\prime} ) }
 \nonumber\\
  ={} & - \frac{1}{\mathcal{Z} }{\rm tr} \left[ e^{ - \beta \hat{{H}} }   {\cal{T}}   
 \left\{\hat{c}_{\sigma} ( \tau ) \hat{c}^{\dagger}_{\sigma} ( \tau^{\prime} ) \right\}
 \right] .
 \end{align}
Using Eq.~\eqref{eq:cX} to express the canonical annihilation and creation
operators in terms of X-operators we obtain in frequency space
 \begin{align}
 G_{\sigma} (  \omega )  ={} & G_0^{0 \sigma, \sigma 0} (  \omega ) +
 G_0^{\bar{\sigma} 2, 2 \bar{\sigma}} (  \omega )
 \nonumber
 \\
  ={} & \frac{ x_0 + x_{\sigma}}{ i \omega + \epsilon_0 - \epsilon_{\sigma}}
 +   \frac{ x_{\bar{\sigma}}  + x_{2}}{ i \omega +\epsilon_{\bar{\sigma}} -
 \epsilon_2 }.
 \end{align}
In the absence of a magnetic field we recover the well-known 
result \cite{Hubbard63,Georges96}
 \begin{align}
 G ( \omega )  ={} &  \frac{1 - \frac{{n}}{2} }{ i \omega + \mu}  + \frac{ \frac{{n}}{2}}{ i \omega + \mu - U }
 \nonumber
 \\
  	={}&  \frac{ \frac{ 1 + x}{2} }{ i \omega + \mu}  + \frac{ \frac{1-x}{2}}{ i \omega + \mu - U },
 \end{align}
where 
 \begin{equation}
 {n} = \sum_{\sigma} \langle c^{\dagger}_{\sigma} c_{\sigma} \rangle
 = \sum_{\sigma} ( x_{\sigma} + x_2 ) = x_{\uparrow} + x_{\downarrow} + 2 x_2
 \end{equation}
is the lattice filling (i.e., the average particle number per lattice site) 
and in the second line we have expressed the lattice filling
in terms of the hole doping $x = 1-n$.
Note that a half-filled lattice corresponds to $n =1$ and $\mu = U/2$.
In general, the lattice filling $n$ and the hole doping $x$ are
related to the chemical potential via
 \begin{equation}
 n  =   1-x = \frac{ 2 e^{ \beta \mu} + 2 e^{ \beta ( 2 \mu - U )}}{ 1  + 2 e^{ \beta \mu }  + e^{ \beta ( 2 \mu - U ) } }.
 \end{equation}

\section{FERMIONIC FOUR-POINT VERTEX OF THE HUBBARD ATOM AT
$U = \infty$}   

\label{app:4point}

\renewcommand{\theequation}{C\arabic{equation}}

\setcounter{equation}{0}

In this appendix we derive the expressions for 
the fermionic four-point vertex of the $t$ model for vanishing hopping 
given in Eqs.~\eqref{eq:Gamma4a} and \eqref{eq:Gamma4b} of Sec.~\ref{sec:hidden}.
To derive the effective interaction between the projected fermions
in the $t$ model for vanishing hopping, 
we need the connected part of the four-point function of the projected
fermionic annihilation and creation operators
$\tilde{c}_{\sigma} ( \tau ) =  X^{0 \sigma} ( \tau )$
and 
$\tilde{c}^{\dagger}_{\sigma} ( \tau ) =  X^{ \sigma 0} ( \tau )$. 
\begin{widetext}
Let us introduce the following notation,  \begin{align}
G_0^{ \tilde{c}_{\sigma_1} \tilde{c}_{\sigma_2}  \tilde{c}^{\dagger}_{\sigma_2}  
\tilde{c}^{\dagger}_{\sigma_1}}
 ( \tau_1 , \tau_2 ; \tau_2^{\prime} , \tau_1^{\prime} )
 = {} & \braket{{\cal{T}} \left\{ \tilde{c}^{\dagger}_{\sigma_1} ( \tau_1^{\prime}  )
 \tilde{c}^{\dagger}_{\sigma_2} ( \tau_2^{\prime}  ) 
 \tilde{c}_{\sigma_2} ( \tau_2 ) \tilde{c}_{\sigma_1} ( \tau_1 ) \right\} }
_{\rm connected} 
\nonumber\\
 ={}&\braket{ {\cal{T}} \left\{ X^{\sigma_1 0} ( \tau_1^{\prime}  )
 X^{\sigma_2 0 } ( \tau_2^{\prime}  ) 
 X^{0 \sigma_2} ( \tau_2 )  X^{0 \sigma_1} ( \tau_1 ) \right\} }_{\rm connected} \nonumber  \\
={}& 
\braket{ {\cal{T}} \left\{ X^{\sigma_1 0} ( \tau_1^{\prime}  )
	X^{\sigma_2 0 } ( \tau_2^{\prime}  ) 
	X^{0 \sigma_2} ( \tau_2 )  X^{0 \sigma_1} ( \tau_1 ) \right\} }
\nonumber\\
&	
- \braket{{\cal{T}} \left\{ X^{\sigma_1 0} ( \tau_1^{\prime}  )   X^{0 \sigma_1} ( \tau_1 ) \right\}}
   \braket{ {\cal{T}} \left\{ 
   	X^{\sigma_2 0 } ( \tau_2^{\prime}  ) 
   	X^{0 \sigma_2} ( \tau_2 ) \right\} }\nonumber\\
&
+ \braket{{\cal{T}} \left\{ X^{\sigma_1 0} ( \tau_1^{\prime}  )   X^{0 \sigma_2} ( \tau_2 ) \right\}}
   \braket{ {\cal{T}} \left\{ 
   	X^{\sigma_2 0 } ( \tau_2^{\prime}  ) 
   	X^{0 \sigma_1} ( \tau_1 ) \right\} }.
 \end{align}
We now employ Eqs.~\eqref{eq:fouriertransform} and \eqref{eq:permutationsum} to explicitly calculate the four-point function. For parallel spin ($\sigma_1 = \sigma_2 = \sigma),$ we obtain  in frequency space,
 \begin{align}
 &  \tilde{G}_0^{\tilde{c}_\sigma \tilde{c}_{\sigma} \tilde{c}^{\dagger}_\sigma \tilde{c}^{\dagger}_{\sigma} }
 ( \omega_1 , \omega_2 ; \omega_2^{\prime} , \omega_1^{\prime} )  
  =  \beta \delta_{ \omega_1^{\prime} + \omega_2^{\prime} , \omega_2 + \omega_1 }
{G}_0^{\tilde{c}_\sigma \tilde{c}_{\sigma} \tilde{c}^{\dagger}_\sigma \tilde{c}^{\dagger}_{\sigma} }
 ( \omega_1 , \omega_2 ; \omega_2^{\prime} , \omega_1^{\prime} )  
 \nonumber
 \\
  = {}& \int_0^{\beta} d \tau_1  \int_0^{\beta} d \tau_2
 \int_0^{\beta} d \tau_2^{\prime} \int_0^{\beta} d \tau_1^{\prime} 
e^{ i ( \omega_1 \tau_1 + \omega_2 \tau_2 - 
 \omega_2^{\prime} \tau_2^{\prime} - \omega_1^{\prime} \tau_1^{\prime} ) }  
\left\langle {\cal{T}} \left\{ X^{\sigma 0} ( \tau_1^{\prime}  )
 X^{\sigma 0 } ( \tau_2^{\prime}  ) 
 X^{0 \sigma} ( \tau_2 )  X^{0 \sigma} ( \tau_1 ) \right\} \right\rangle_{\rm connected}
 \nonumber
 \\
  ={}& \beta \delta_{ \omega_1^{\prime} + \omega_2^{\prime} ,  \omega_2 + \omega_1 }
 \left[ \beta \delta_{\omega_1^{\prime}  , \omega_1 } - \beta \delta_{ \omega_2^{\prime} , \omega_1 } \right]
 \frac{ n_{\bar{\sigma}} ( 1 - n_{\bar{\sigma}} )}{ (i \omega_1^{\prime} + \mu ) ( i \omega_2^{\prime} + \mu )}.
 \end{align}
Actually, this expression is only valid for vanishing magnetic field so that we should set
$n_{\uparrow } = n_{\downarrow} = n /2$.
On the other hand, for the anti-parallel spin ($\sigma_2 = - \sigma_1 = - \sigma = \bar{\sigma}),$ we obtain
\begin{align} 
& \tilde{G}_0^{\tilde{c}_\sigma \tilde{c}_{\bar{\sigma}} \tilde{c}^{\dagger}_{\bar{\sigma}} 
\tilde{c}^{\dagger}_{\sigma} }
 ( \omega_1 , \omega_2 ; \omega_2^{\prime} , \omega_1^{\prime} ) 
=   \beta \delta_{ \omega_1^{\prime} + \omega_2^{\prime} , \omega_2 + \omega_1 }    G_0^{\tilde{c}_\sigma \tilde{c}_{\bar{\sigma}} \tilde{c}^{\dagger}_{\bar{\sigma}} 
\tilde{c}^{\dagger}_{\sigma} }
 ( \omega_1 , \omega_2 ; \omega_2^{\prime} , \omega_1^{\prime} ) 
 \nonumber
 \\
 ={}&  
     \int_0^{\beta} d \tau_1  \int_0^{\beta} d \tau_2
 \int_0^{\beta} d \tau_2^{\prime} \int_0^{\beta} d \tau_1^{\prime} 
e^{ i ( \omega_1 \tau_1 + \omega_2 \tau_2 - 
 \omega_2^{\prime} \tau_2^{\prime} - \omega_1^{\prime} \tau_1^{\prime} ) }  
\left\langle {\cal{T}} \left\{ X^{\sigma 0} ( \tau_1^{\prime}  )
 X^{\bar{\sigma} 0 } ( \tau_2^{\prime}  ) 
 X^{0 \bar{\sigma}} ( \tau_2 )  X^{0 \sigma} ( \tau_1 ) \right\} \right\rangle_{\rm connected}
 \nonumber
 \\
  ={}& 
 \frac{ \beta \delta_{ \omega_1^{\prime} + \omega_2^{\prime} , \omega_2 + \omega_1 }}{
 ( i \omega_1^{\prime} + \mu )( i \omega_2^{\prime} + \mu )( i \omega_2 + \mu )
( i \omega_1 + \mu )}
\Biggl\{  \left( 1 -\frac{n}{2} \right)
 ( i \omega_1^{\prime} + i \omega_2^{\prime} + 2 \mu ) 
 \nonumber
 \\
  & \hspace{56mm}
 - \left[  \left( \frac{n^2}{4}  -x_2 \right)   \beta \delta_{ \omega_1^{\prime} , \omega_1 } 
  +    \left(  \frac{n}{2}  - x_2 \right)    \beta \delta_{ \omega_2^{\prime} , \omega_1 } 
\right]
 (i \omega_1 + \mu ) ( i \omega_2 + \mu )
\Biggr\}.
\end{align}
The corresponding initial conditions of the four-point vertices are
 \begin{align}
 \Gamma_0^{ \bar{\psi} \bar{\psi} \psi \psi }
 (  \omega_1^{\prime} \sigma , \omega_2^{\prime} \sigma ; \omega_2 \sigma , \omega_1 \sigma)
  ={} &  - \left[ \frac{ (1 - \frac{n}{2} )^4 }{ ( i \omega_1^{\prime} + \mu) ( i \omega_2^{\prime} + \mu)    
 ( i \omega_2 + \mu ) ( i \omega_1 + \mu) } 
\right]^{-1}   G_0^{\tilde{c}_\sigma \tilde{c}_{\sigma} \tilde{c}^{\dagger}_\sigma \tilde{c}^{\dagger}_{\sigma} }
 ( \omega_1 , \omega_2 ; \omega_2^{\prime} , \omega_1^{\prime} )  
 \nonumber
 \\
 ={} & -
 \beta ( \delta_{ \omega_1^{\prime}  , \omega_1 } - \delta_{ 
\omega_2^{\prime} , \omega_1 } )
 \frac{ \frac{n}{2} }{ (1 - \frac{n}{2} )^3} (i \omega_1 + \mu ) ( i \omega_2 + \mu ),
\end{align}
and
\begin{align}
  \Gamma_0^{ \bar{\psi} \bar{\psi} \psi \psi }
 (  \omega_1^{\prime} \sigma , \omega_2^{\prime} \bar{\sigma}; \omega_2 \bar{\sigma} , \omega_1 \sigma)
   ={} &   - \left[ \frac{ (1 - \frac{n}{2}   )^4 }{ ( i \omega_1^{\prime} + \mu) ( i \omega_2^{\prime} + \mu)   
 ( i \omega_2 + \mu ) ( i \omega_1 + \mu) } \right]^{-1}   G_0^{\tilde{c}_\sigma \tilde{c}_{\bar{\sigma}} \tilde{c}^{\dagger}_{\bar{\sigma}} \tilde{c}^{\dagger}_{\sigma} }
 ( \omega_1 , \omega_2 ; \omega_2^{\prime} , \omega_1^{\prime} )  
 \nonumber
 \\
 ={}& - \frac{  i \omega_1^{\prime} + i \omega_2^{\prime} + 2 \mu }{ (1 -  \frac{n}{2}  )^3}
 +    \frac{\beta}{ (1 -  \frac{n}{2}  )^4}  
 \left[  \left(   \frac{n^2}{4}   -x_2 \right)    
  \delta_{ \omega_1^{\prime} ,  \omega_1 } 
  +    \left(  \frac{n}{2}  - x_2  \right)    \delta_{ \omega_2^{\prime} , \omega_1 } 
\right] (i \omega_1 + \mu ) ( i \omega_2 + \mu ).
 \end{align}
For the special frequency combination needed in the FRG flow equation~\eqref{eq:sigmaflowt}, this reduces to
 \begin{equation}
 	\label{eq:gammapar}
 \Gamma_0^{ \bar{\psi} \bar{\psi} \psi \psi }
 (  \omega \sigma , \omega^{\prime} \sigma ; \omega^{\prime} \sigma , \omega \sigma)
  =  - \frac{ \beta  \frac{n}{2} }{ ( 1 - \frac{n}{2} )^3}  ( 1 - \delta_{\omega , \omega^{\prime}} ) ( i \omega + \mu ) ( i \omega^{\prime} + \mu ),
 \end{equation}
and 
 \begin{equation}
 	\label{eq:gammaantipar}
\Gamma_0^{ \bar{\psi} \bar{\psi} \psi \psi }
 (  \omega \sigma , \omega^{\prime} \bar{\sigma} ; \omega^{\prime} \bar{\sigma} , \omega \sigma)
  =   - \frac{ i \omega + i \omega^{\prime} + 2 \mu }{ ( 1 - \frac{n}{2} )^3 }
 + \frac{\beta}{ (1 -  \frac{n}{2}  )^4}  
 \left[  \left(   \frac{n^2}{4}   -x_2 \right)     
  +    \left(  \frac{n}{2}  - x_2  \right)    \delta_{ \omega , \omega^{\prime} } 
\right] (i \omega + \mu ) ( i \omega^{\prime} + \mu ).
 \end{equation}
For $U = \infty$ where $x_2 =0$ the four-point vertices~\eqref{eq:gammapar} and~\eqref{eq:gammaantipar} reduce to the expressions given in Eqs.~\eqref{eq:Gamma4a} and~\eqref{eq:Gamma4b} of the main text.
\end{widetext}


\begin{thebibliography}{99}
\bibitem{footnotenotation}
	In our notation the symbol $\sum_{ \langle ij \rangle }$ represents summation over district pairs of nearest neighbors, whereas in the notation used by
	Wiegmann \cite{Wiegmann88} and also by F\"{o}rster \cite{Foerster89}
	the symbol $\sum_{ij}$ counts each pair of nearest neighbors twice. 
	Hence, our $J/2$ corresponds to the $\tilde{J} $ introduced by 
	F\"{o}rster so that the supersymmetric point is at $t = \tilde{J} = J/2$.
\bibitem{footnotecanonical}
	With the definition $\epsilon_{ ab} = \epsilon_b - \epsilon_a$ the projected fermionic annihilation operator $\tilde{c}_{\sigma} ( \tau ) = X^{ 0\sigma} ( \tau )$
	introduced in Eq.~\eqref{eq:cprojected} has in the atomic limit the
	time-evolution
	$ \tilde{c}_{\sigma} ( \tau ) = e^{ (\epsilon_0 - \epsilon_{\sigma} ) \tau } \tilde{c}_{\sigma} (0) 
	= e^{ - \epsilon_{0 \sigma } \tau }  \tilde{c}_{\sigma} (0 )
	= e^{ \mu \tau } \tilde{c}_{\sigma} (0 )$. This is formally identical with the 
	time-evolution of a canonical fermionic annihilation operator $c_{\sigma} ( \tau ) = e^{ - ( \epsilon - \mu ) \tau } c_{\sigma} (0) $ which is determined by the  ``grand canonical'' Hamiltonian
	$ \sum_{\sigma} ( \epsilon - \mu ) c^{\dagger}_{\sigma} c_{\sigma}$.
\end{thebibliography}
\end{document}